\documentclass[twocolumn]{aastex631}

\hypersetup{linkcolor=cyan,citecolor=cyan,filecolor=cyan,urlcolor=cyan}

\shorttitle{SHEL: STIS}
\shortauthors{Allen et al.}
\begin{document}

\title{HST SHEL: Enabling Comparative Exoplanetology with HST/STIS}

\author[0000-0002-0832-710X]{Natalie H. Allen}\altaffiliation{NSF Graduate Research Fellow}
\affiliation{Department of Physics and Astronomy, Johns Hopkins University, 3400 N. Charles Street, Baltimore, MD 21218, USA}
\correspondingauthor{Natalie H. Allen}
\email{nallen19@jhu.edu}

\author[0000-0001-6050-7645]{David K. Sing}
\affiliation{Department of Physics and Astronomy, Johns Hopkins University, 3400 N. Charles Street, Baltimore, MD 21218, USA}
\affiliation{Department of Earth and Planetary Sciences, Johns Hopkins University, 3400 N. Charles Street, Baltimore, MD 21218, USA}

\author[0000-0001-9513-1449]{N\'estor Espinoza}
\affiliation{Space Telescope Science Institute, 3700 San Martin Drive, Baltimore, MD 21218, USA}
\affiliation{Department of Physics and Astronomy, Johns Hopkins University, 3400 N. Charles Street, Baltimore, MD 21218, USA}

\author[0000-0002-2432-8946]{Richard O'Steen}
\affiliation{Space Telescope Science Institute, 3700 San Martin Drive, Baltimore, MD 21218, USA}

\author[0000-0002-6500-3574]{Nikolay K. Nikolov}
\affiliation{Space Telescope Science Institute, 3700 San Martin Drive, Baltimore, MD 21218, USA}

\author[0000-0003-4408-0463]{Zafar Rustamkulov}
\affiliation{Department of Earth and Planetary Sciences, Johns Hopkins University, 3400 N. Charles Street, Baltimore, MD 21218, USA}

\author[0000-0001-5442-1300]{Thomas M. Evans-Soma}
\affiliation{School of Information and Physical Sciences, University of Newcastle, Callaghan, NSW, Australia}
\affiliation{Max Planck Institute for Astronomy, K\"{o}nigstuhl 17, D-69117 Heidelberg, Germany}

\author[0000-0002-1056-3144]{Lakeisha M. Ramos Rosado}
\affiliation{Department of Physics and Astronomy, Johns Hopkins University, 3400 N. Charles Street, Baltimore, MD 21218, USA}

\author[0000-0003-4157-832X]{Munazza K. Alam}
\affiliation{Space Telescope Science Institute, 3700 San Martin Drive, Baltimore, MD 21218, USA}

\author[0000-0003-3204-8183]{Mercedes L\'opez-Morales}
\affiliation{Center for Astrophysics ${\rm \mid}$ Harvard {\rm \&} Smithsonian, 60 Garden St, Cambridge, MA 02138, USA}

\author[0000-0002-7352-7941]{Kevin B. Stevenson}
\affiliation{Johns Hopkins APL, 11100 Johns Hopkins Rd, Laurel, MD 20723, USA}

\author[0000-0003-4328-3867]{Hannah R. Wakeford}
\affiliation{University of Bristol, HH Wills Physics Laboratory, Tyndall Avenue, Bristol, UK}

\author[0000-0002-2739-1465]{Erin M. May}
\affiliation{Johns Hopkins APL, 11100 Johns Hopkins Rd, Laurel, MD 20723, USA}

\author[0000-0002-9158-7315]{Rafael Brahm}
\affiliation{Facultad de Ingeniería y Ciencias, Universidad Adolfo Ibáñez, Av. Diagonal las Torres 2640, Peñalolén, Santiago, Chile}
\affiliation{Millennium Institute for Astrophysics, Chile}
\affiliation{Data Observatory Foundation, Chile}

\author[0009-0004-8891-4057]{Marcelo Tala Pinto}
\affiliation{Millennium Institute for Astrophysics, Chile}
\affiliation{Facultad de Ingeniería y Ciencias, Universidad Adolfo Ibáñez, Av. Diagonal las Torres 2640, Peñalolén, Santiago, Chile}



\begin{abstract}
The Hubble Space Telescope (HST) has been our most prolific tool to study exoplanet atmospheres. As the age of JWST begins, there is a wealth of HST archival data that is useful to strengthen our inferences from JWST. Notably, HST/STIS and its 0.3-1 $\mu$m wavelength coverage extends past JWST's 0.6 $\mu$m wavelength cutoff and holds an abundance of potential information: alkali (Na, K) and molecular (TiO, VO) species opacities, aerosol information, and the presence of stellar contamination. However, time series observations with HST suffer from significant instrumental systematics and can be highly dependent on choices made during the transit fitting process. This makes comparing transmission spectra of planets with different data reduction methodologies challenging, as it is difficult to discern if an observed trend is caused by differences in data reduction or underlying physical processes. Here, we present the Sculpting Hubble's Exoplanet Legacy (SHEL) program, which aims to build a consistent data reduction and light curve analysis methodology and associated database of transmission spectra from archival HST observations. In this paper, we present the SHEL analysis framework for HST/STIS and its low-resolution spectroscopy modes, G430L and G750L. We apply our methodology to four notable hot Jupiters: WASP-39\,b, WASP-121\,b, WASP-69\,b, and WASP-17\,b, and use these examples to discuss nuances behind analysis with HST/STIS. Our results for WASP-39\,b, WASP-121\,b, and WASP-17\,b are consistent with past publications, but our analysis of WASP-69\,b differs and shows evidence of either a strong scattering slope or stellar contamination. The data reduction pipeline and tutorials are available on Github.
\end{abstract}



\section{Introduction} \label{sec:intro}
The Hubble Space Telescope (HST) has been indispensable in our analysis of the atmospheres of exoplanets. Since the first detection of an exoplanet atmosphere \citep[][]{Charbonneau_2002}, HST has peered into dozens of atmospheres and discovered a wealth of information using the method of transmission spectroscopy \citep[e.g. ][]{Vidal-Madjar_2003, Pont_2008, Deming_2013, Kreidberg_2014, Knutson_2014, Sing_2016, Wakeford_2017, Evans_2017, Bourrier_2018, Benneke_2019, Wakeford_2020, Mikal_Evans_2023}, in which small changes in the opacity of a planet as a function of wavelength during transit allow for the characterization of a planet's atmosphere. However, now that we have characterized a sufficiently large sample of individually interesting planets, the next chapter in exoplanet exploration is comparative exoplanetology -- a more holistic view of exoplanets and their atmospheres through population-wide characteristics and trends. By carrying out comparative studies for multiple planets, we can begin to discern underlying population-level atmospheric trends and unravel the mysteries of these faraway worlds. \\

Due to the small atmospheric signals in transit (typically tens to hundreds of parts per million) produced by changes in opacity in an exoplanet atmosphere, the data reduction and light curve analysis techniques play a significant role in the outcome of exoplanet atmosphere studies. Different ways of reducing and analyzing data can yield significantly different results for a single planet, much less different planets. This is especially true for HST, which has considerable instrumental systematics and whose orbital configuration results in gaps in time series observations due to Earth occultations. Both of these factors make interpretation of transmission spectra from HST sensitive to analysis technique and model parameters. Therefore, to make meaningful population-level inferences about exoplanet atmospheres, it is critical to have uniform and homogeneous data reduction techniques for exoplanet transit observations to mitigate spectral analysis-driven differences in a given transmission spectrum. This is the motivation behind the Sculpting Hubble's Exoplanet Legacy program (SHEL; HST Proposal 16634, PI David Sing). The goal of SHEL is to analyze all archival transit observations of exoplanets with HST. There have been other efforts towards uniform comparative exoplanetology in the past \citep[i.e.][]{Sing_2016, Tsiaras_2018, Changeat_2022, Estrela_2022}, but SHEL is the first effort using  multiple instruments, directly and uniformly rederived prior system parameters through custom joint fits of available photometry and radial velocity (RV) data, and updated handing of systematics, e.g. enabling the use of %handling using modern techniques and advances that allow for more complete use of the data, especially 
the first orbit which has historically been discarded. \\ 

%As the name implies, the timing for comparative exoplanetology with HST is especially apt as we enter the new domain of exoplanet atmosphere observations with JWST. \\

%The first instrument that we focus on is 
Here we present the first results for the Space Telescope Imaging Spectrograph (STIS). In particular, we focus on the low resolution modes STIS/G430L and STIS/G750L, which cover $\sim 0.29-0.57 \mu$m and $\sim 0.53-1.0 \mu$m, respectively, at a resolving power of R $\simeq 500$. Note that the wavelength coverage of these two gratings overlap, such that it can be checked for any misalignment or offset between the data products. This wavelength range provides information on a variety of important processes and opacity sources in exoplanet atmospheres, such as clouds and hazes, atomic features like alkali metals Na {\sc i} and K {\sc i}, and molecules like the exotic TiO and VO at high temperatures. These wavelengths are also strongly affected by the presence of stellar contamination \citep{Sing_2011, Pont_2013, Rackham_2018, Rackham:2019}, and can be used to identify systems where this effect is significant. This wavelength range overlaps with multiple JWST NIRSpec modes, notably PRISM ($\sim 0.6-5.3 \mu$m), and NIRISS/SOSS ($\sim 0.6-2.8 \mu$m). This allows the two observatories to work in perfect synergy, as the overlap region can again be used to check consistency between observations. The first look from the Transiting Exoplanet Early Release Science team has shown that HST and JWST show close agreement \citep{Feinstein_2023, Rustamkulov_2023}, such that data from HST can confidently be used with that from JWST even for modes without wavelength overlap. Therefore, observations from this wavelength region will be in high demand as new observations of previous HST targets are observed. \\

Over the past few decades, we have built up a considerable archive of exoplanet transit and RV observations, especially for the most popular targets such as those amenable for atmospheric characterization with space-based telescopes. When considered simultaneously, these observations constrain an exoplanet system's orbital and planetary parameters much more precisely than when fitting individual observations. Since the time series gaps in HST transit observations can make the determination of some of these system parameters difficult, the constraints from these additional observations are useful to increase our inference ability with the HST observations. Therefore, we also present our joint fit framework, which simultaneously fits all available high quality transit and RV observations (see \autoref{tab:sources} for the data used for our example fits) to uniformly derive the system parameters for all of our SHEL targets. These derived parameters are then available for use when fitting for the HST transit depths, and allows us to more accurately fit the HST observations than when just considering the HST data or taking constraints from single studies in the literature. \\

This paper is presented as follows. We begin with a description of the data used for testing our analysis methodology in \autoref{sec:data}. We describe our joint fits of photometric and RV data to rederive the parameters of each system in a uniform manner in \autoref{sec:joint_fits}, and our data reduction pipeline in \autoref{sec:stis_pipeline}. We discuss our light curve fitting methods in \autoref{sec:lc_fit}, including a brief summary of our extensive systematics method testing (which is explained more thoroughly in \autoref{app:systematics}). Then, we show an in-depth example result using our method in \autoref{sec:w69b} for WASP-69\,b. We detail some lessons learned during these tests in \autoref{sec:lessons} and finish with our conclusions in \autoref{sec:conclusion}.\\

\section{Data used for testing}\label{sec:data}
As the main purpose of this work is to determine a robust data reduction and analysis methodology using the most up-to-date methodologies for transmission spectroscopy observed with HST/STIS, we use a few notable planets to build, test, and refine our methods: WASP-39\,b, WASP-121\,b, WASP-69\,b, and WASP-17\,b. These targets were chosen as they are generally representative of the type of data obtained by HST/STIS for Hot Jupiters, have all had an analysis of their HST/STIS spectrum previously published, and are all JWST Cycle 1 GTO, ERS and/or GO targets. Here, we briefly describe these planets and their previous space-based transmission spectroscopy studies. \\

WASP-39\,b is a hot gas giant ($R=1.27\,\,R_J, \,\, M=0.28 \,\, M_J, \,\, T_{eq}=1116 \,\, K$) in a 4.055 day orbit around a late G-type star \citep{faedi_2011}. Observations of WASP-39\,b were taken as part of HST ID 12473 (PI David Sing). Three transits were observed with STIS's low resolution capabilities, two with the STIS/G430L grating and one with the STIS/G750L grating, on 09 Feb 2014, 13 Feb 2014, and 17 Mar 2014. These STIS data have been previously reduced in \citet{Sing_2016}, and were restudied and put in context with additional HST/WFC3 data in \citet{Wakeford_2018}. The planet is exceptionally well-characterized, including as the subject of the Transiting Exoplanet Early Release Science Program, in which its atmosphere was observed with four JWST modes \citep{Rustamkulov_2023, Alderson_2023, Feinstein_2023, Ahrer_2023}, and was additionally observed with MIRI through DD 2783 (PI Diana Powell).  \\

WASP-121\,b is an inflated gas giant ($R=1.753 \,\, R_J, \,\, M = 1.183 \,\, M_J$) that lies firmly in the ultrahot planet regime ($T_{eq} \simeq 2400 \,\, K$) due to its short 1.27 day orbit around a F6 star \citep{Delrez_2016}. Observations of WASP-121\,b were taken as part of the HST PanCET program (ID 14767, PI David Sing). Three transits were observed with STIS's low resolution capabilities, two with the STIS/G430L grating and one with the STIS/G750L grating, on 24 Oct 2016, 06 Nov 2016, and 12 Nov 2016. These STIS data were previously published in \citet{Evans_2018}, and WASP-121\,b has since been the target of a JWST phase curve with NIRSpec/G395H \citep{Mikal_Evans_2023}, which included the coverage of one transit and two eclipses, and will have an additional JWST NIRISS/SOSS phase curve as part of GTO 1201 (PI David Lafreniere) and eclipse with JWST MIRI/LRS with GO 2961 (PI Paul Molliere). \\

WASP-69\,b is an inflated warm gas giant ($R=1.057 \,\, R_J, \,\, M = 0.26 \,\, M_J, \,\, T_{eq} = 963 \,\, K$) in a 3.87 day orbit around a K5 star \citep{Anderson_2014}. Observations of WASP-69\,b were taken as part of the HST PanCET program (ID 14767, PI David Sing). Three transits were observed with STIS's low resolution capabilities, two with the STIS/G430L grating and one with the STIS/G750L grating, on 14 May 2017, 26 May 2017, and 04 Oct 2017. These STIS data were previously published in \citet{Estrela_2021}. It will be observed in eclipse with JWST MIRI/LRS and two NIRCam modes through GTO 1185 (PI Thomas Greene), and in transit with MIRI/LRS and NIRSpec/G395H through GO 3712 (PI Patricio Cubillos). \\

WASP-17\,b is a low density hot gas giant ($R=1.991\,\,R_J, \,\, M = 0.486 \,\, M_J, \,\, T_{eq} = 1770 \,\, K$) in a retrograde 3.7 day period around a F6 star \citep{Anderson_2010, Triaud_2010}. Observations of WASP-17\,b were taken as part of HST ID 12473 (PI David Sing). Three transits were observed with STIS's low resolution capabilities, two with the STIS/G430L grating and one with the STIS/G750L grating, on 08 Jun 2012, 15 Mar 2013, and 19 Mar 2013. These STIS data were also originally analyzed in \citet{Sing_2016}, and were put into full HST and Spitzer context in \citet{Alderson_2022}. WASP-17\,b will be heavily studied with observations in transit and eclipse with MIRI/LRS, NIRISS/SOSS, and NIRSpec/G395H through GTO 1353 (PI Nikole Lewis), for which the first observation in transit with MIRI/LRS discovered evidence for quartz (SiO$_2$) clouds \citep{Grant_2023}. \\

\section{Joint Fits} \label{sec:joint_fits}
%TESS has been transformative for the analysis of exoplanet systems, both in discovery and characterization. For our purpose in analyzing HST transmission spectroscopy observations, TESS can play a crucial role by providing high quality, high cadence transit observations of systems all across the sky, which allows for a precise determination of system parameters. Due to the gaps in the time series data, these parameters are often not well constrained through HST observations alone. As the choice of priors for these system parameters can be important for the result of transmission spectroscopy transit fits, taking advantage of this data from a variety of observatories is critical in building our uniform analysis methodology. \\

The first step of SHEL consists of rederiving system parameters in a uniform and precise way. For that, we perform joint fits of all available high quality, well-documented photometric transit and RV data. This is a combination of sources from the literature and data from the publicly available TESS archive. We carry out these fits using \textit{juliet} \citep{juliet}. For $p=R_p/R_s$, impact parameter $b$, period $P$ and time of transit center $T_0$, we put priors restricted to those from recent literature sources, although the radius ratio $p$ is fit for separately from each instrument and allowed to vary 10\%. We also fit for limb darkening parameters $q_1$ and $q_2$ \citep[the quadratic limb darkening law parameterization from][]{Kipping_2013}, for which we set a uniform prior from 0 to 1 and fit individually for each instrument. We fit for the RV amplitude $K$ and systemic RV per instrument $\mu$ with wide priors appropriate to the data. We also fit for orbital eccentricity with the parameterization $\sqrt{e} \,\, cos(\omega \pi/180)$ and $\sqrt{e} \,\, sin(\omega \pi/180)$\footnote{$secosomega$ and $sesinomega$, respectively, in \textit{juliet}} for eccentricity $e$ and argument of periastron $\omega$. These are defined between -1 and 1, which we set as the prior range, and we use them based on the discussion in \cite{Eastman_2013} and \cite{juliet}. If the fit eccentricity values are consistent within $3\sigma$ with 0, we will assume the orbit is circular when using these priors with the HST data. Of the four targets presented as examples in this work, WASP-39\,b, WASP-121\,b, and WASP-17\,b are fully consistent with circular orbits according to this criterion, while WASP-69\,b is on the borderline, but as previous in-depth studies have shown it has a circular orbit \citep[e.g.][]{Wallack_2019}, we adopt the same for our analysis. We note that we do not include eclipse observations in our joint fits, as our short period hot Jupiter targets of interest are largely on approximately circular orbits \citep[see e.g.][with further evidence from our own joint fit results]{Fortney_2021}, and thus the additional constraint on eccentricity from eclipse timing is unlikely to considerably change the determination between circular or elliptical orbits in our study. However, we note that our framework can be easily adapted to add in eclipse observations and their associated parameter constraining ability, especially for considering targets with known non-circular orbits.  \\

Any two of $P$, the stellar density $\rho$, and the scaled semi-major axis $a/R_s$ can be used to solve for the third, as $\rho$ can be combined with $P$ and Kepler's third law to obtain $a/R_s$ and vice-versa. $\rho$ can be defined precisely using the methodology from \citet{brahm1, brahm2}, and so if we have priors from this technique available we fit for $P$ and $\rho$, but if stronger priors are available on $a/R_s$, we use that and $P$ instead. If photometric data are available in a non-detrended state, we detrend the data in the same way as it is described in the literature. For TESS data, we detrend with respect to time using a Gaussian Process (GP) with an approximate Matern kernel from \textit{celerite} \citep{celerite}. This has two associated parameters, $GP\_sigma$ which is the amplitude of the GP, and $GP\_rho$ which is the time/length-scale of the GP. We follow \cite{PatelEspinoza2022}, and fit those hyperparameters using out-of-transit data, and then fix those for the joint fits using only the in-transit data. For datasets that are detrended with more than one regressor, we instead use a multi-dimensional squared-exponential kernel from \textit{george} \citep{hodlr}. This has the same amplitude parameter $GP\_sigma$ as the previous kernel, and then one additional parameter $GP\_alpha$ for each regressor, which is the inverse squared length-scale of said parameter's component of the GP. All GP parameters are allowed to vary over many orders of magnitude to not constrain their fitting (i.e. log uniform from 1e-6 to 1e6). Finally, each transit gets an extra parameter $mflux$, which is the offset relative flux (normal distribution around 0 with $\sigma = 0.1$), and each instrument (transit or RV) gets an extra jitter term $sigma\_w$ (log uniform between 0.1 and 10,000 for transits, and between 0.001 and 100 for RVs). \\

\begin{figure*}
    \centering
    \includegraphics[width=\textwidth]{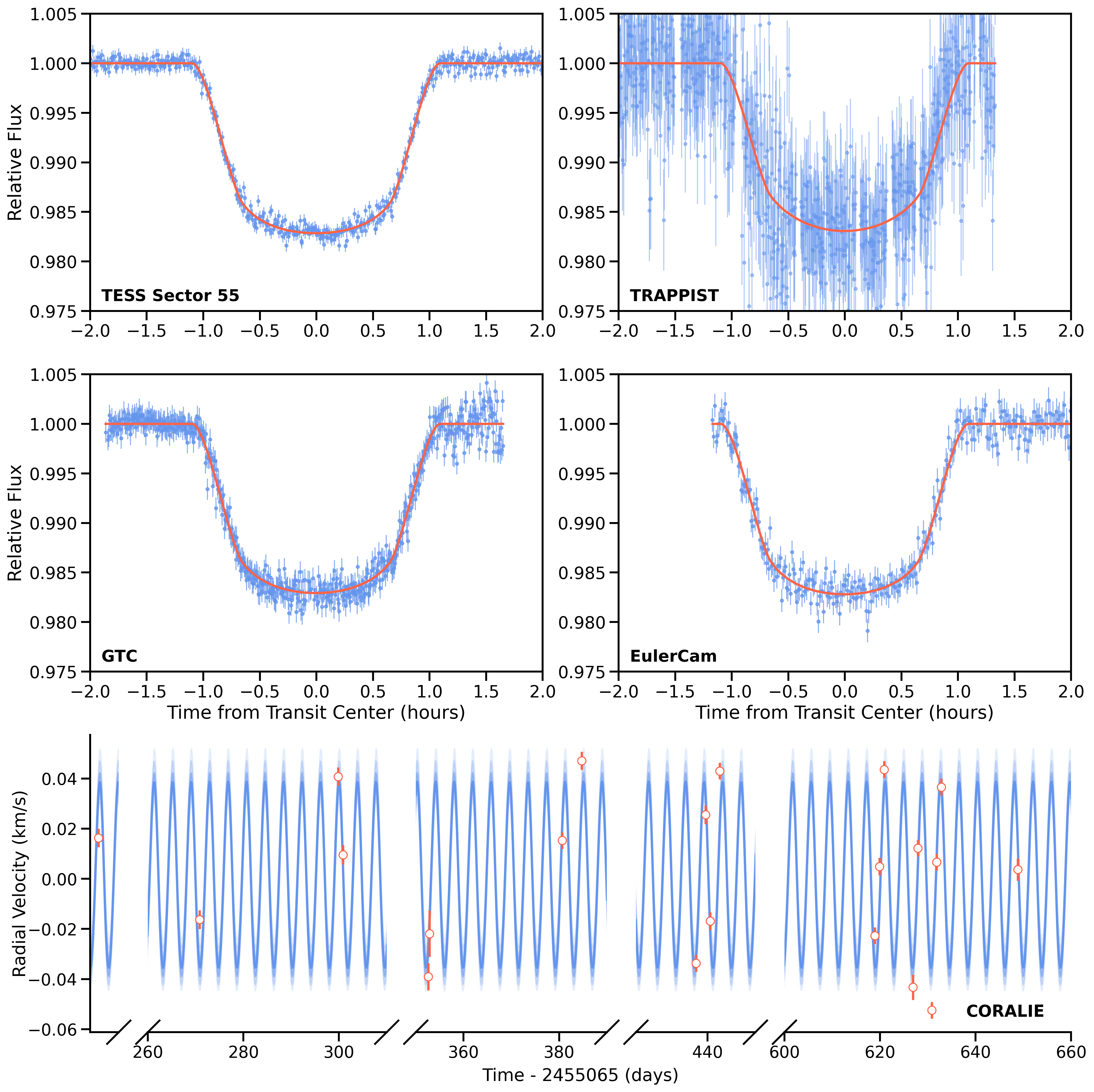}
    \caption{An example of our joint fit for WASP-69\,b. All the priors used for this fit, and their corresponding best-fit values, are given in \autoref{tab:joint_fit}. }
    \label{fig:joint}
\end{figure*}

To show an example of how we carry out these fits, we detail here the joint fits for our four test planets. WASP-39\,b's joint fit is unique, as we simply use the result of the joint fit from Carter \& May et al., (submitted), which jointly fit the JWST white light curves from NIRSpec/PRISM \citep{Rustamkulov_2023}, NIRSpec/G395H \citep{Alderson_2023}, NIRISS/SOSS \citep[Orders 1 and 2, ][]{Feinstein_2023}, and NIRCam/F322W2 and F210M \citep{Ahrer_2023}, photometric data from NGTS and TESS Sector 51 \citep{Ahrer_2023}, and the RVs from SOPHIE and CORALIE \citep{faedi_2011}, but has somewhat different priors and methods from the other fits as a result (notably, this is the only fit that assumes zero eccentricity, and also uses linear systematics detrending, shown as $theta$ in the table). For WASP-121\,b, we jointly fit the transit portion of the JWST NIRSpec/G395H phase curve \citep[NRS1 and NRS2, ][]{Mikal_Evans_2023}, TESS Sectors 7, 33, 34, and 61, and RVs from CORALIE and HARPS \citep{Bourrier_2020}. For WASP-69\,b, we jointly fit the GTC/OSIRIS transit \citep{Murgas_2020}, TESS Sector 55, transits from TRAPPIST and EulerCam \citep{Anderson_2014}, and RVs from CORALIE \citep{Anderson_2014}. For WASP-17\,b, we jointly fit the transit from EulerCam \citep{Anderson_2010}, transits from the 1.5 m Danish Telescope at ESO La Silla \citep{Southworth_2012}, TESS Sectors 12 and 38, and RVs from CORALIE \citep{Anderson_2010}, HARPS \citep{Anderson_2010}, and MIKE \citep{Bayliss_2010}. \autoref{tab:sources} summarizes all of the observations used in our joint fits. The priors and fit parameter values for our test planets are given in \autoref{tab:joint_fit}, and our joint fit for WASP-69\,b is shown as an example in \autoref{fig:joint}. A database containing our fits for planets observed with HST as completed, along with all the data used as inputs for the fits, is available on Github\footnote{\url{https://github.com/rosteen/SHEL_project/}}.\\

%  Total, the WASP-39 b is a x parameter fit, WASP-121 b is a 45 parameter fit, WASP-69 b is a 39 parameter fit, and WASP-17 b is a x parameter fit.

\begin{deluxetable*}{c l c c}
    \tablecaption{Observations used for our joint fits. Sources are provided where we use previously published data. All TESS data was downloaded directly from the TESS archive.\label{tab:sources}}
    \tablehead{ \colhead{Planet} &
    \colhead{Observatory} & \colhead{Type} & \colhead{Source}}
    \startdata
    WASP-39\,b & JWST | NIRSpec/PRISM & Transit & \citet{Rustamkulov_2023} \\
    & JWST | NIRSpec/G395H & Transit & \citet{Alderson_2023} \\
    & JWST | NIRISS/SOSS & Transit & \citet{Feinstein_2023} \\
    & JWST | NIRCam/F322W2 and F210M & Transit & \citet{Ahrer_2023} \\
    & NGTS & Transit & \citet{Ahrer_2023} \\
    & TESS (Sector 51) & Transit & -- \\
    & OHP 1.93-m | SOPHIE & RV & \citet{faedi_2011} \\
    & Euler 1.2-m | CORALIE & RV & \citet{faedi_2011} \\
    \hline
    WASP-121\,b & JWST | NIRSpec/G395H & Transit & \citet{Mikal_Evans_2023} \\
    & TESS (Sectors 7, 33, 34, 61) & Transit & -- \\
    & Euler 1.2-m | CORALIE & RV & \citet{Bourrier_2020} \\
    & ESO 3.6-m | HARPS & RV & \citet{Bourrier_2020} \\
    \hline
    WASP-69\,b & GTC | OSIRIS & Transit & \citet{Murgas_2020} \\
    & TESS (Sector 55) & Transit & -- \\
    & TRAPPIST & Transit & \citet{Anderson_2014} \\
    & EulerCam & Transit & \citet{Anderson_2014} \\
    & Euler 1.2-m | CORALIE & RV & \citet{Anderson_2014} \\
    \hline
    WASP-17\,b & EulerCam & Transit & \citet{Anderson_2010} \\
    & Danish 1.5-m & Transit & \citet{Southworth_2012} \\
    & TESS (Sectors 12 and 38) & Transit & -- \\
    & Euler 1.2-m | CORALIE & RV & \citet{Anderson_2010} \\
    & ESO 3.6-m | HARPS & RV & \citet{Anderson_2010} \\
    & Magellan Clay | MIKE & RV & \citet{Bayliss_2010} \\
    \enddata

\end{deluxetable*}

\begin{deluxetable*}{c c c c c}
    \tabletypesize{\footnotesize}
    \tablecaption{Details of our joint fits for WASP-39\,b, WASP-121\,b, WASP-69\,b, and WASP-17\,b, both priors and final best-fit values. Priors given as $N(\mu,\sigma^2)$ are a normal distribution with mean $\mu$ 
    and variance $\sigma^2$, $U(a,b)$ are a uniform distribution between $a$ and $b$, and $L(a,b)$ are a log-uniform distribution between log($a$) and log($b$). Full posterior results can be found on Github. \label{tab:joint_fit}}
    \tablehead{ \colhead{ } &
    \colhead{WASP-39\,b} & \colhead{WASP-121\,b} & \colhead{WASP-69\,b} & \colhead{WASP-17\,b}}
    \startdata
     \multicolumn{5}{c}{\textbf{Priors}} \\
     \hline
     \multicolumn{5}{l}{\textbf{System Parameters}} \\
     $P_1$ [d] & $N(4.05527655,0.7\text{e-}^2)$ & $N(1.274925,2\text{e-}7^2)$ & $N(3.86814,2\text{e-}6^2)$ & $N(3.73543, 8\text{e-}6^2)$ \\
     $T_{0,1}$ [BJD] & $N(2459791.615201,0.1^2)$ & $N(2458119.72074,2\text{e-}4^2)$ & $N(2457176.17789,2\text{e-}4^2)$ & $N(2454592.80154, 5\text{e-}5^2)$ \\
     $b_1$ & $N(0.45, 0.1^2)$ & $N(0.1,0.05^2)$ & $N(0.69,0.05^2)$ & $N(0.4, 0.04^2)$ \\
     $(a/R_s)_1$ & - & $N(3.81, 0.01^2)$ & $N(12.0, 0.5^2)$ & - \\
     $\rho_1$ & $L(100.,10000)$ & - & - & $N(527.695, 83.292^2)$ \\
     $K_1$ [m/s] & $U(0, 200)$ & $U(-1000, 1000)$ & $U(-1000, 1000)$ & $U(-600, 600)$ \\
     $\sqrt{e} \,\, sin(\omega \pi/180)$ & - & $U(-1.0,1.0)$ & $U(-1.0,1.0)$ & $U(-1.0,1.0)$ \\
     $\sqrt{e} \,\, cos(\omega \pi/180)$ & - & $U(-1.0,1.0)$ & $U(-1.0,1.0)$ & $U(-1.0,1.0)$ \\
     \multicolumn{5}{l}{\textbf{Transit Instrument Specific Parameters}$^a$} \\
     $p_1$ & $U(0.0,0.3)$ & $N(0.12,0.012^2)$ & $N(0.13,0.013^2)$ & $N(0.13,0.013^2)$ \\
     $q_1$ & $U(0.0,1.0)$ & $U(0.0,1.0)$ & $U(0.0,1.0)$ & $U(0.0,1.0)$ \\
     $q_2$ & $U(0.0,1.0)$ & $U(0.0,1.0)$ & $U(0.0,1.0)$ & $U(0.0,1.0)$ \\
     $mflux$ & $N(0.0,0.1^2)$ & $N(0.0,0.1^2)$ & $N(0.0,0.1^2)$ & $N(0.0,0.1^2)$ \\
     $sigma\_w$ & $L(0.1,10000)$ & $L(0.1,10000)$ & $L(0.1,10000)$ & $L(0.1,10000)$ \\
     $GP\_sigma^b$ & $L(1\text{e-}3,1\text{e}2)$ & $L(1\text{e-}6,1\text{e}6)$ & $L(1\text{e-}6,1\text{e}6)$ & $L(1\text{e-}6,1\text{e}6)$\\
     $GP\_rho^b$ & $L(1\text{e-}2,1\text{e}2)$ & $L(1\text{e-}3,1\text{e}3)$ & $L(1\text{e-}3,1\text{e}3)$ & $L(1\text{e-}3,1\text{e}3)$\\
     $GP\_alpha^c$ & - & - & $L(1\text{e-}3,1\text{e}3)$ & -\\
     $theta^d$ & $U(-10,10)$  & - & - & - \\
     \multicolumn{5}{l}{\textbf{RV Instrument Specific Parameters}$^e$} \\
     $\mu$ [m/s] & $U(-100, 100)$ & $U(-100, 100)$ & $U(-200, 200)$ & $U(-120, 120)$ \\
     $sigma\_w$ & $L(1,100)$ & $L(0.001,100)$ & $L(0.001,100)$ & $L(0.001,100)$ \\
     \hline
     \multicolumn{5}{c}{\textbf{Fit Results}$^f$} \\
     \hline
     $P_1$ [d] & $4.0552842518\pm{3\text{e-}6}$ & $1.2749248114 \pm 4\text{e-}8$ & $3.8681386499^{+3\text{e-}7}_{-2\text{e-}7} $ & $3.7354858282\pm 1\text{e-}7$ \\
     $T_{0,1}$ [BJD] & $2459791.6120684328\pm 1\text{e-}5$ & $2458119.7207217668 \pm 5 \text{e-}5 $ & $2457176.1776311793^{+1\text{e-}4}_{-2\text{e-}4}$ & $2454592.8011189527 \pm 1\text{e-}4$ \\
     $b_1$ & $0.450 \pm 0.002$ & $0.189^{+0.011}_{-0.012}$ & $0.682^{+0.007}_{-0.006}$ & $0.268 \pm 0.044$ \\
     $(a/R_s)_1$ & $11.391 \pm 0.012$ & $3.811\pm 0.008$ & $11.796^{+0.193}_{-0.171}$ & $7.050^{+0.122}_{-0.160}$ \\
     $\rho_1$ & $1699.955^{+5.454}_{-5.465}$ & $644.257^{+4.024}_{-4.256}$ & $2075.348^{+103.573}_{-88.710}$ & $475.128^{+25.180}_{-31.525}$ \\
     $K_1$ [m/s] & $36.18^{+4.49}_{-4.37}$ & $180.69^{+10.72}_{-10.87}$ & $37.40^{+2.70}_{-2.79}$ & $53.92^{+2.91}_{-2.80}$ \\
     $e_1$ & - & $0.04\pm 0.02$ & $0.06\pm 0.02$ & $0.03 \pm 0.02$\\
     $\omega_1$ & - & $22.39^{+8.22}_{-20.91}$ & $34.25^{+21.71}_{-18.23}$ & $85.65^{+29.41}_{-51.56}$ \\
     \hline 
     \multicolumn{5}{c}{\textbf{Final Prior on \textit{p}}$^g$} \\
     \hline
      & $N(0.14,0.05^2)$ & $N(0.12,0.05^2)$ & $N(0.13,0.05^2)$ & $N(0.12,0.05^2)$ \\
     \enddata
     \tablenotetext{a}{Each of these transit instrument specific parameters is defined for each instrument and campaign in the fit, although the $q_1$ and $q_2$ are shared between common observations using the same instrument (i.e. all TESS data). The instruments used for each fit are given in in \autoref{sec:joint_fits}.}
     \tablenotetext{b}{Approximate Matern kernel is used to detrend: WASP-39\,b -- NIRCam/F322W2 and F210M, NIRISS/SOSS Orders 1 and 2, and TESS sector 51; WASP-121\,b -- TESS sectors 7, 33, 34, and 61; WASP-69\,b -- TESS sector 55, EulerCam, and TRAPPIST; WASP-17\,b -- TESS sectors 12 and 38, EulerCam, and Danish 1.5 m.}
     \tablenotetext{c}{Multi-dimensional squared-exponential kernel is only used for the WASP-69\,b fit, for the GTC data with 3 regressors.}
     \tablenotetext{c}{Linear detrending is only used for the WASP-39\,b fit, for which two parameters are used for each of the the NIRSpec/G395H detector's data, and one parameter is used for the NIRSpec/PRISM data.}
     \tablenotetext{e}{Each of these RV instrument specific parameters is defined for each instrument and campaign in the fit. The instruments used for each fit are given in \autoref{sec:joint_fits}.}
     \tablenotetext{f}{Only the results for the system parameters are shown in the Fit Results---all others are available on Github.}
     \tablenotetext{g}{We leave much larger prior range on the radius ratio $p$ of $\pm 0.05$ in our STIS fits than the results of our joint fit actually suggest in order to capture the variations in transit depth that indicate opacity sources in the planetary atmosphere. We use the average of all the separately fit \textit{p} values as our normal distribution's mean value. }
\end{deluxetable*}

\section{STIS Data Reduction Pipeline}\label{sec:stis_pipeline}
Our STIS pipeline is broadly adapted from an extensively used IDL data reduction pipeline \citep[i.e.][]{Sing_2011, Nikolov_2014, Sing_2016}. In this section we will go through the steps in which we clean and extract the spectrum. This methodology is tested using STIS data from our four test exoplanets detailed above. The pipeline is publicly available on Github\footnote{\url{https://github.com/natalieallen/stis_pipeline}}. \\

Starting with the flat-fielded products as downloaded from MAST (*\_flt.fits), which contain each HST orbit's data, we extract each of the time series science exposures (header ``SCI"), along with their associated error (header ``ERR") and data quality (header ``DQ") frames. If the orbit contains any 1 second exposures, which are sometimes taken first in a given time series in an attempt to minimize systematics \citep{Sing_2015}, those are ignored in this step, as they are not taken for science purposes. At this step, we also extract the information from each time series' associated engineering jitter file (*\_jit.fits), which is the spacecraft pointing data averaged over 3 second intervals\footnote{For information on each of the HST/STIS file types used, see \url{https://hst-docs.stsci.edu/stisdhb/chapter-2-stis-data-structure/2-2-types-of-stis-files}}. This is much more finely sampled than our science files, so we take the average of each jitter vector included in this file over the time span of our science exposure. There are sometimes spurious large values included in these files as an error, so our pipeline automatically finds and averages over points with a value greater than $10^{30}$ prior to taking the full time average. \\

Now that we have our time series exposures, we begin to clean the data. First, we do a quick trace of the center of the spectrum as a function of wavelength. This is just to use as an input to the cleaning function, and is imperfect since the data are still raw and thus has outliers that affect the tracing algorithm, but is sufficient for our needs. We find the centroid of the spectrum in each column and then fit the centroid values with a second order Chebyshev polynomial to smooth it. We note that the residuals between the centroid values and the second order Chebyshev polynomial are normally distributed, and thus we are confident the fit is appropriate for the trace. This trace is done for each time series exposure. The cleaning function itself is split into multiple steps, and are done in subsets of 3-5 exposures in each orbit. First, using the DQ frames, we mark all pixels in each exposure with a flag of 16, which are ``pixels having dark rate $> 5 \sigma$ times the median dark level"\footnote{For information on the data quality flags, see \url{https://hst-docs.stsci.edu/stisdhb/chapter-2-stis-data-structure/2-5-error-and-data-quality-array}}, and are the major data quality flag of note in the STIS data. This steps accounts for the pixels for which we should doubt the output values due to previous calibration data. \\

The second step, which is why we do the cleaning in subsets, uses difference images. In this method, a ``difference image" for an exposure is created by subtracting each of the other exposures in a temporally-local subset of frames from the target frame, and then taking the median of this set of subtracted frames \citep[for a more in-depth discussion of this technique, see][]{Nikolov_2014}. Assuming equal flux from the science target between exposures, this difference image should purely be due to spurious signals like cosmic rays causing excess flux. In this difference image, we reject values along each row in windows of 20 pixels greater than $5 \sigma$ from the median window value. \\

While the difference images account for changes in time, problems in the detector that are constant in time like hot and cold pixels will not be seen. Thus, we do an additional step to find and mark these cases. We do this ourselves rather than relying on DQ flags, as our level of tolerance for this behavior may be more strict than that of the automatic pipeline. To do this, we fit a spline to each column of each exposure to build a smoothed frame. We then create a 5x5 box around each pixel, take the median values of the spline image in this box, and reject the center pixel if it is greater than $3 \sigma$ from this value. \\

The final step is to find any potential problematic pixels that have been missed thus far in a spline-fitting cleaning. We again fit a spline to each column in each exposure to create our smoothed frame. Then, for each column, we take the normalized average of the surrounding 2 columns in each direction, and then reject values greater than a given $\sigma$ away from this median spline along the column of interest, where $\sigma$ is the standard deviation of the residuals between the spline and the column. Due to how highly peaked the STIS spectrum is, the spline has a difficult time with fitting the peak of the spectrum but not fitting any potential bad pixels' profile (which makes this cleaning useless if it occurs). Thus, we set different $\sigma$ reject cutoffs depending on the distance of the pixel from the spectral center trace. For most of the frame, there is a $3 \sigma$ cutoff, but the center 3-4 pixels is $2\mathrm{x}= 6 \sigma$, and the center 1-2 pixels is $4\mathrm{x}= 12 \sigma$. \\

Now that we have marked all of the ``bad" pixels, we proceed to replace them. To do this, for a given marked pixel we simply take the average of the neighboring pixels in the same row in the surrounding columns. We note that this assumes that the trace is not tilted locally, which we are confident is a safe assumption as our trace smoothly varies over less than a pixel in y value across the length of the detector in all observations of our test planets. If one of those pixels is also marked, we account for this by taking the average of the two previous pixels in the same row instead. In addition, there are a few columns in the detector that contain dozens of marked bad pixels from the DQ frame that can be seen by eye in the image. For these columns, we overwrite the entire column with the average of the surrounding columns. We note that while replacing flagged pixels by interpolation is not strictly correct, it is widely used by the community for these types of spectroscopic analyses \citep[e.g.][]{Nikolov_2014, Tamburo_2018, Alam_2020}. In addition, the low level of bad pixels, especially in high flux regions of the detector, means that our analysis is not sensitive to our choice of bad pixel replacement method. We verified this by testing the light curves obtained by instead replacing the bad pixels not associated with DQ flags by the value of that pixel in the median frame used for the difference images method for one of our datasets, and replacing the DQ flag pixels with NaN values (since their median frame values will always be ``bad" as well), for which the resulting normalized light curves were consistent to within errors. However, we always suggest visual investigation of the data, and in the case of datasets with a large number of marked bad pixels, it is possible that the method of pixel replacement should be investigated and considered more deeply.\\

\begin{figure*}
\gridline{\fig{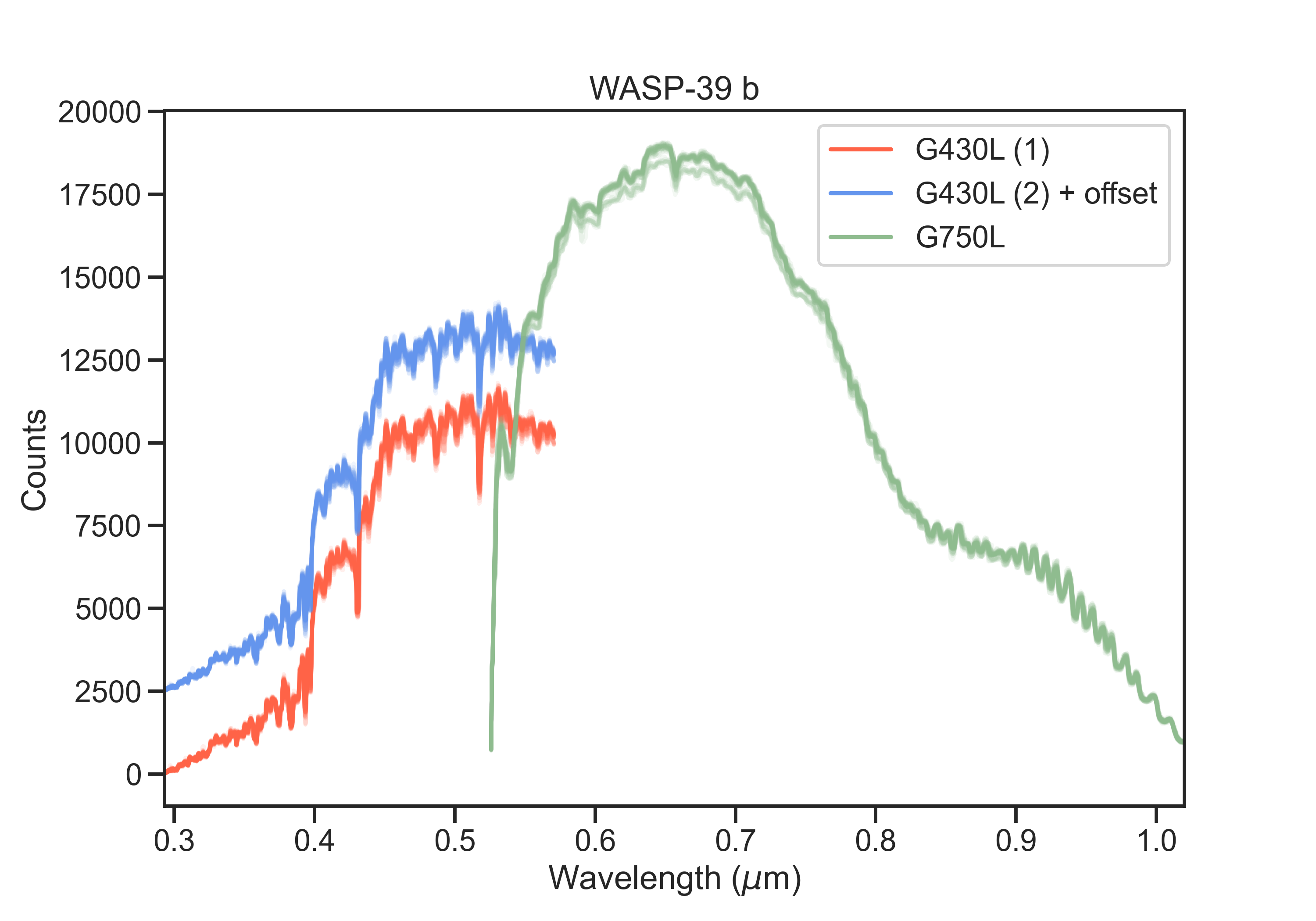}{0.49\textwidth}{}
\fig{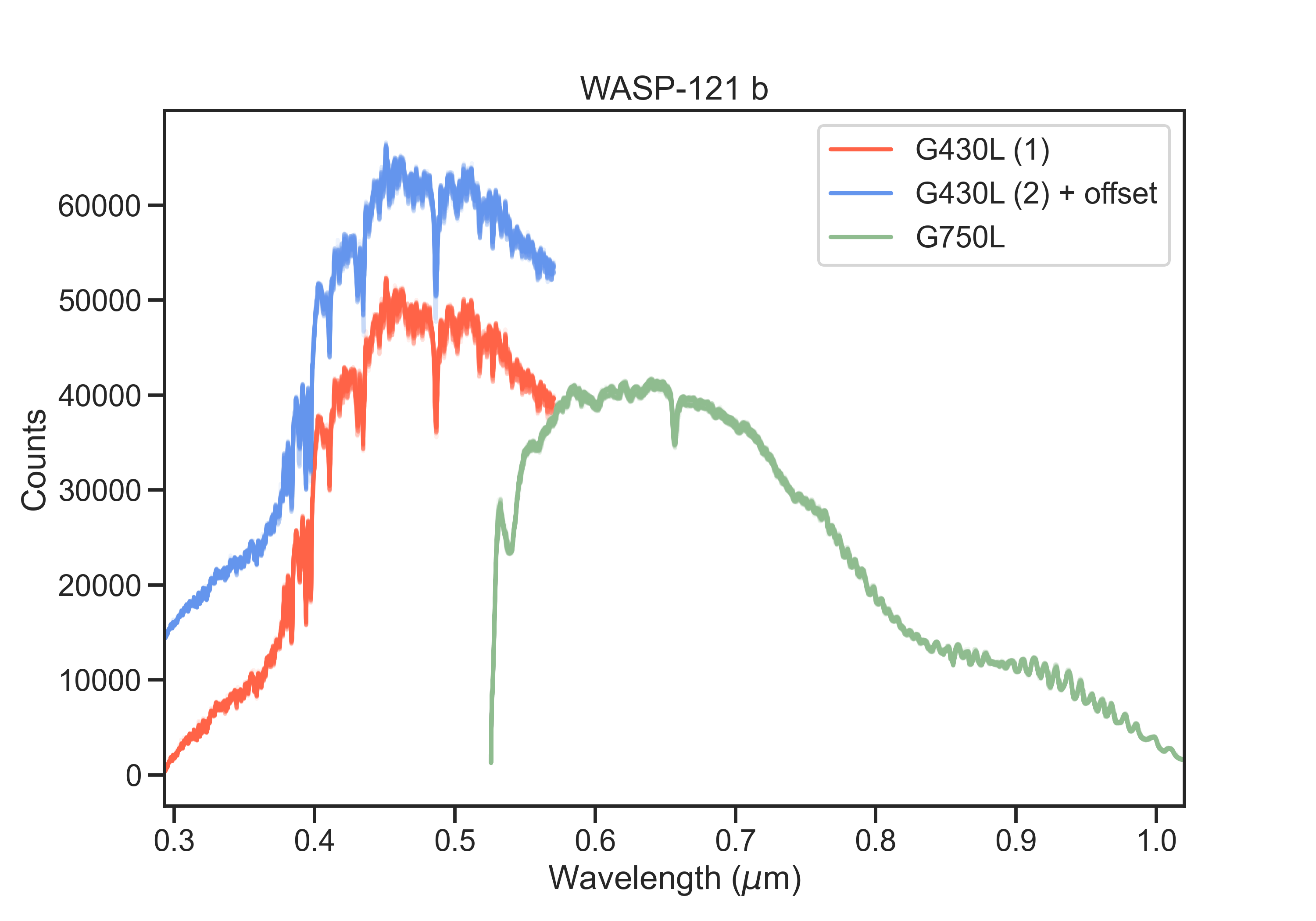}{0.49\textwidth}{}}
\gridline{\fig{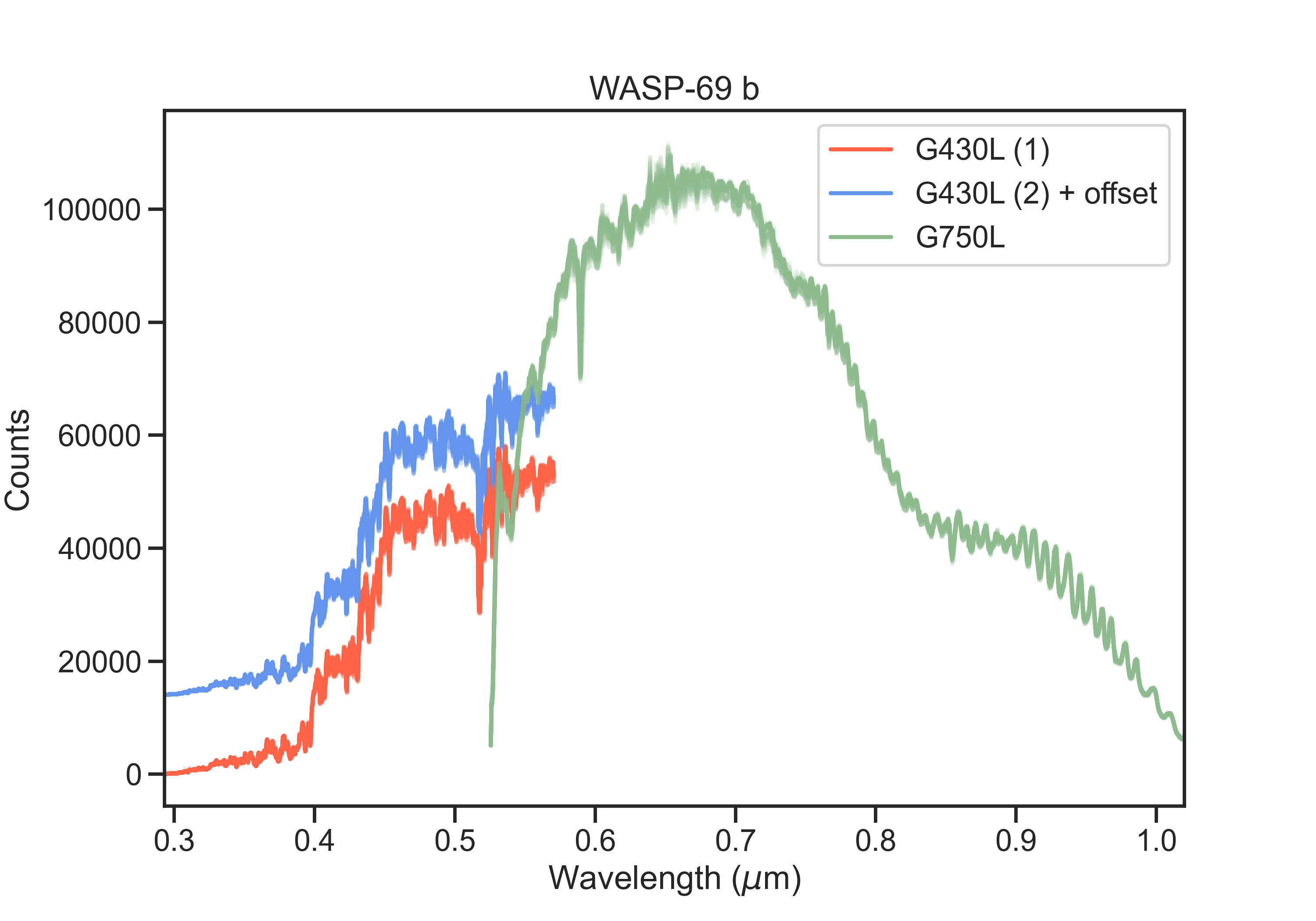}{0.49\textwidth}{}
\fig{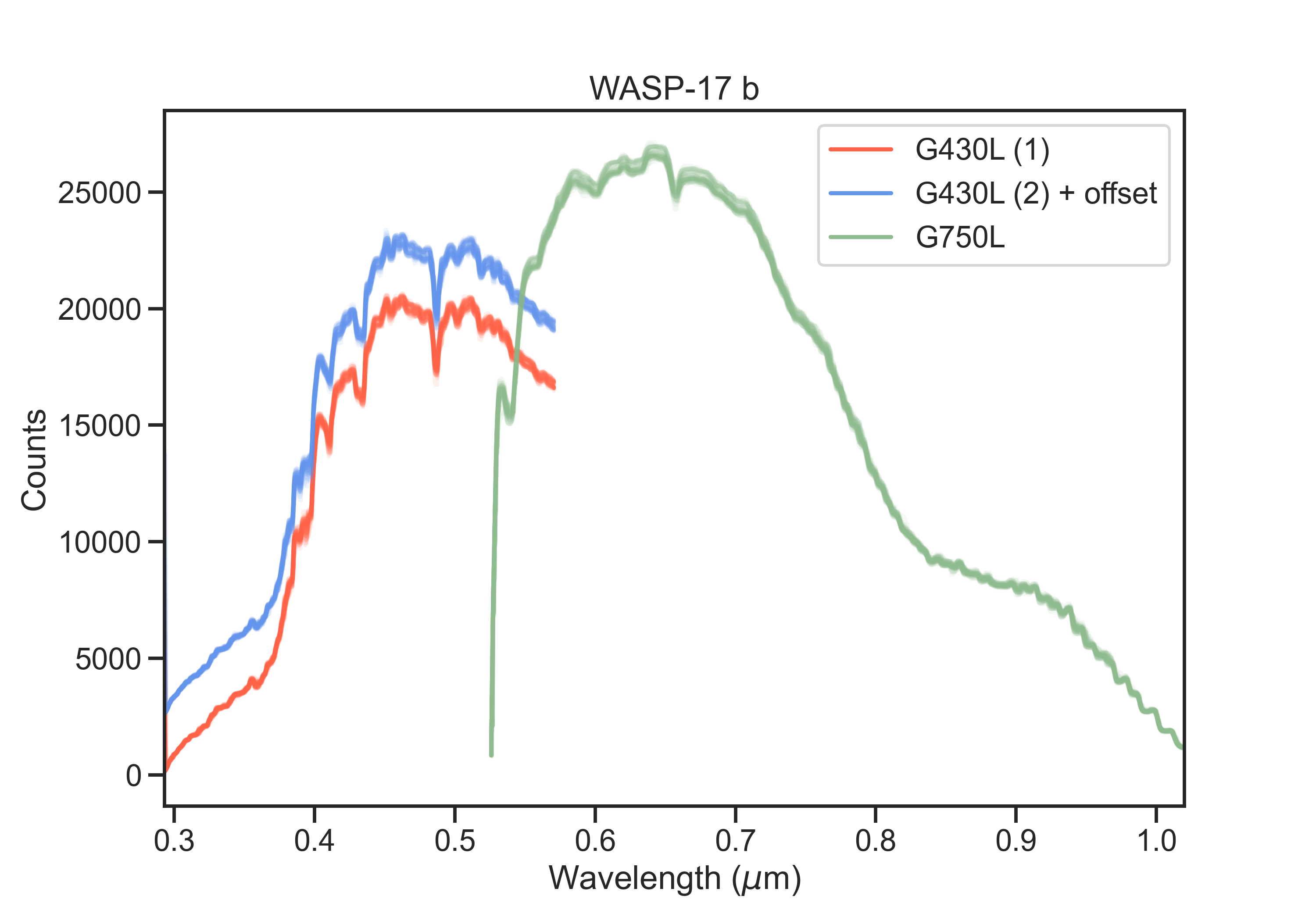}{0.49\textwidth}{}}
\caption{The extracted spectra from G430L transit 1 (red), G430L transit 2 (blue, offset vertically for comparison) and G750L (green) for WASP-39\,b (upper left), WASP-121\,b (upper right), WASP-69\,b (lower left), and WASP-17\,b (lower right). Each figure is showing the entire time series of extracted spectra stacked on top of each other.}\label{fig:extracted_spec}
\end{figure*}

With our exposures cleaned, we perform the same spectral tracing procedure as described previously. We store the trace fit parameters (the location on the detector, plus the two variables associated with the shape of the trace) for subsequent use detrending systematics in the light curve fitting step. Then, we proceed to the spectral extraction. To do this, we use optimal extraction \citep{Marsh_1989} with the implementation as described in \cite{Brahm_2017}, which fits the profile of the spectrum with a series of polynomials (taking advantage of the expected smoothness of spectra) to give each pixel in the science frame a weight with which to perform a weighted extraction. We use a large aperture of 15 pixels with the optimal extraction, as the method will set the weights of pixels that do not contain significant flux to not contribute and we can therefore be sure we are not losing any flux through the use of an aperture too small for the spectrum. This procedure will naturally correct for any remaining bad pixels flexibly. %While typically the optimal extraction algorithm gives weights and sums together all pixels containing flux, which is necessary in the low signal to noise regime for which it was originally created, for our purposes we set a strict aperture radius of 6.5 pixels, which has been shown to be ideal in past studies (CITE). 
We then do a simple median filter on the resulting extracted spectrum to get rid of any potential final outliers. We note that any of the numerical values stated in this section can be easily changed in our pipeline for other studies and their needs, but that these are the best parameters we found to work with multiple different data sets simultaneously. \\
The stacked extracted spectrum time series, two G430L transits and one G750L transit, for each of the four planets is shown in \autoref{fig:extracted_spec}. \\ 

\section{Light Curve Fitting}\label{sec:lc_fit}
\subsection{White Light Curve Fitting}\label{sec:wlc_fit}
With the cleaned, extracted spectra obtained in the last section, we begin to analyze our time series light curve. By summing the entirety of the flux extracted for each exposure, we obtain the white light curve. We note that at this stage, we convert the header specified JD$_{\mathrm{UTC}}$ to BJD$_{\mathrm{TBD}}$ using \textit{barycorrpy} (\citealp{Kanodia_2018}, Python routines for the framework from \citealp{Eastman_2010}). We perform light curve fits with \textit{juliet} \citep{juliet}. Our fit contains both a basic transit model and a systematics model, the latter of which was determined with a significant amount of testing, briefly described in \autoref{sec:sys_determination} and in much more detail in \autoref{app:systematics}. Most of the parameters and their associated priors for our transit fit are obtained by our joint fit of all available previous data for the target of interest, detailed in \autoref{sec:joint_fits}. We fix all the system parameters from \autoref{tab:joint_fit} besides $R_p/R_s$, and additionally fit for $mflux$, $sigma\_w$, and systematics detrending parameter coefficients for each transit. This includes the time of transit center $T0$, though we note that we ran additional tests where $T0$ was allowed to vary $\pm 0.05$ that did not significantly affect our resulting light curve fits in comparison to the fixed case. \\

Limb darkening is an especially important consideration for HST light curves, as the required phasing due to the orbits of HST can leave gaps in the time series during ingress and egress, which increases the already significant uncertainty in the correct handling of the effect of limb darkening. %We cannot in general fit for the limb darkening coefficients due to these breaks in the time series observations. 
We fix our limb darkening parameters to those from theory in order to stay internally consistent, especially in the case of observations with poor coverage of the ingress and egress of the transit. We use the quadratic limb darkening coefficients from \textit{ExoTiC-LD} \citep{grant_2022}. For our sampling, we use \textit{dynesty} and its dynamic nested sampling method to directly sample the posterior for an accurate determination of parameter errors. We note that our pipeline also has the option to use \textit{pymultinest}. \\

%, but its method of Bayesian evidence maximization is disfavored in this case, as our objective is consistent and repeatable fits, along with accurate parameter posterior determination. \\

\subsection{Spectroscopic Light Curve Fitting}\label{sec:slc_fit}
With the results from the white light curve fit, we then move on to the spectroscopic light curve fit, in which we split the full spectrum into a set of wavelength bins and carry out the same fitting. The process then follows the same as for the white light curve fits: all system parameters are fixed besides R$_{p}/$R$_{s}$, and we additionally fit for $mflux$, $sigma\_w$, and systematics detrending parameter coefficients. We ran the same test for fixing $T0$ or allowing it to vary that was done for the white light curve fits, and again found that our results were insensitive to this choice. The limb darkening parameters are calculated from \textit{ExoTiC-LD} for each bin individually and also fixed. We use the same \textit{dynesty} dynamic nested sampler for the spectroscopic light curve fits. \\

\subsubsection{Determination of Systematics Methodology}\label{sec:sys_determination}

\begin{deluxetable*}{C C C C C C C C C}[htb!]
    \tablecaption{Fit white light curve $R_p/R_s$ for our final linear and GP detrending methods for WASP-39\,b, WASP-121\,b, WASP-69\,b, and WASP-17\,b.\label{tab:wlc_fits}}
    \tablecolumns{9}
    \tablehead{ & \multicolumn{2}{c}{WASP-39\,b} & \multicolumn{2}{c}{WASP-121\,b} & \multicolumn{2}{c}{WASP-69\,b} & \multicolumn{2}{c}{WASP-17\,b}\\
    Mode & Linear & GP & Linear & GP & Linear & GP & Linear & GP}

    \startdata
    \textrm{G430L} & $0.1448 \pm 0.0010$ & $0.1459 \pm 0.0007$ & $0.1201\pm 0.0018$ & $0.1221\pm 0.0018$ & $0.1319\pm 0.0010$ & $0.1349\pm0.0006$ & $0.1258 \pm 0.0011$ & $0.1269 \pm 0.0013$  \\
    \textrm{G750L} & $0.1447 \pm 0.0008$& $0.1444 \pm 0.008$& $0.1196 \pm 0.0007$ & $0.1197\pm 0.0005$& $0.1251 \pm 0.0017$ & $0.1223 \pm 0.0011$& $0.1233 \pm 0.0016$ & $0.1212 \pm 0.0019$ \\
    \enddata
\end{deluxetable*}

\begin{figure*}
\gridline{\fig{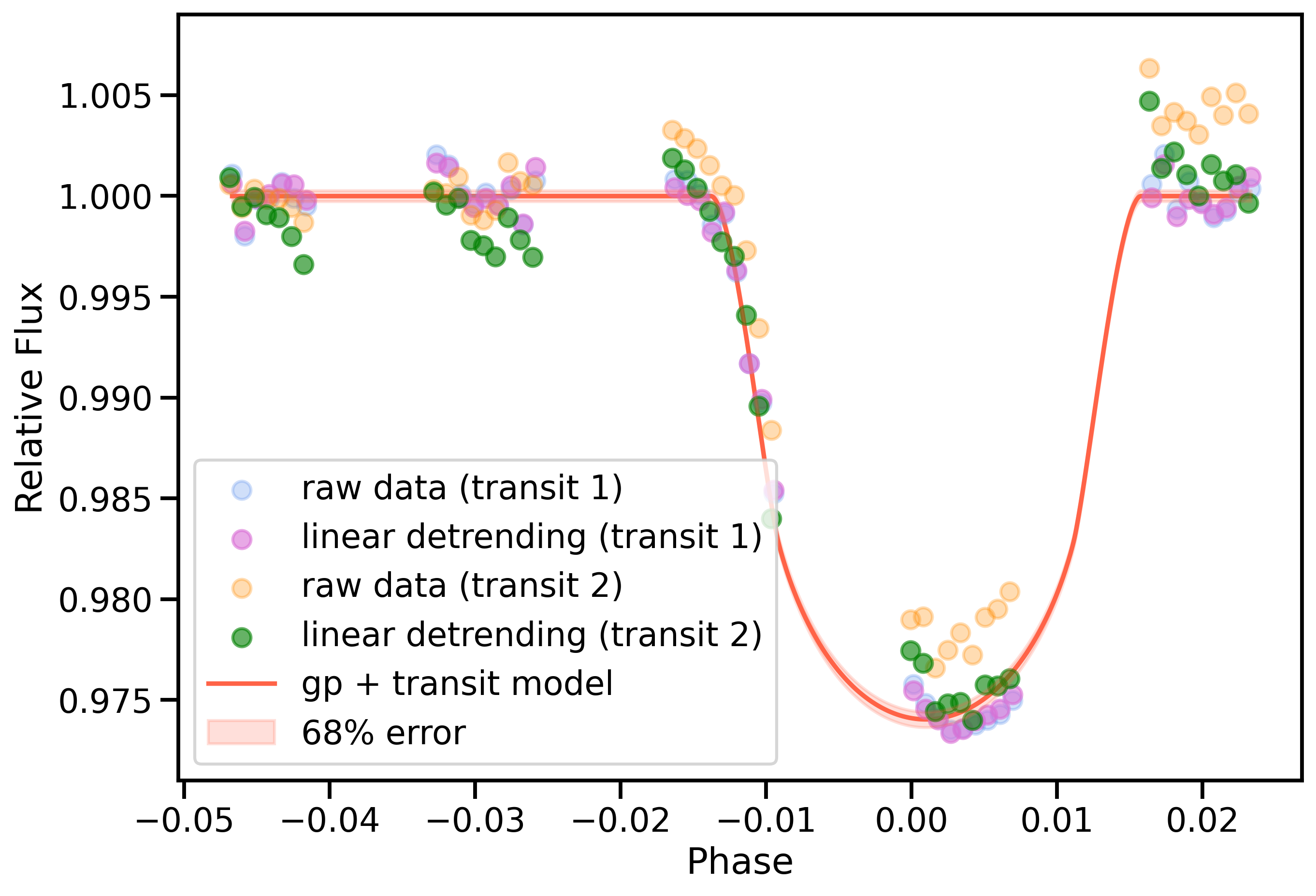}{0.245\textwidth}{}
          \fig{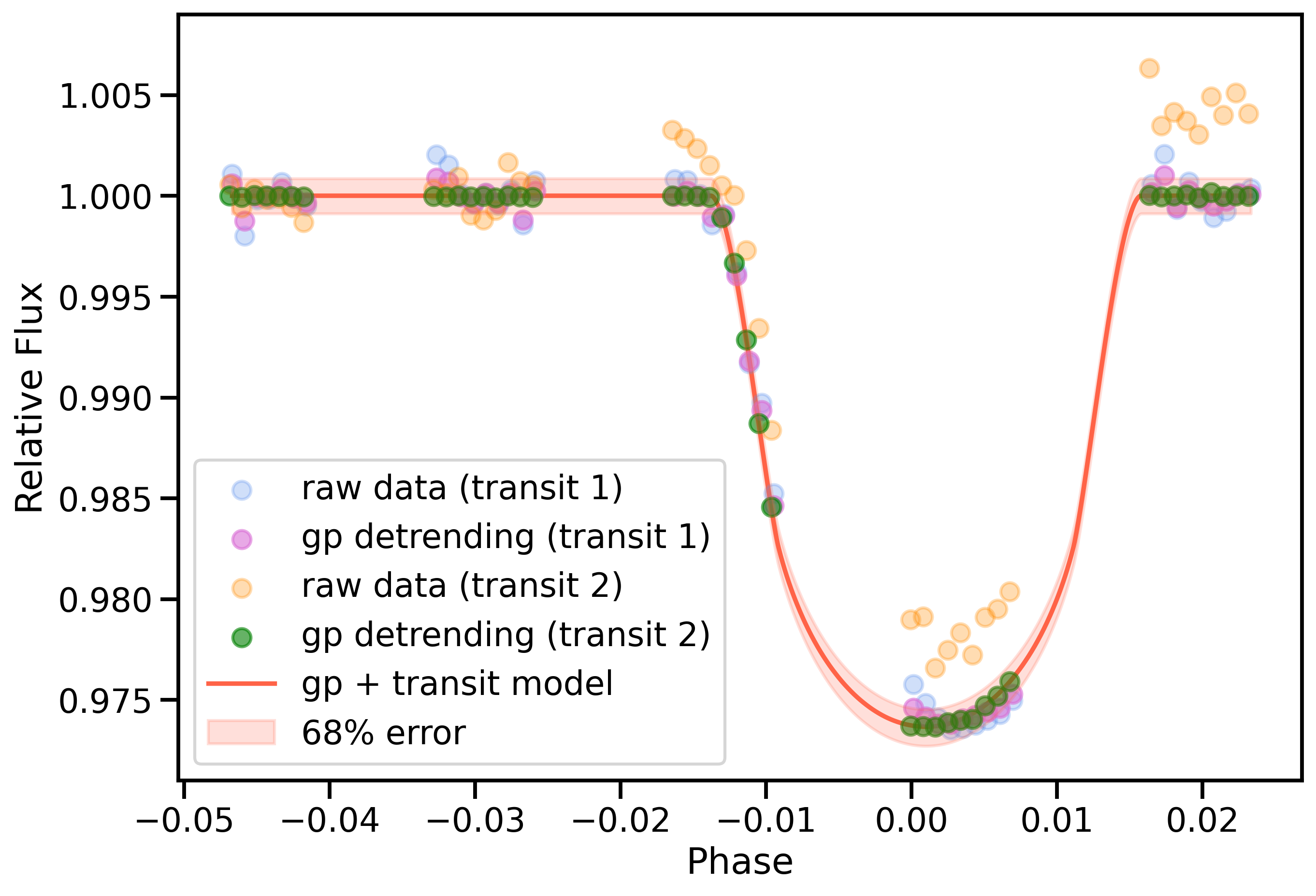}{0.245\textwidth}{}
          \fig{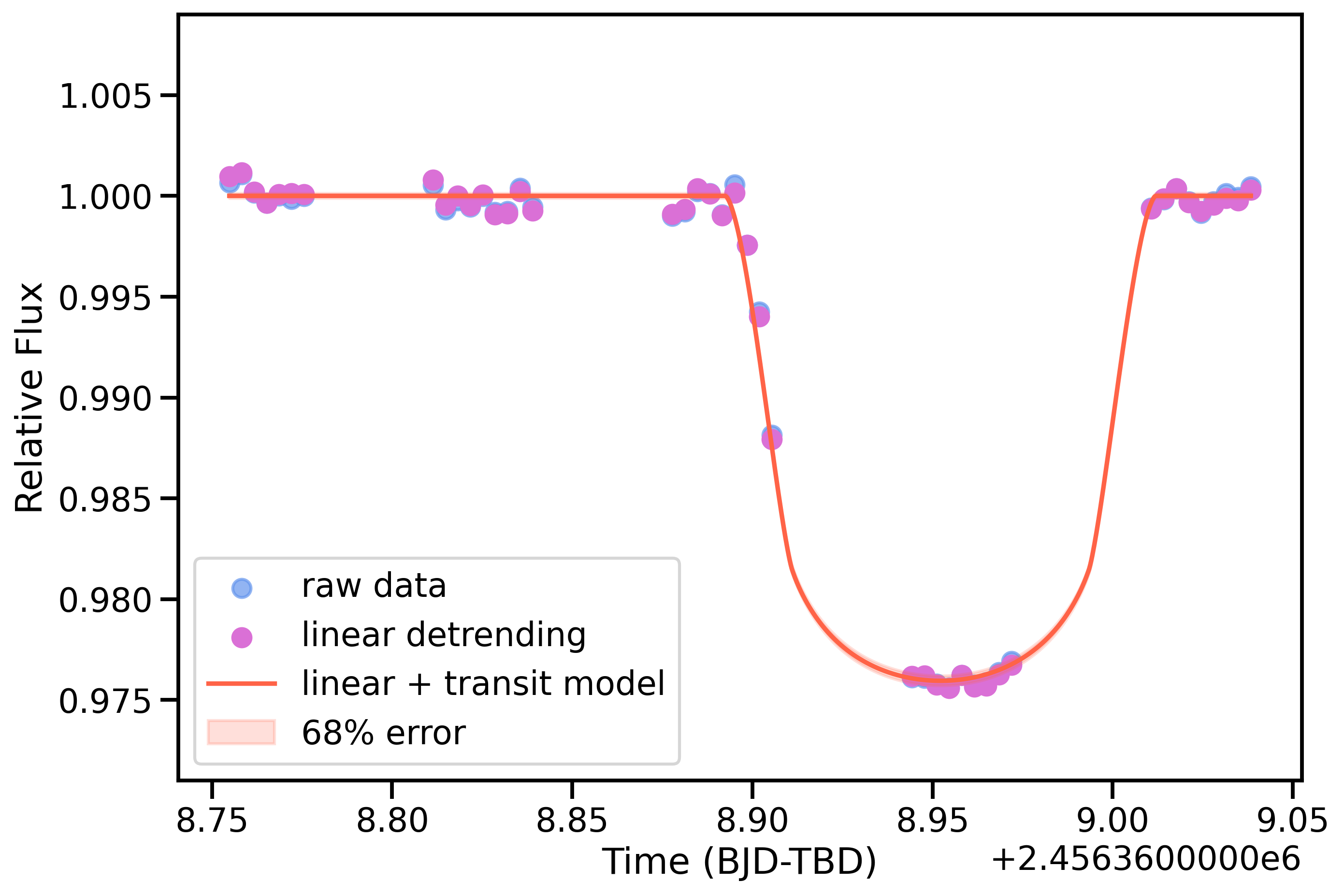}{0.245\textwidth}{}
          \fig{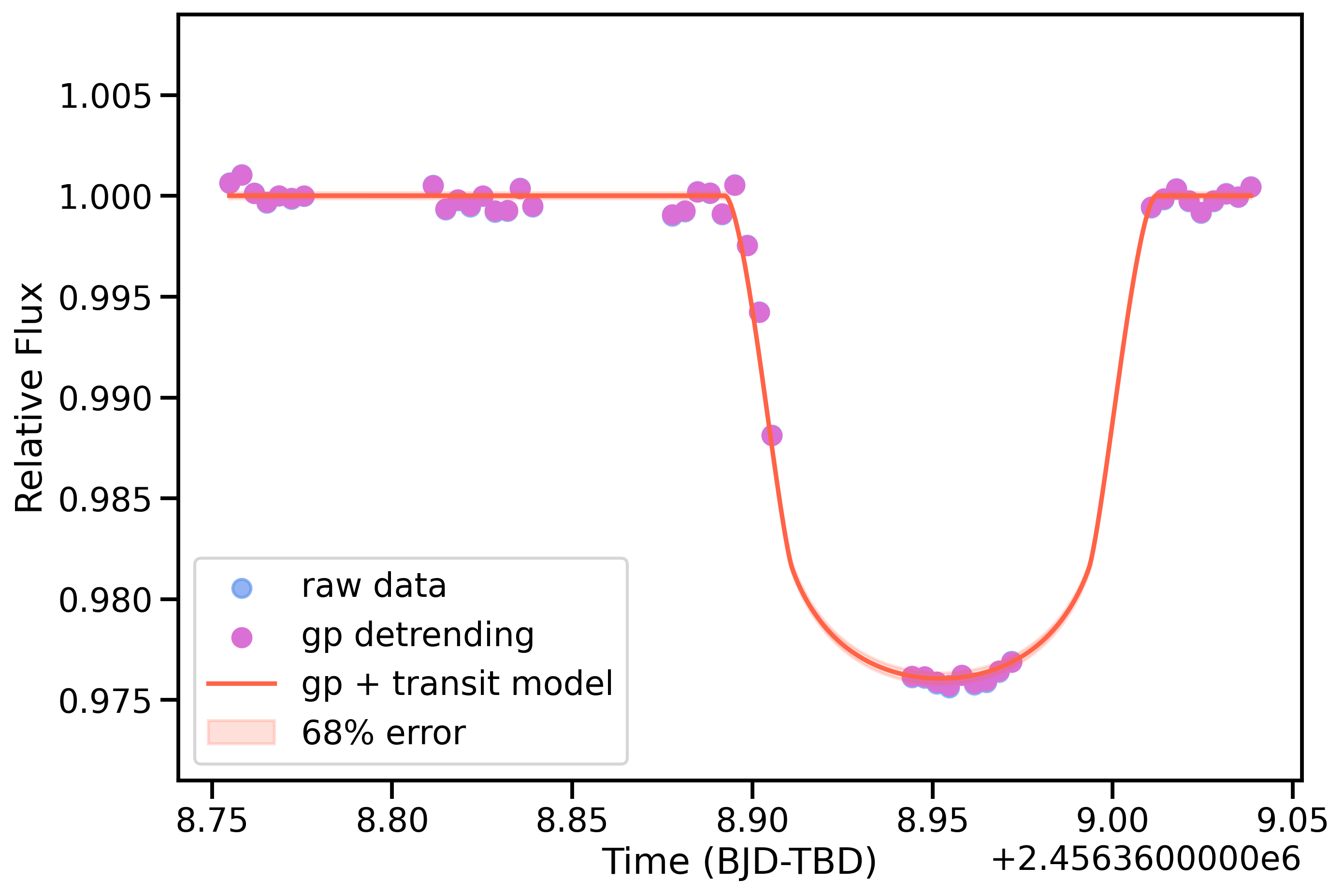}{0.245\textwidth}{}}
\gridline{\fig{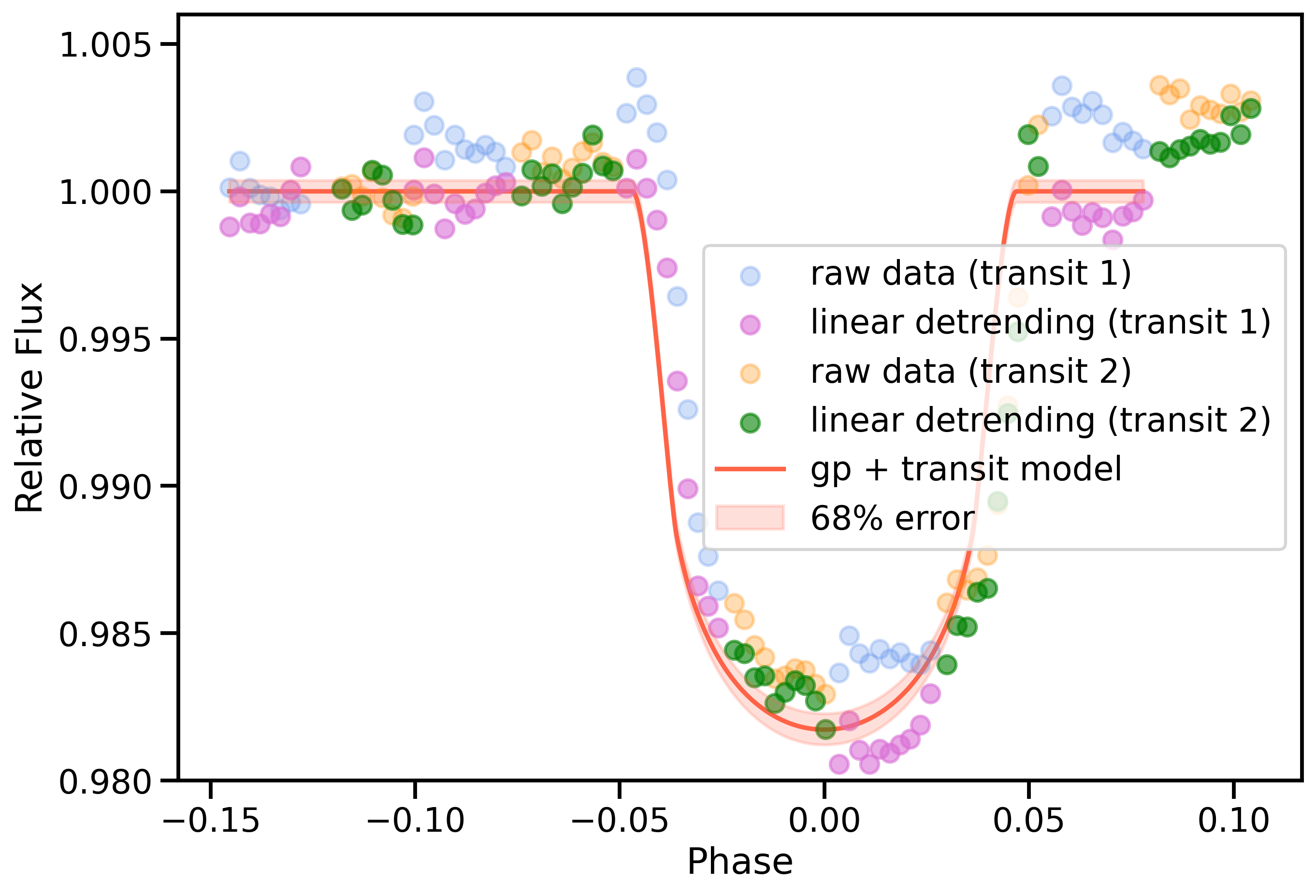}{0.245\textwidth}{}
          \fig{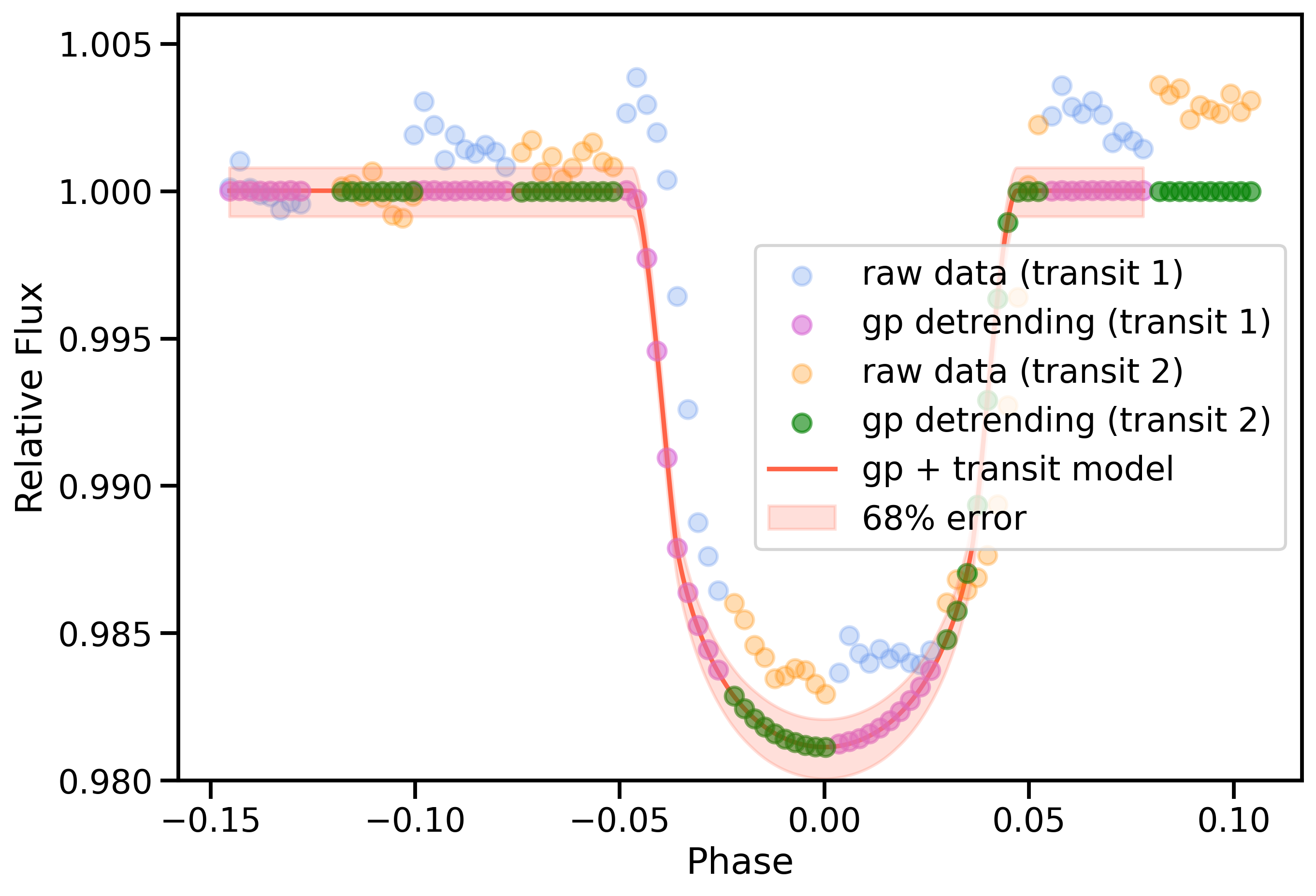}{0.245\textwidth}{}
          \fig{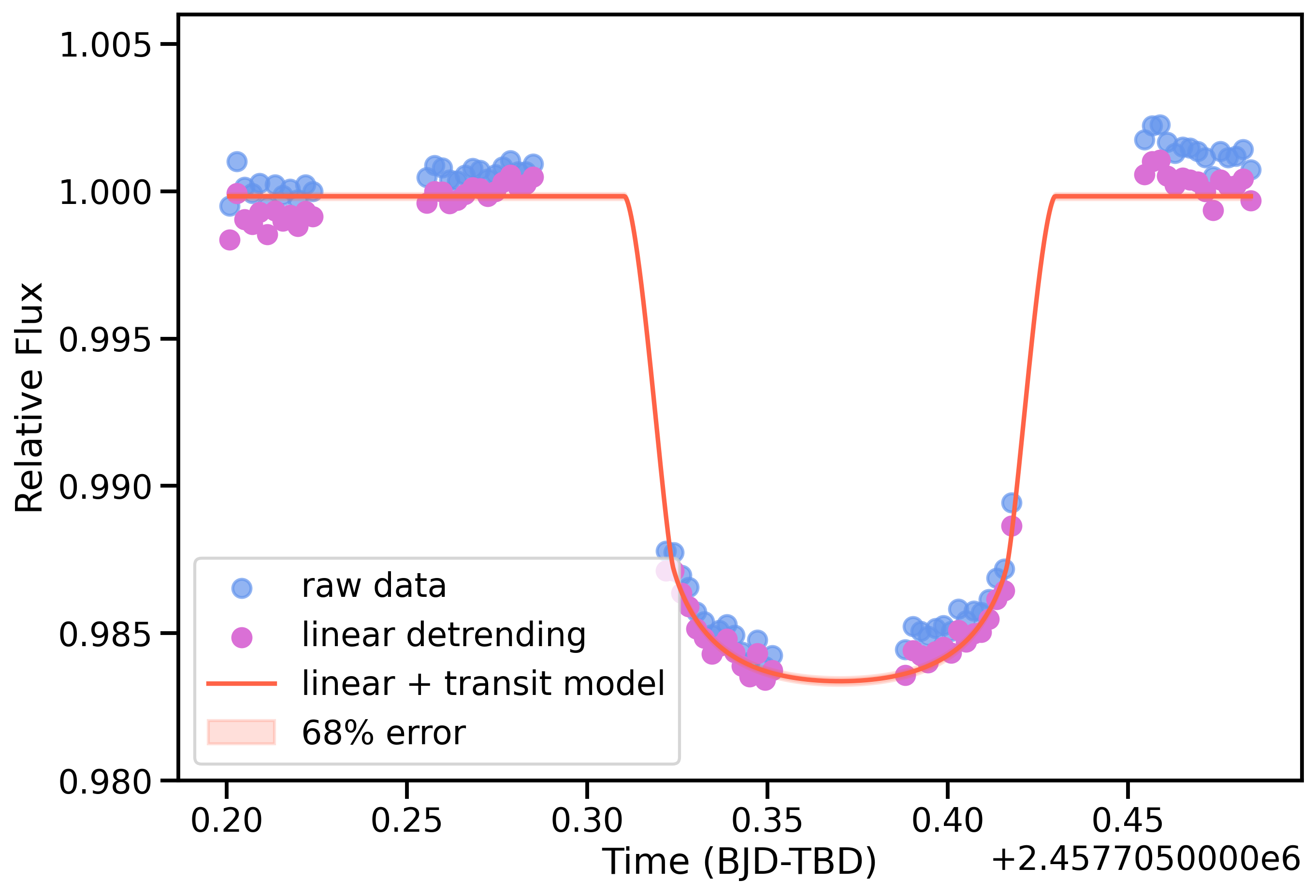}{0.245\textwidth}{}
          \fig{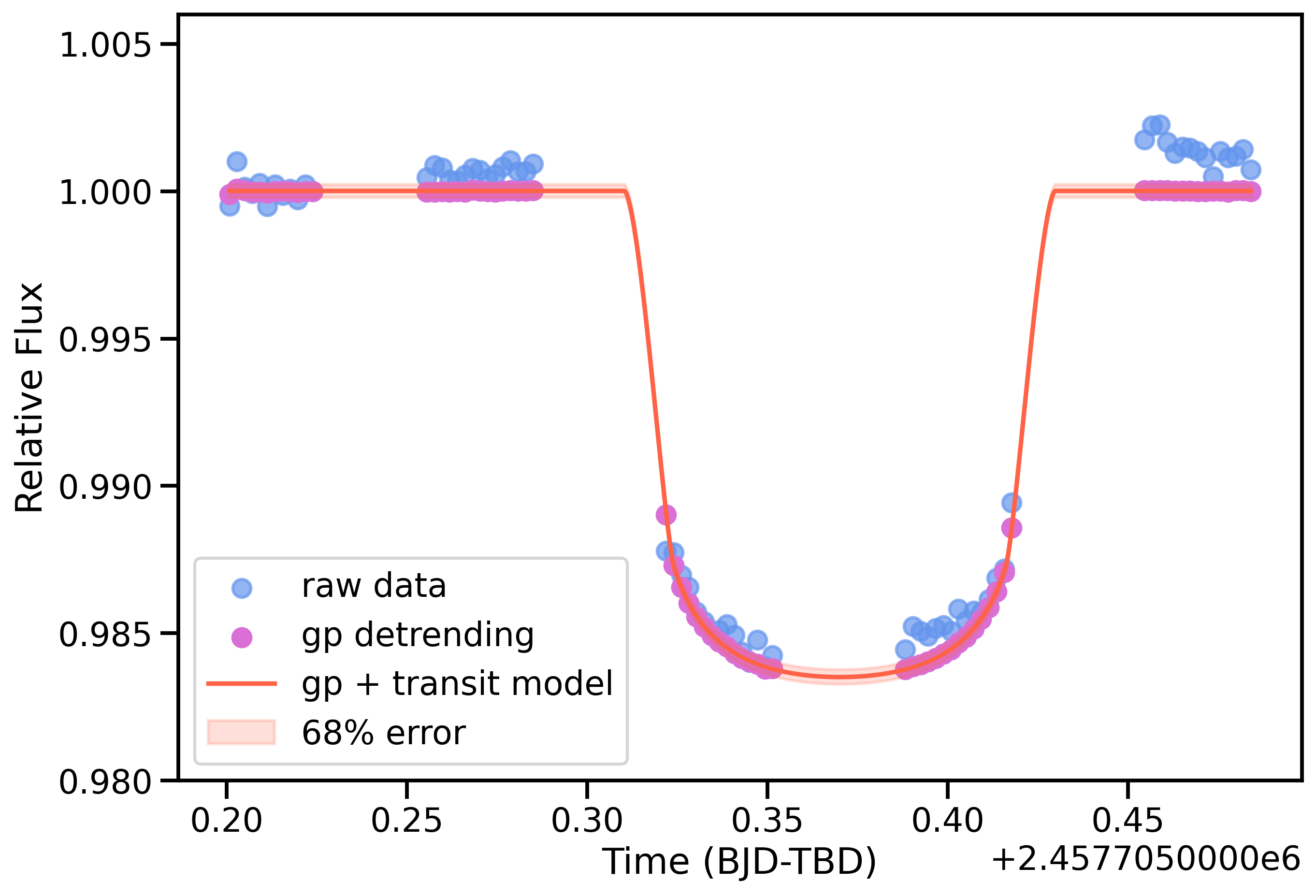}{0.245\textwidth}{}}
\gridline{\fig{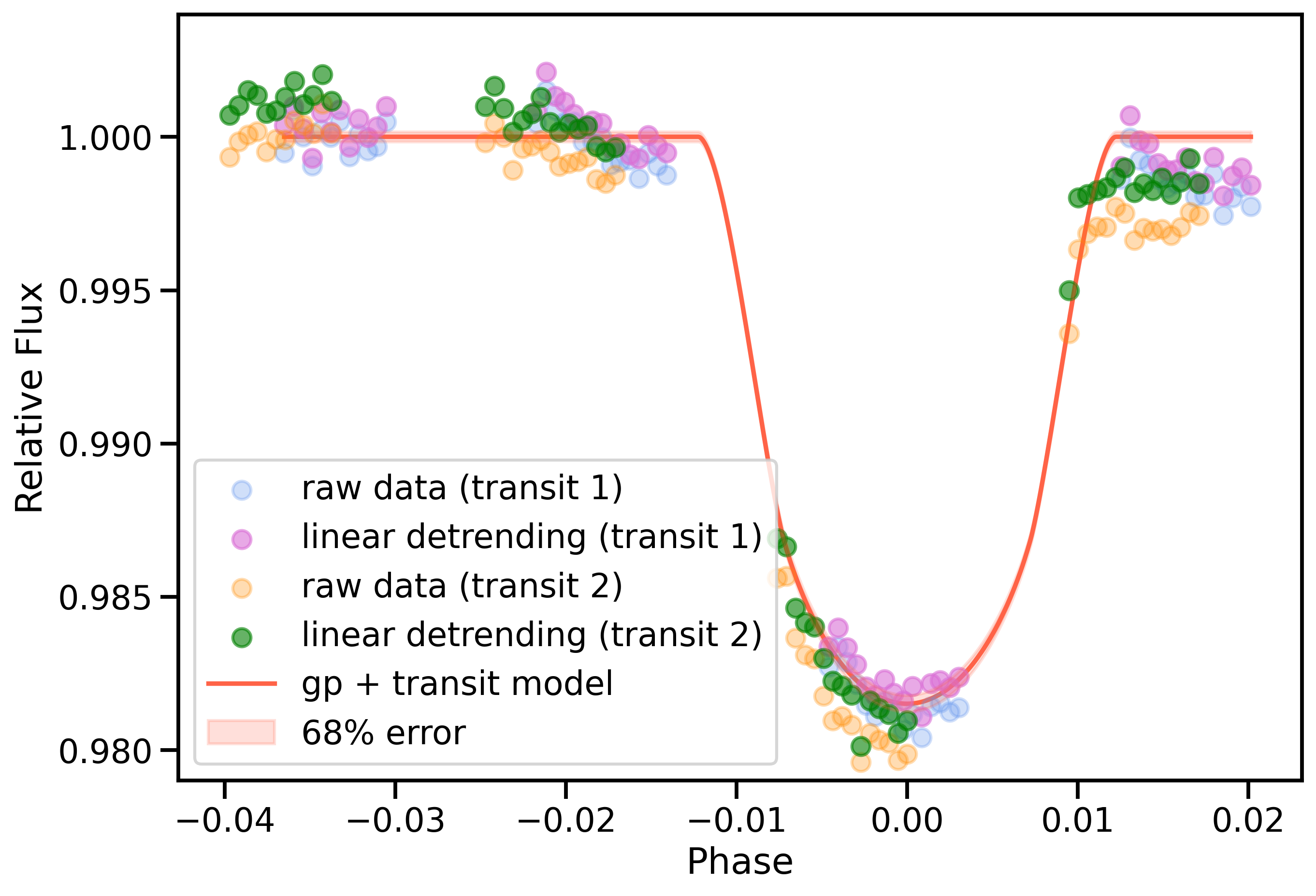}{0.245\textwidth}{}
          \fig{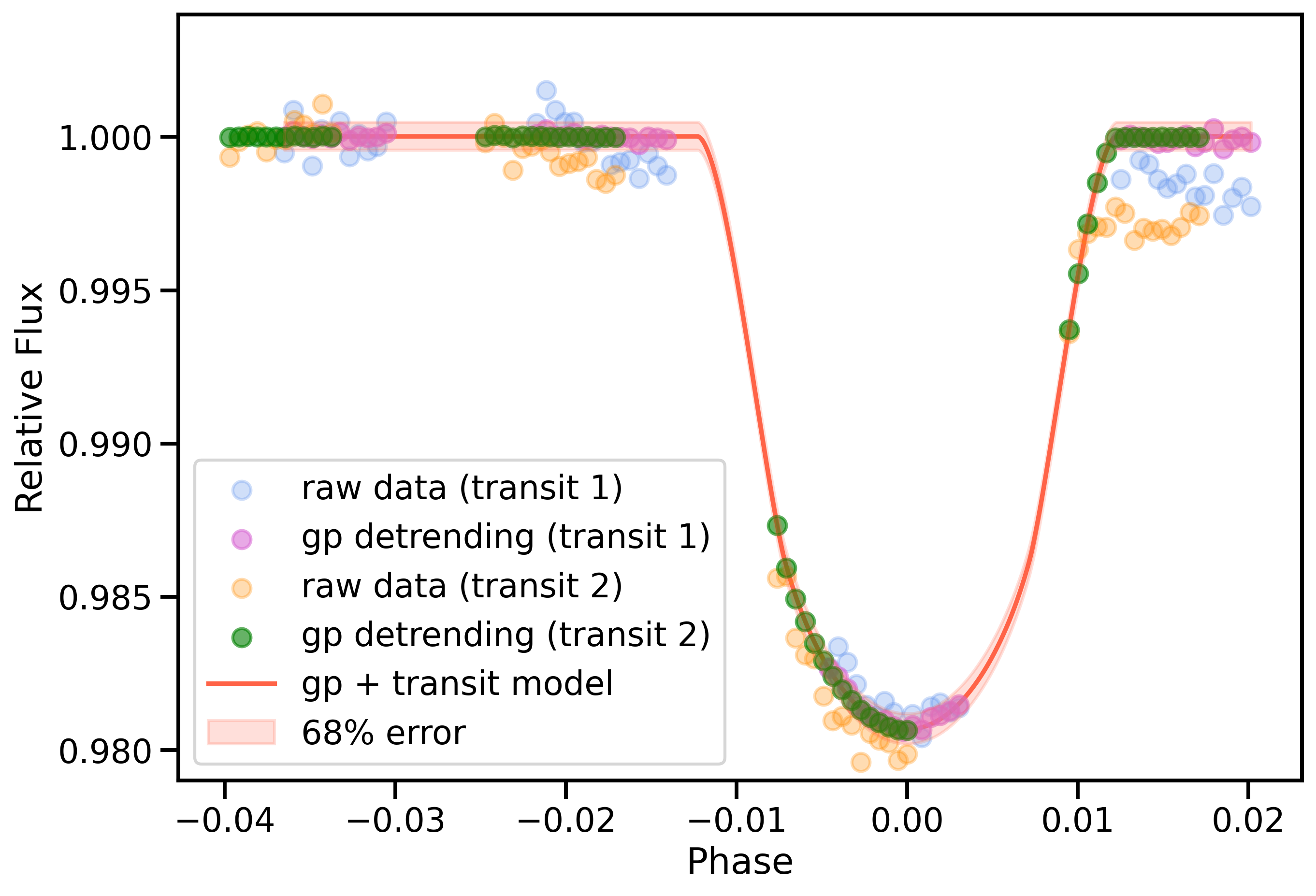}{0.245\textwidth}{}
          \fig{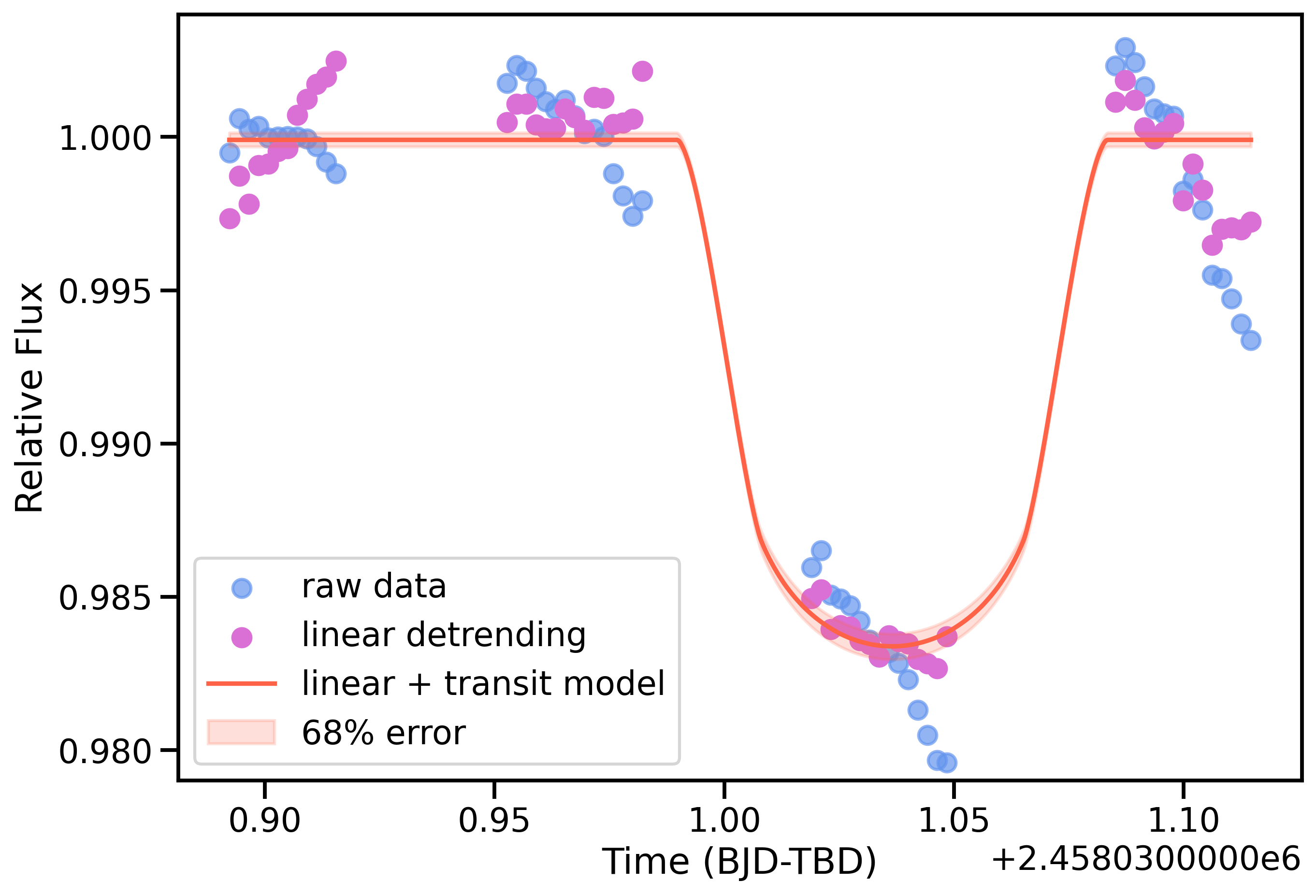}{0.245\textwidth}{}
          \fig{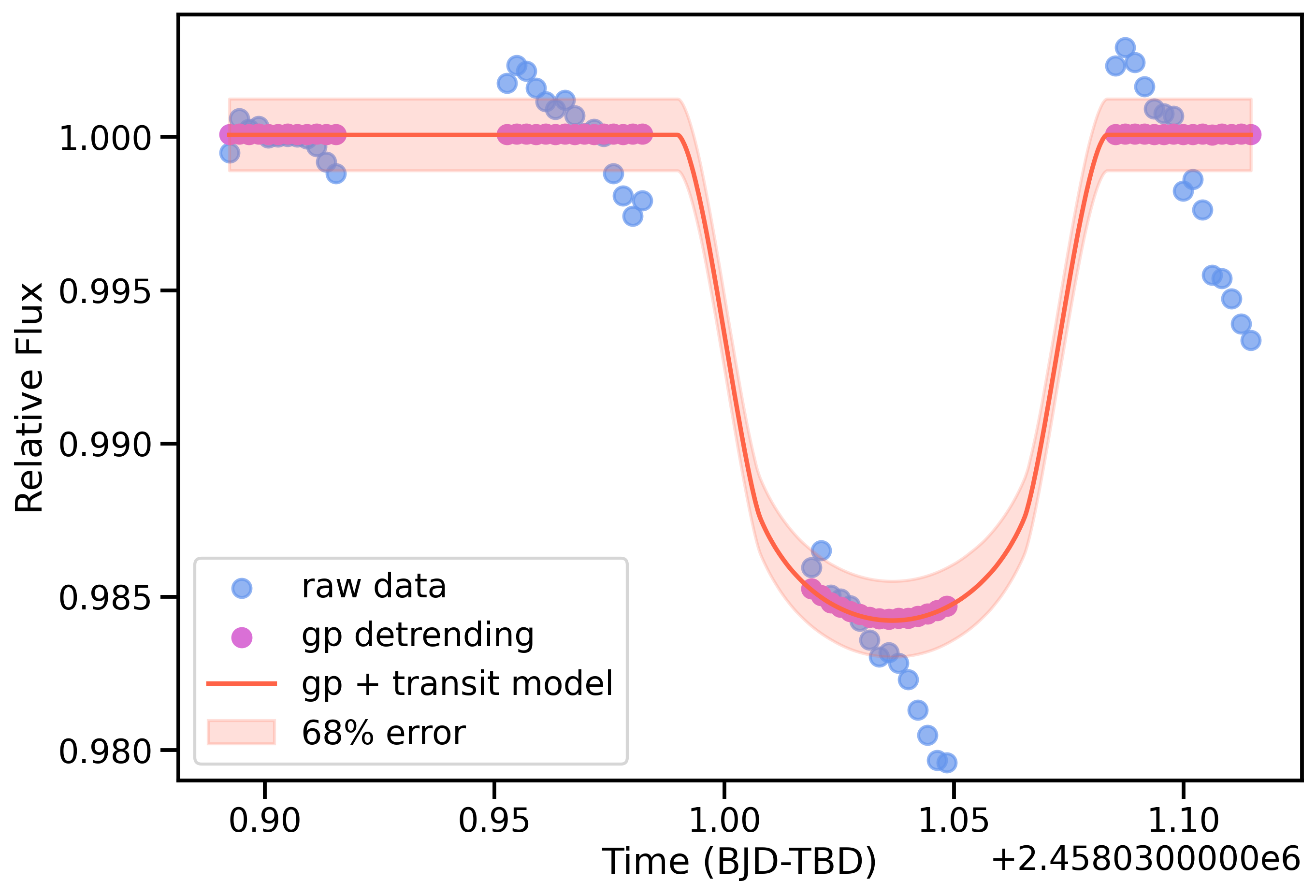}{0.245\textwidth}{}}
\gridline{\fig{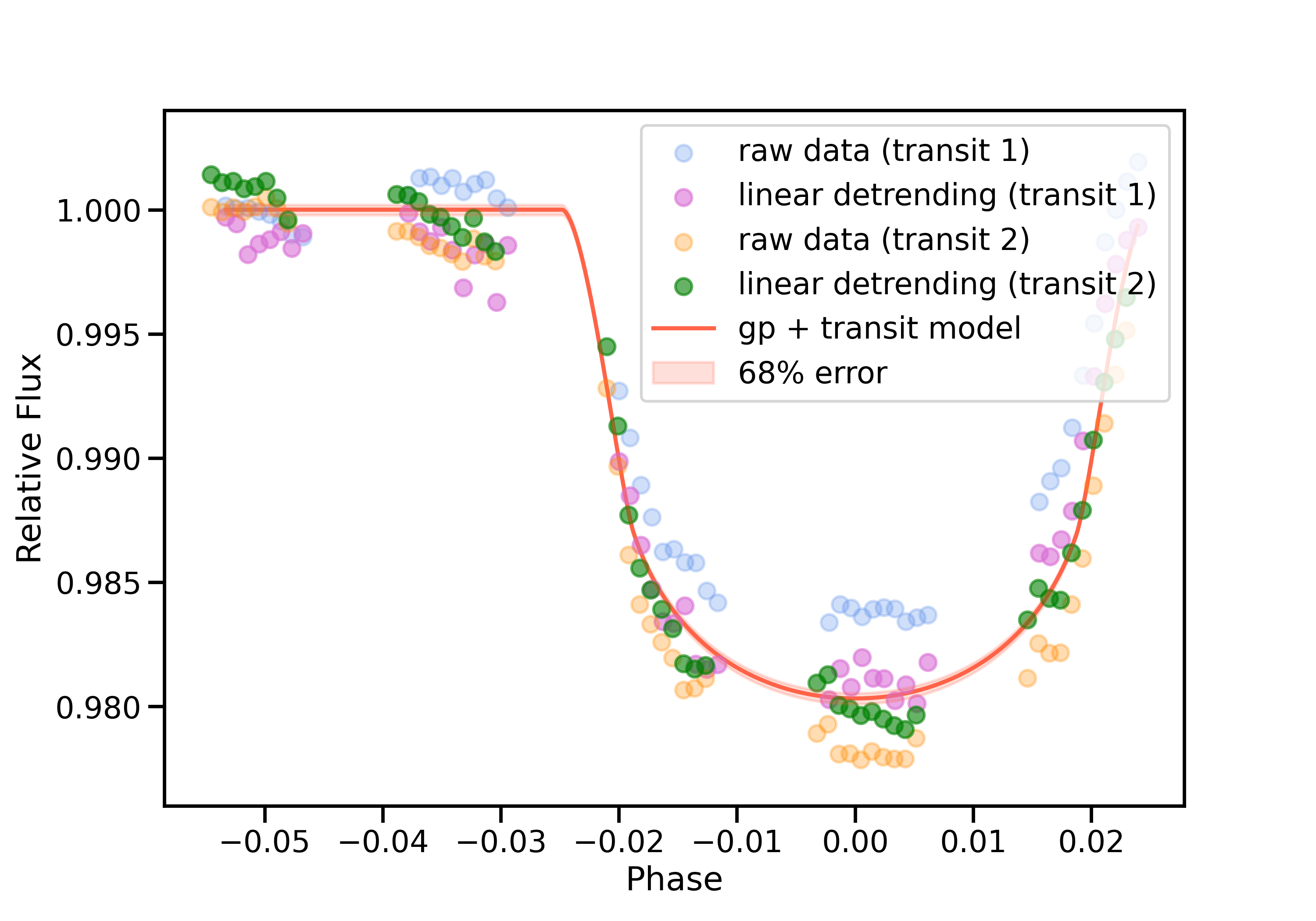}{0.245\textwidth}{}
          \fig{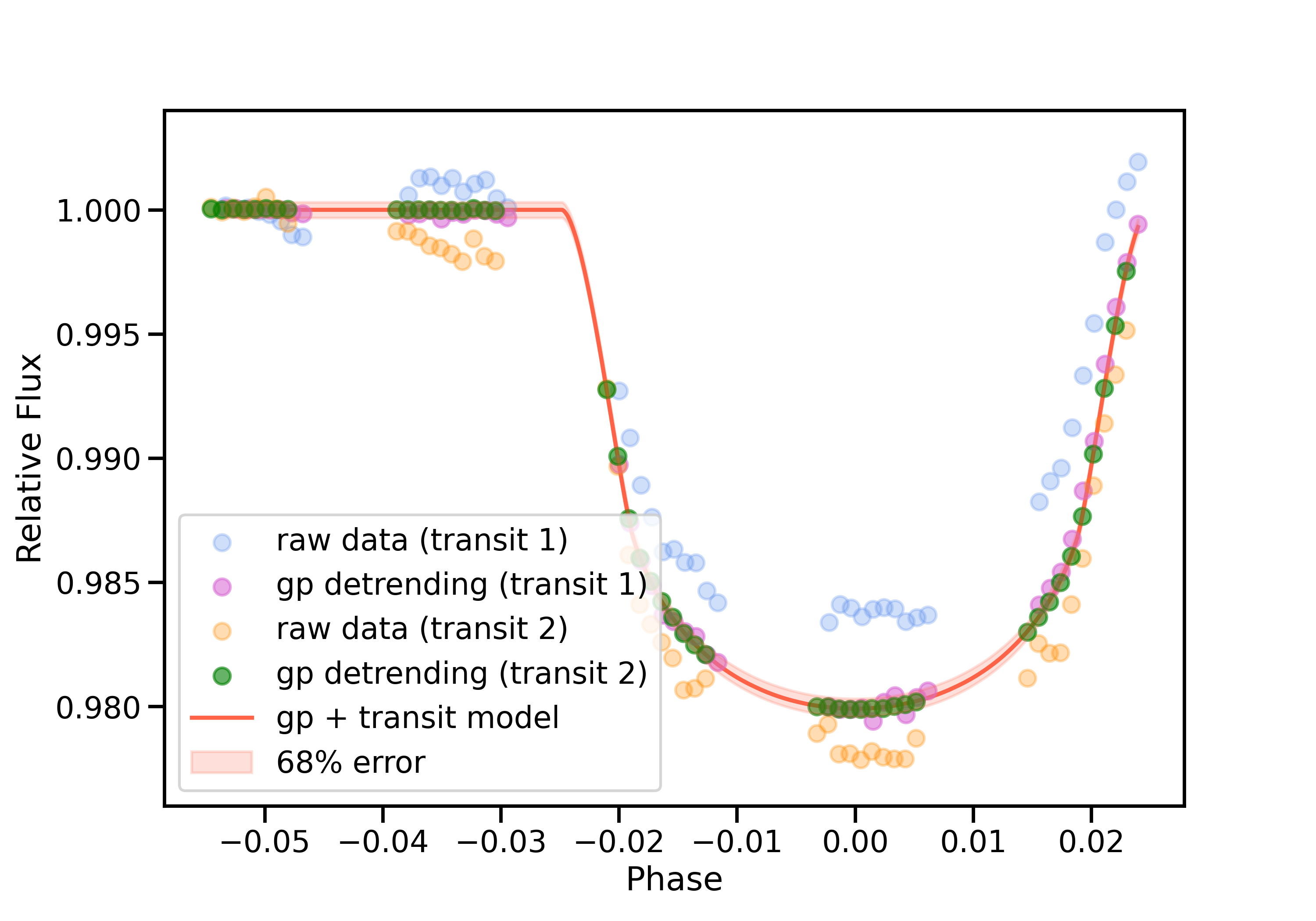}{0.245\textwidth}{}
          \fig{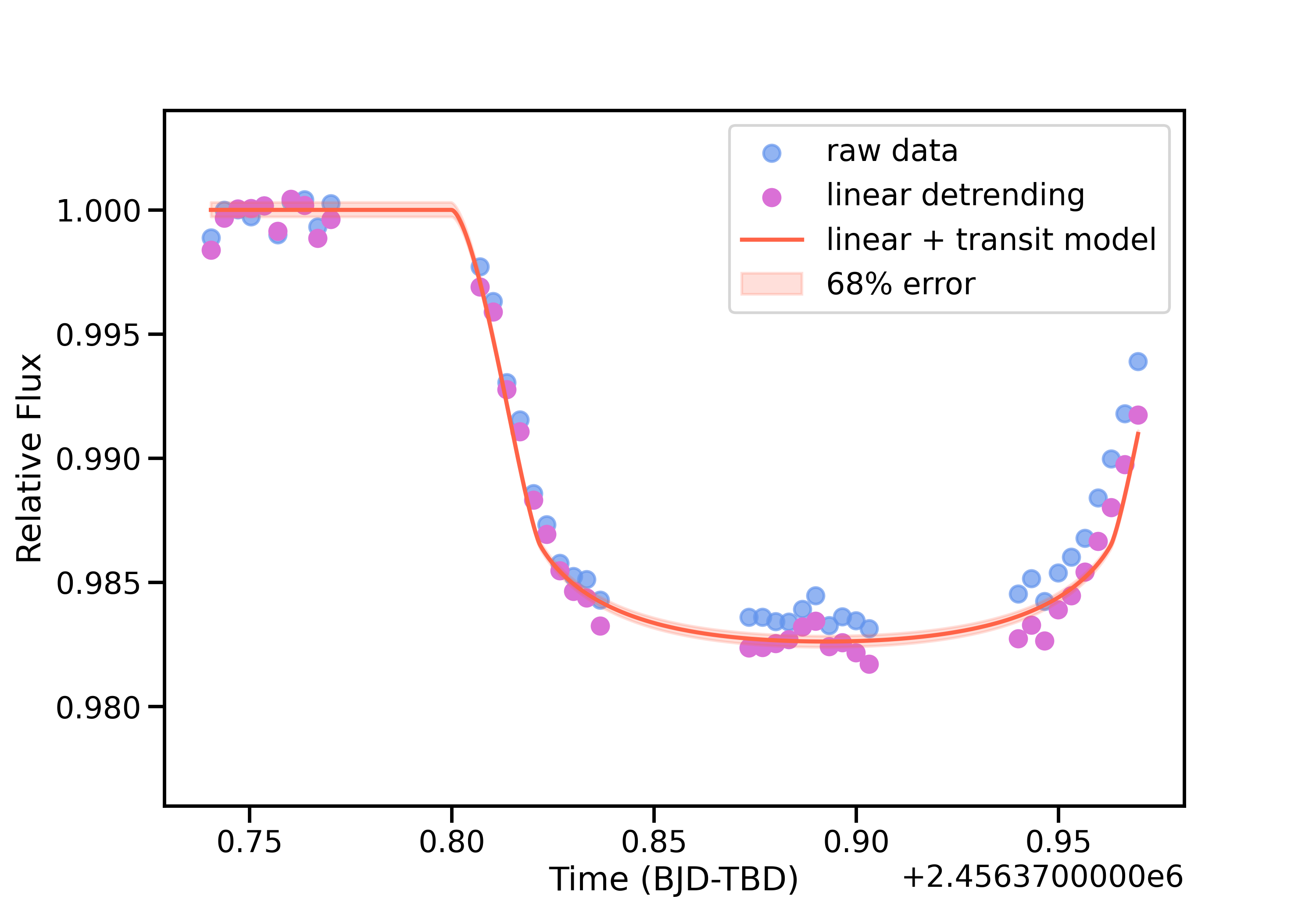}{0.245\textwidth}{}
          \fig{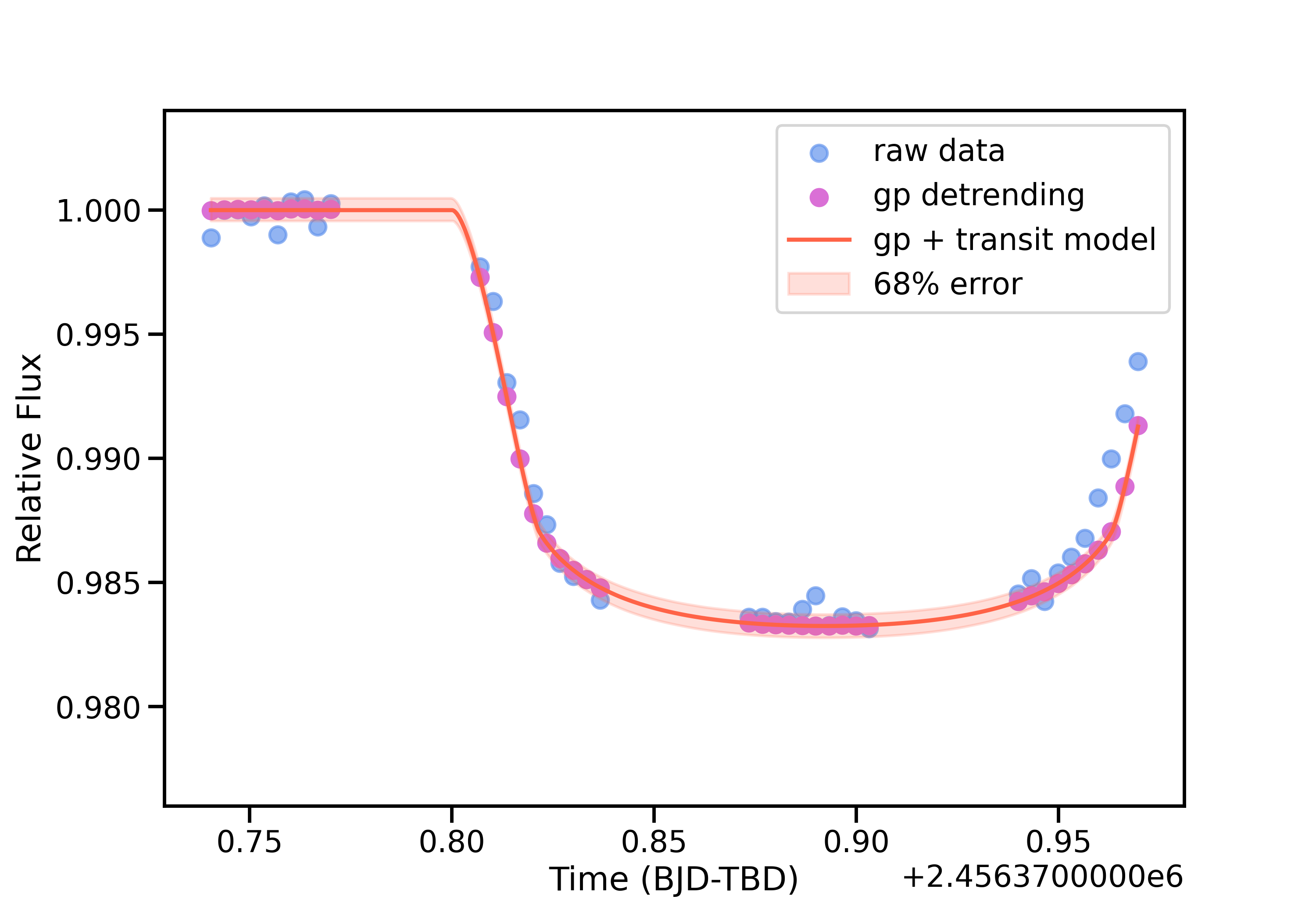}{0.245\textwidth}{}}
    \caption{The white light curve fits for each of our four planets: first row WASP-39\,b, second row WASP-121\,b, third row WASP-69\,b, fourth row WASP-17\,b. From left to right: joint linear detrending G430L, joint GP detrending G430L, linear detrending G750L, GP detrending G750L. All linear detrending is done with 2 PCs as regressors, and all GP detrending is done with 5 PCs as regressors. All plots of the same planet are shown on a common y-axis scale.}
    \label{fig:wlc}

\end{figure*}

Time series observations with HST suffer from significant systematics, which can be seen clearly at the white light curve and spectroscopic light curve stages (see e.g. \autoref{fig:wlc}). To account for this, we use the information obtained from the engineering jitter data, as well the parameters from our spectral tracing. Following \cite{sing_2019}, we use the following six jitter vectors that were found to be most strongly correlated with the data: RA, DEC, Latitude, Longitude, V2\_roll, V3\_roll. We then have three additional detrending vectors from the spectral trace -- the y-value of the start of the trace, and the two coefficients of the Chebyshev polynomial fit. Finally, following past success we also include the HST orbital phase to the fourth power \citep[e.g.][]{Brown_2001, Sing_2008, Nikolov_2015, Alam_2018}. We standardize all of these detrending vectors by first subtracting the mean and then dividing by the standard deviation of the vector values. \\

To determine the best detrending method, we carry out a series of tests on a few example transmission spectroscopy datasets and compare the resulting fits. To handle our systematics, there are two typically used methods which we test in our framework -- a gaussian process (GP) model and a linear model. We will use an exponential prior for our GP detrending vector coefficients, and a log normal prior for our linear coefficients. The exponential prior strongly penalizes unfavorable regressors, which prevents overfitting with the flexible GP \citep[see e.g.][]{stat_book}. The most simple test is by using both a GP and linear model to detrend the data with the 10 jitter vectors described above. For our model comparison, we will use Bayesian evidence to fairly penalize the use of additional parameters. \\

\begin{figure*}
\gridline{\fig{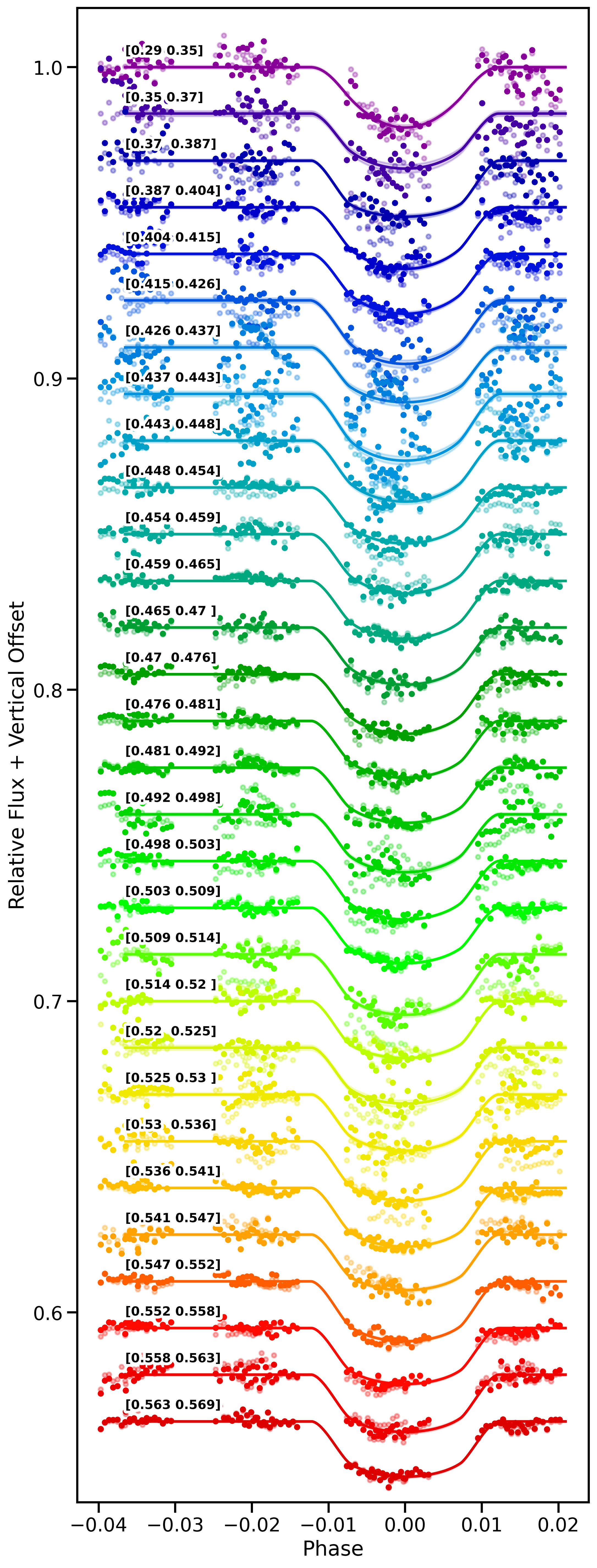}{0.4\textwidth}{}
          \fig{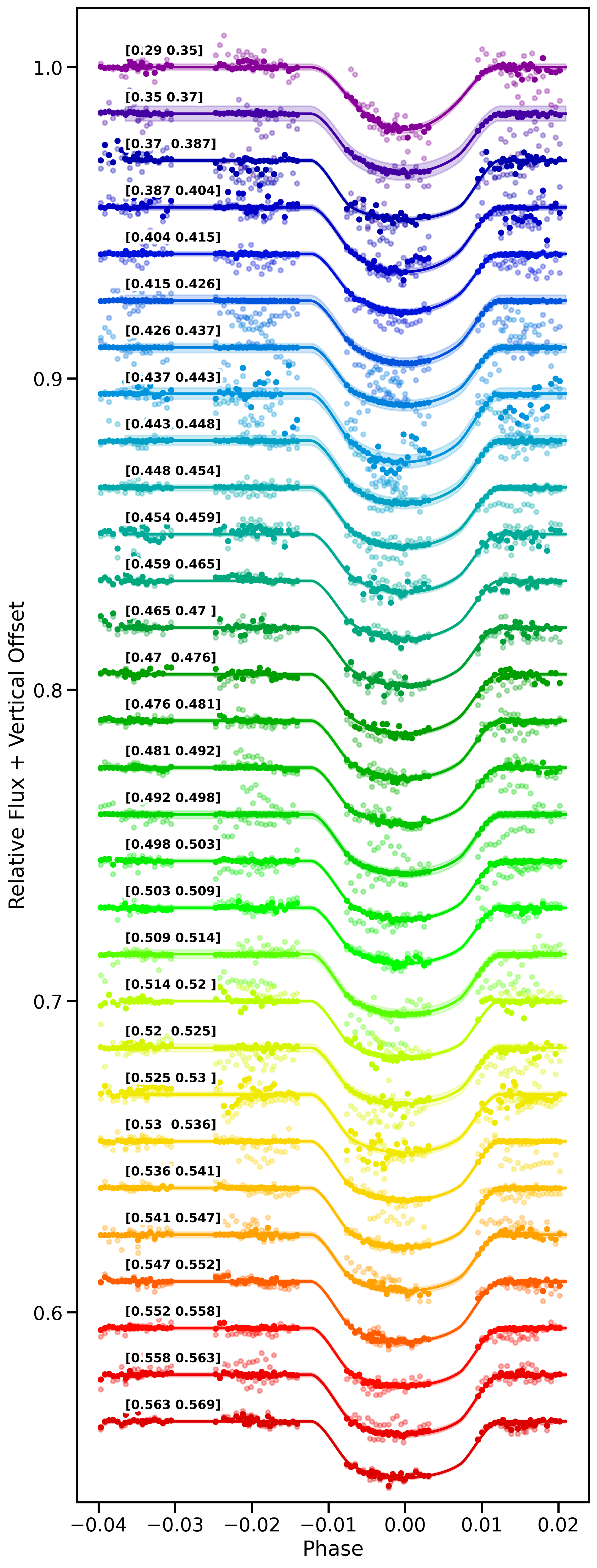}{0.4\textwidth}{}}
    \caption{The spectroscopic light curve fits for WASP-69\,b, from G430L (two transits, phased together). The left is the linear detrending using 2 PCs as regressors, and the right is the GP detrending using 5 PCs as regressors. The raw data points are shown as more transparent color version of the final detrended points for reference, and each wavelength bin is offset vertically. The beginning and end values of the wavelengths included in the bin are shown above the pre-transit baseline of each spectroscopic light curve (in $\mu$m). }
    \label{fig:slc_430}
\end{figure*}

\begin{figure*}
    \gridline{\fig{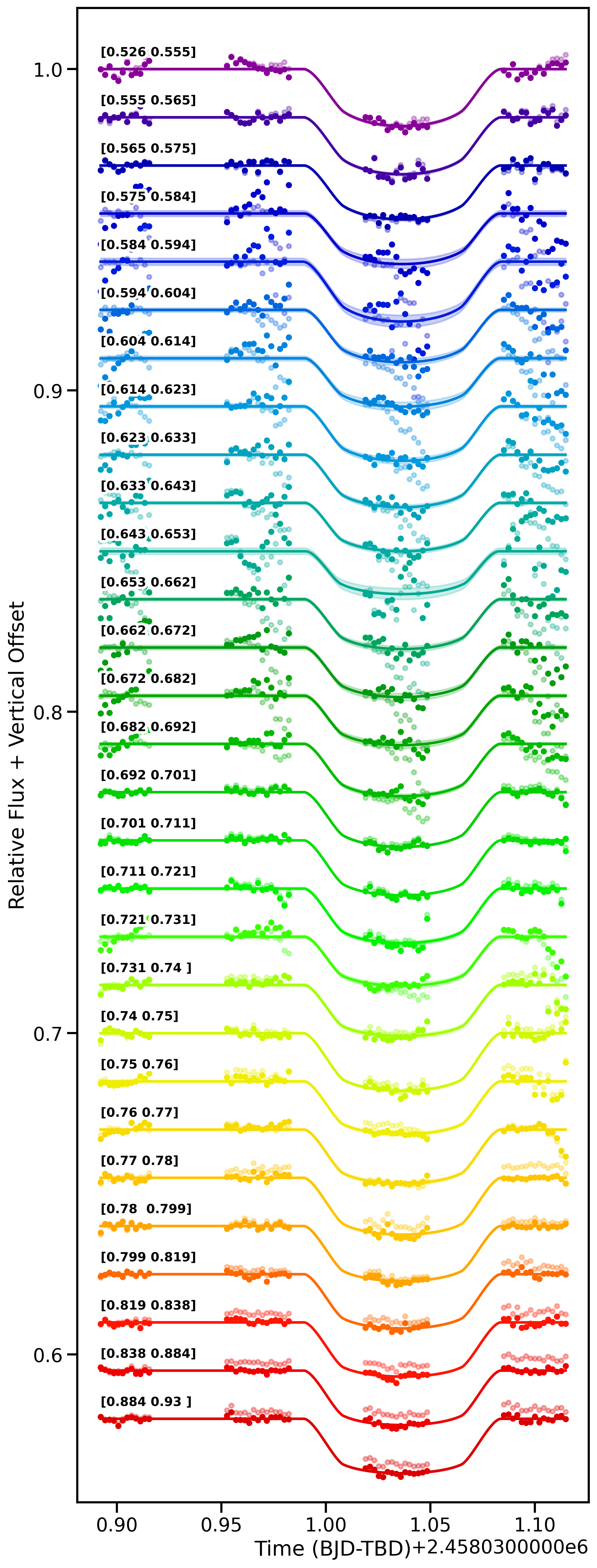}{0.4\textwidth}{}
          \fig{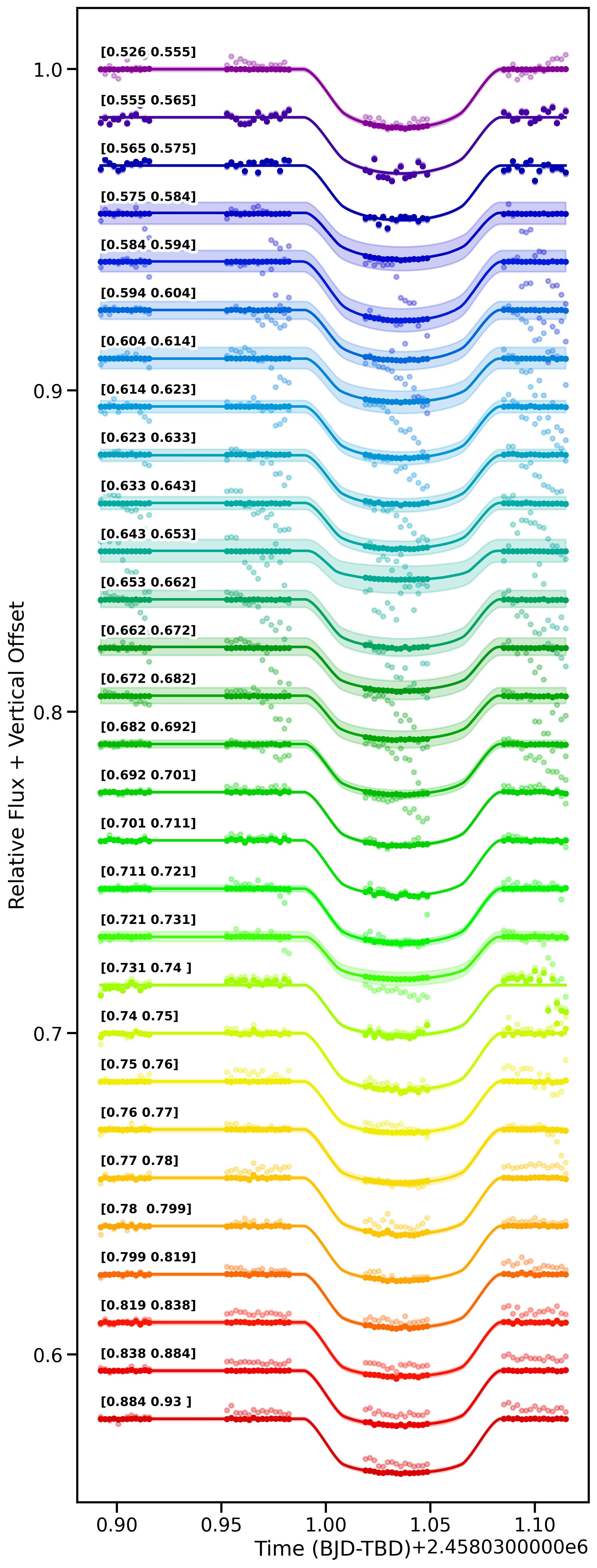}{0.4\textwidth}{}}
    \caption{The same as \autoref{fig:slc_430}, but for G750L (single transit).}
    \label{fig:slc_750}
\end{figure*}

\begin{figure*}
    \centering
\gridline{\fig{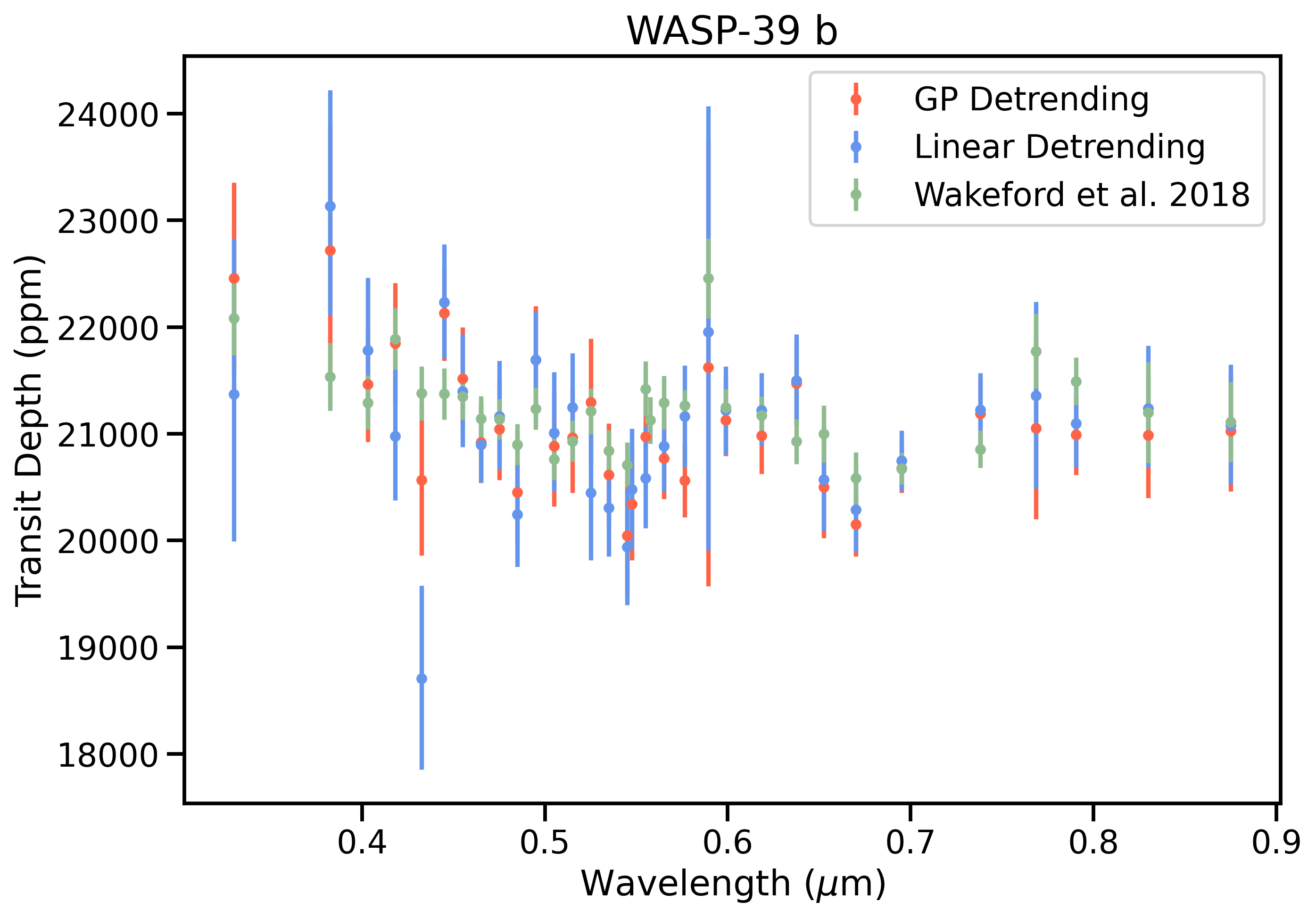}{0.49\textwidth}{}
          \fig{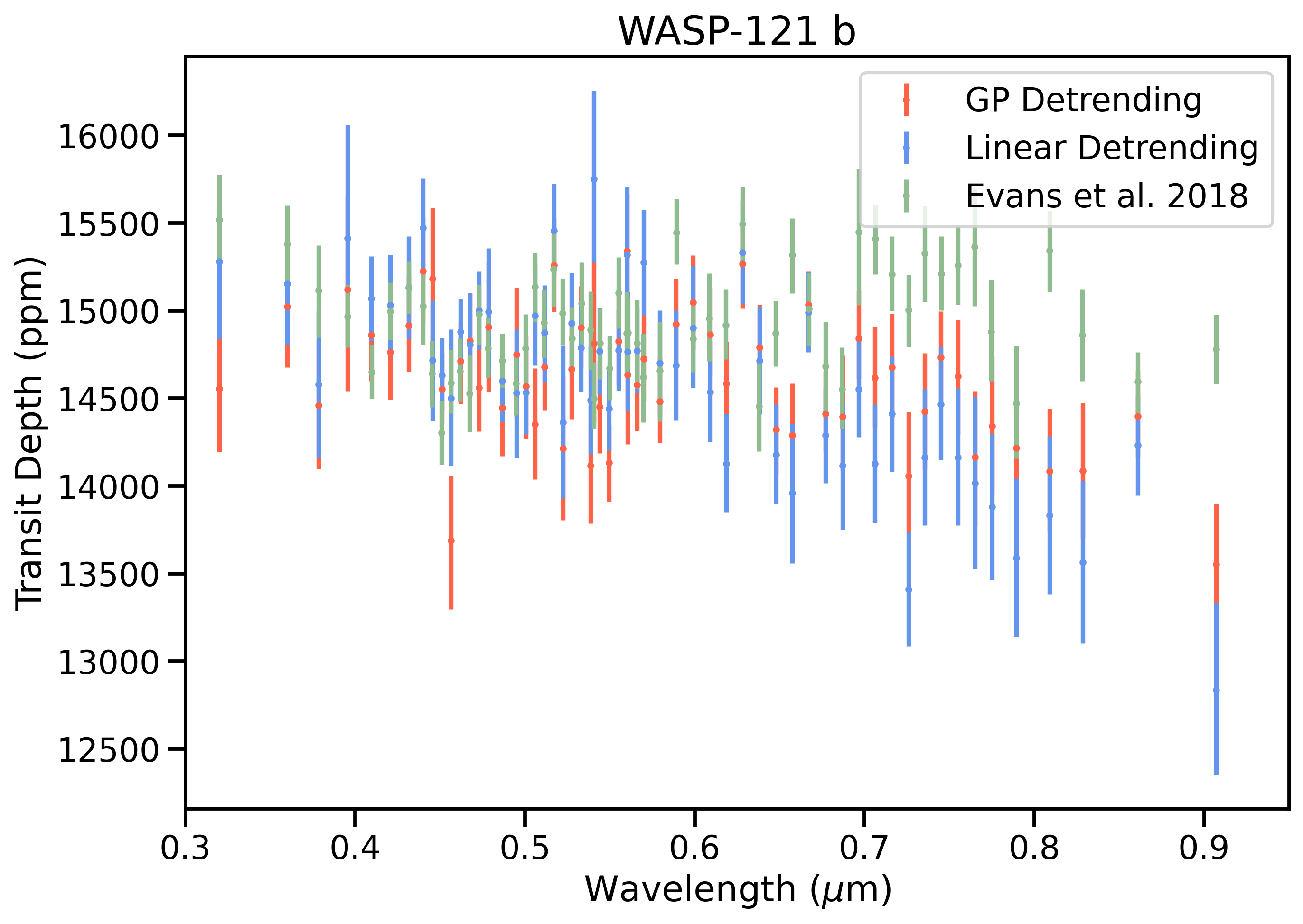}{0.49\textwidth}{}}
\gridline{\fig{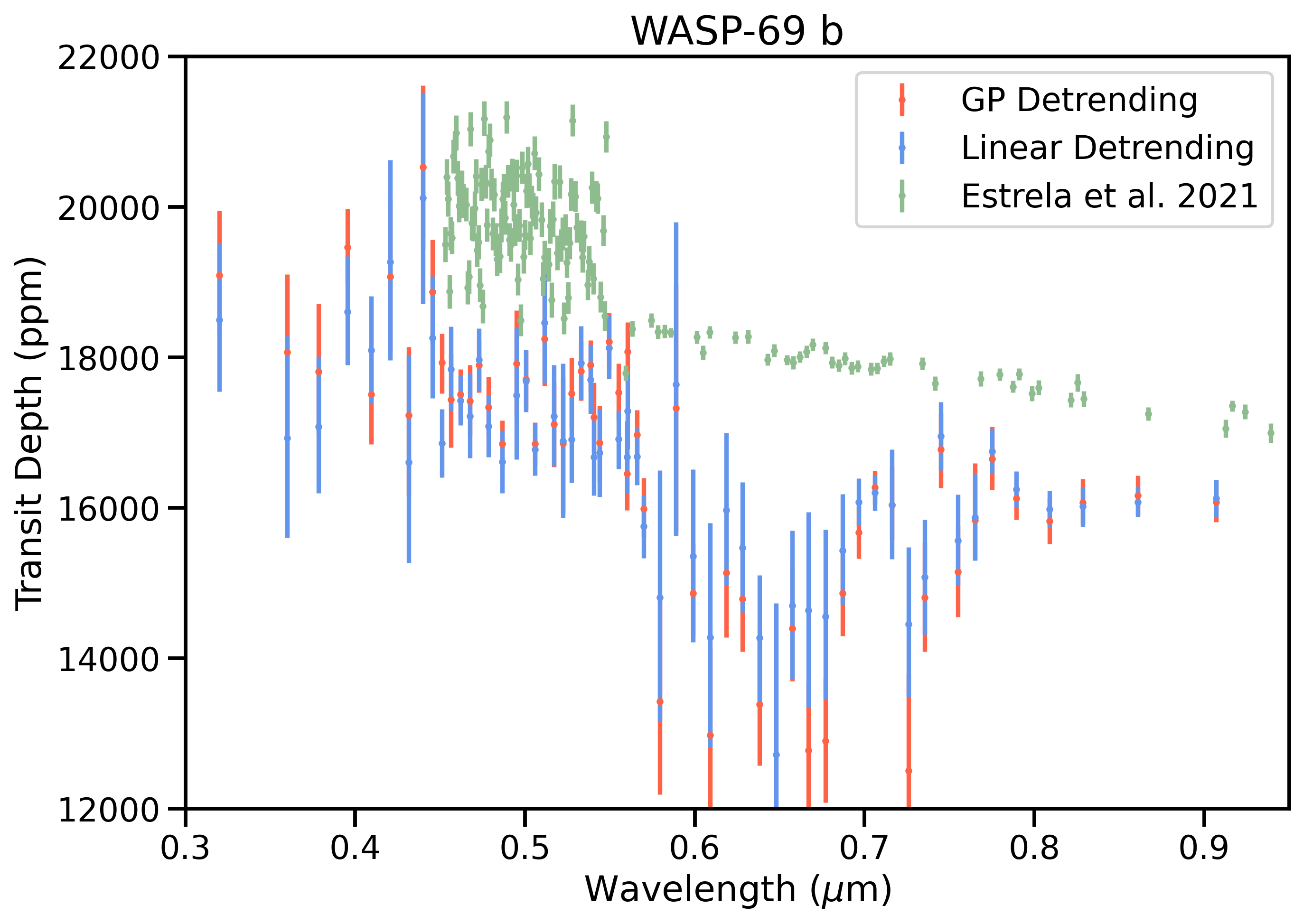}{0.49\textwidth}{}
          \fig{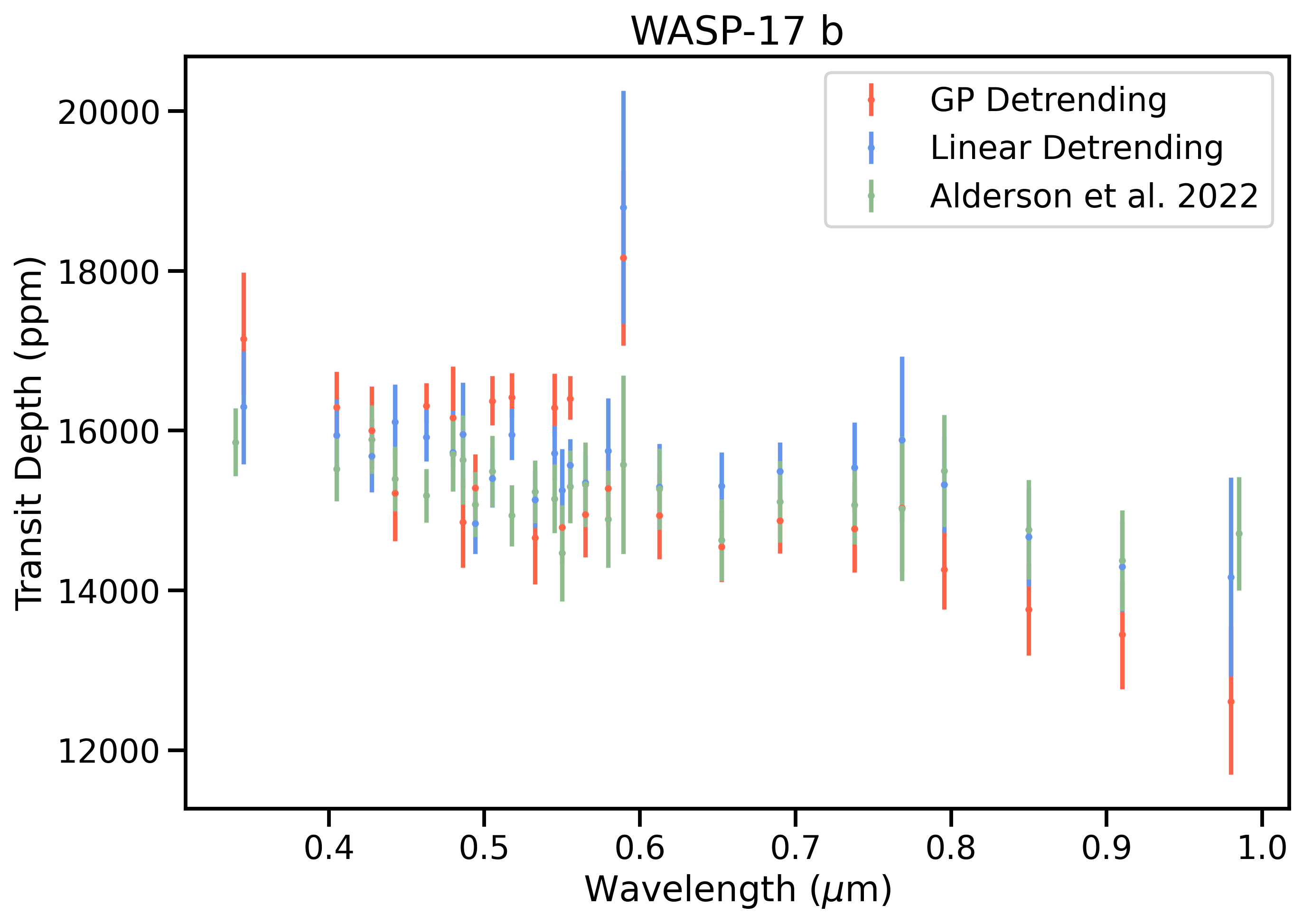}{0.49\textwidth}{}}
    \caption{The final spectra for each of our four test planets, with both linear detrending (with 2 PCs as regressors) and GP detrending (with 5 PCs as regressors). Note that the y-axis scale of WASP-69\,b has been restricted to hide some large outliers due to the large systematics present in the G750L data (clearly seen in the white light curves, see \autoref{fig:wlc}), but the full scale can be seen in \autoref{fig:w69_full} in \autoref{app:w69_full}. Also shown are past results from the literature for comparison: \citet{Wakeford_2018} for WASP-39\,b, \citet{Evans_2018} for WASP-121\,b, \citet{Estrela_2021} for WASP-69\,b, and \citet{Alderson_2022} for WASP-17\,b. Note that our results show close agreement with those for WASP-39\,b, WASP-121\,b (after offset, see \autoref{fig:w121_offset}), and WASP-17\,b, but significant difference for WASP-69\,b.}
    \label{fig:final_spec}
\end{figure*}

These input vectors contain a large amount of information about various processes that impact the observations and their quality, but are highly correlated with each other, especially with respect to HST orbital phase. In addition, the previously described penalizing behavior of exponential priors has only been shown for uncorrelated input vectors. Additionally, the use of 10 systematics parameters is a nuisance with regards to the fitting, as it significantly increases the number of free parameters in the fit, especially when doing joint fits of multiple observations simultaneously. Therefore, we perform principal component analysis (PCA) to create a set of uncorrelated vectors from this initial set and test those as input parameters, both to conform with the expectation of the exponential priors and to attempt to cut down on the dimensionality of the fit. After renormalizing the resulting principal components (PCs), we test with both the GP and linear systematics methods from 1 to 10 PCs as inputs, which is the total number of PCs created from 10 input vectors. Following testing (see \autoref{app:systematics}), we found that the best systematics model for our data and purposes is 5 PCs for GP detrending, or 2 PCs for linear detrending. Using these best systematics models, we show our white light curve fits in \autoref{fig:wlc}, an example of our spectroscopic light curve fits for WASP-69\,b in \autoref{fig:slc_430} and \autoref{fig:slc_750} (for G430L and G750L, respectively -- the spectroscopic light curve fits for WASP-39\,b, WASP-121\,b, WASP-17\,b are shown in \autoref{app:slc}), and our final spectra for each of our test planets in \autoref{fig:final_spec}. The $R_p/R_s$ associated with our final white light and spectroscopic light curve fits are given in \autoref{tab:wlc_fits} and \autoref{tab:wlc_fits}, respectively. Jupyter notebooks used for generating these fits are available on Github\footnote{\url{https://github.com/natalieallen/stis_pipeline}}.  \\

%\startlongtable
\movetabledown=20mm
\begin{longrotatetable}
\begin{deluxetable*}{C C C C C C C C C C C C}
    \tabletypesize{\scriptsize}
    \tablecaption{Same as \autoref{tab:wlc_fits}, but for the spectroscopic light curves.\label{tab:slc_fits}}
    \tablecolumns{12}
    \tablehead{\multicolumn{3}{c}{WASP-39\,b} & \multicolumn{3}{c}{WASP-121\,b} & \multicolumn{3}{c}{WASP-69\,b} & \multicolumn{3}{c}{WASP-17\,b}\\
    Bin ($\mu$m) & Linear & GP & Bin ($\mu$m)  & Linear & GP & Bin ($\mu$m)  & Linear & GP & Bin ($\mu$m)  & Linear & GP}
    \startdata
        0.29-0.37 & $0.1499\pm0.0030$ & $0.1462 \pm 0.0050$ & 0.29-0.35 & $0.1206 \pm 0.0016$ & $0.1236 \pm 0.0019$ & 0.29-0.35 & $0.1382 \pm 0.0031$ & $0.1360 \pm 0.0037$ & 0.29-0.39 & $0.1310 \pm 0.0032$ & $0.1277 \pm 0.0028$\\
        0.37-0.395 & $0.1507\pm0.0039$ & $0.1521 \pm 0.0036$ & 0.35-0.37 & $0.1226 \pm 0.0014$ & $0.1231 \pm 0.0015$ & 0.35-0.37 & $0.1344 \pm 0.0040$ & $0.1301 \pm 0.0052$ & 0.39-0.42 & $0.1276 \pm 0.0018$ & $0.1263 \pm 0.0018$ \\
        0.395-0.411 & $0.1465\pm0.0018$ & $0.1475 \pm 0.0023$ & 0.37-0.387 & $0.1202 \pm 0.0016$ & $0.1207 \pm 0.0017$ & 0.37-0.387 & $0.1335 \pm 0.0034$ & $0.1307 \pm 0.0036$ & 0.42-0.435 & $0.1265 \pm 0.0022$ & $0.1252 \pm 0.0018$ \\
        0.411-0.425 & $0.1478\pm0.0019$ & $0.1448 \pm 0.0021$ & 0.387-0.404 & $0.1230 \pm 0.0024$ & $0.1242 \pm 0.0026$ & 0.387-0.404 & $0.1395 \pm 0.0018$ & $0.1364 \pm 0.0027$ & 0.435-0.45 & $0.1234 \pm 0.0024$ & $0.1269 \pm 0.0018$ \\
        0.425-0.44 & $0.1434\pm0.0025$ & $0.1367 \pm 0.0031$ & 0.404-0.415 & $0.1219 \pm 0.0011$ & $0.1228 \pm 0.0010$ & 0.404-0.415 & $0.1323 \pm 0.0025$ & $0.1345 \pm 0.0027$ & 0.45-0.475 & $0.1277 \pm 0.0011$ & $0.1262 \pm 0.0014$ \\
        0.44-0.45 & $0.1487\pm0.0015$ & $0.1491 \pm 0.0019$ & 0.415-0.426 & $0.1215 \pm 0.0012$ & $0.1226 \pm 0.0012$ & 0.415-0.426 & $0.1381 \pm 0.0037$ & $0.1388 \pm 0.0049$ & 0.475-0.484 & $0.1271 \pm 0.0025$ & $0.1254 \pm 0.0021$ \\
        0.45-0.46 & $0.1467\pm0.0016$ & $0.1463 \pm 0.0019$ & 0.426-0.437 & $0.1221 \pm 0.0012$ & $0.1230 \pm 0.0012$ & 0.426-0.437 & $0.1313 \pm 0.0041$ & $0.1289 \pm 0.0055$ & 0.484-0.488 & $0.1219 \pm 0.0023$ & $0.1263 \pm 0.0027$ \\
        0.46-0.47 & $0.1446 \pm 0.0013$ & $0.1446 \pm 0.0012$ & 0.437-0.443 & $0.1234 \pm 0.0012$ & $0.1244 \pm 0.0011$ & 0.437-0.443 & $0.1433 \pm 0.0038$ & $0.1418 \pm 0.0050$ & 0.488-0.5 & $0.1236 \pm 0.0017$ & $0.1218 \pm 0.0016$ \\
        0.47-0.48 & $0.1451 \pm 0.0017$ & $0.1455 \pm 0.0018$ & 0.443-0.448 &$0.1232 \pm 0.0018$ & $0.1213 \pm 0.0014$ & 0.443-0.448 & $0.1374 \pm 0.0026$ & $0.1351 \pm 0.0030$ & 0.5-0.51 & $0.1279 \pm 0.0012$ & $0.1241 \pm 0.0015$ \\
        0.48-0.49 & $0.1430 \pm 0.0020$ & $0.1423 \pm 0.0017$ & 0.448-0.454 & $0.1206 \pm 0.0008$ & $0.1210 \pm 0.0009$ & 0.448-0.454 & $0.1339 \pm 0.0015$ & $0.1298 \pm 0.0018$ & 0.51-0.525 & $0.1281 \pm 0.0012$ & $0.1263 \pm 0.0013$ \\
        0.49-0.5 & $0.1473 \pm 0.0017$ & $0.1472 \pm 0.0015$ & 0.454-0.459 & $0.1170 \pm 0.0017$ & $0.1204 \pm 0.0016$ & 0.454-0.459 & $0.1321 \pm 0.0024$ & $0.1336 \pm 0.0021$ & 0.525-0.54 & $0.1211 \pm 0.0024$ & $0.1230 \pm 0.0015$ \\
        0.5-0.51 & $0.1445 \pm 0.0020$ & $0.1449 \pm 0.0020$ & 0.459-0.465 & $0.1213 \pm 0.0010$ & $0.1220 \pm 0.0008$ & 0.459-0.465 & $0.1323 \pm 0.0013$ & $0.1320 \pm 0.0013$ & 0.54-0.55 & $0.1276 \pm 0.0017$ & $0.1254 \pm 0.0014$ \\
        0.51-0.52 & $0.1448 \pm 0.0018$ & $0.1458 \pm 0.0017$ & 0.465-0.47 & $0.1218 \pm 0.0012$ & $0.1217 \pm 0.0012$ & 0.465-0.47 & $0.1320 \pm 0.0018$ & $0.1312 \pm 0.0022$ & 0.55-0.56 & $0.1281 \pm 0.0011$ & $0.1248 \pm 0.0013$ \\
        0.52-0.53 & $0.1459 \pm 0.0020$ & $0.1430 \pm 0.0021$ & 0.47-0.476 & $0.1207 \pm 0.0011$ & $0.1225 \pm 0.0009$ & 0.47-0.476 & $0.1338 \pm 0.0014$ & $0.1341 \pm 0.0016$ & 0.56-0.57 & $0.1223 \pm 0.0022$ & $0.1239 \pm 0.0014$ \\
        0.53-0.54 & $0.1436 \pm 0.0017$ & $0.1425 \pm 0.0016$ & 0.476-0.481 & $0.1221 \pm 0.0015$ & $0.1224 \pm 0.0015$ & 0.476-0.481 & $0.1317 \pm 0.0015$ & $0.1307 \pm 0.0016$ & - & - & - \\
        0.54-0.55 & $0.1416 \pm 0.0020$ & $0.1412 \pm 0.0019$ & 0.481-0.492 & $0.1202 \pm 0.0012$ & $0.1208 \pm 0.0010$ & 0.481-0.492 & $0.1298 \pm 0.0013$ & $0.1289 \pm 0.0016$ & - & - & - \\
        0.55-0.56 & $0.1448 \pm 0.0013$ & $0.1435 \pm 0.0016$ & 0.492-0.498 & $0.1215 \pm 0.0016$ & $0.1205 \pm 0.0015$ & 0.492-0.498 & $0.1339 \pm 0.0026$ & $0.1323 \pm 0.0034$ & - & - & - \\
        0.56-0.57 & $0.1441 \pm 0.0013$ & $0.1445 \pm 0.0016$ & 0.498-0.503 & $0.1207 \pm 0.0012$ & $0.1206 \pm 0.0010$ & 0.498-0.503 & $0.1331 \pm 0.0015$ & $0.1330 \pm 0.0016$ & - & - & - \\
        - & - & - & 0.503-0.509 & $0.1198 \pm 0.0012$ & $0.1224 \pm 0.0012$ & 0.503-0.509 & $0.1298 \pm 0.0011$ & $0.1295 \pm 0.0014$ & - & - & - \\ 
        - & - & - & 0.509-0.514 & $0.1212 \pm 0.0011$ & $0.1220 \pm 0.0011$ & 0.509-0.514 & $0.1351 \pm 0.0023$ & $0.1359 \pm 0.0030$ & - & - & - \\
        - & - & - & 0.514-0.52 & $0.1235 \pm 0.0011$ & $0.1243 \pm 0.0011$ & 0.514-0.52  & $0.1308 \pm 0.0022$ & $0.1312 \pm 0.0026$ & - & - & - \\
        - & - & - & 0.52-0.525 & $0.1192 \pm 0.0018$ & $0.1198 \pm 0.0018$ & 0.52-0.525 & $0.1298 \pm 0.0036$ & $0.1299 \pm 0.0040$ & - & - & - \\
        - & - & - & 0.525-0.53 & $0.1211 \pm 0.0012$ & $0.1222 \pm 0.0012$ & 0.525-0.53 & $0.1324 \pm 0.0018$ & $0.1300 \pm 0.0022$ & - & - & -  \\
        - & - & - & 0.53-0.536 & $0.1221 \pm 0.0010$ & $0.1216 \pm 0.0010$ & 0.53-0.536 & $0.1335 \pm 0.0015$ & $0.1339 \pm 0.0018$ & - & - & -  \\
        - & - & - & 0.536-0.541 & $0.1188 \pm 0.0014$ & $0.1204 \pm 0.0013$ & 0.536-0.541 & $0.1338 \pm 0.0013$ & $0.1330 \pm 0.0018$ & - & - & -  \\
        - & - & - & 0.541-0.547 & $0.1202 \pm 0.0011$ & $0.1215 \pm 0.0010$ & 0.541-0.547 & $0.1299 \pm 0.0020$ & $0.1294 \pm 0.0023$ & - & - & -  \\
        - & - & - & 0.547-0.552 & $0.1189 \pm 0.0009$ & $0.1202 \pm 0.0010$ & 0.547-0.552 & $0.1349 \pm 0.0014$ & $0.1346 \pm 0.0016$ & - & - & -  \\
        - & - & - & 0.552-0.558 & $0.1218 \pm 0.0011$ & $0.1215 \pm 0.0009$ & 0.552-0.558 & $0.1324 \pm 0.0015$ & $0.1300 \pm 0.0015$ & - & - & -  \\
        - & - & - & 0.558-0.563 & $0.1210 \pm 0.0016$ & $0.1215 \pm 0.0014$ & 0.558-0.563 & $0.1344 \pm 0.0015$ & $0.1315 \pm 0.0020$ & - & - & -   \\
        - & - & - & 0.563-0.569 & $0.1207 \pm 0.0011$ & $0.1215 \pm 0.0010$ & 0.563-0.569 & $0.1303 \pm 0.0013$ & $0.1292 \pm 0.0015$ & - & - & -  \\
        \hline
        0.53-0.565 &$0.1426 \pm 0.0019$ & $0.1431 \pm 0.0020$ & 0.526-0.555 & $0.1217 \pm 0.0019$ & $0.1255 \pm 0.0020$ & 0.526-0.555 & $0.1312 \pm 0.0018$ & $0.1291 \pm 0.0021$ & 0.53-0.57 & $0.1216 \pm 0.0019$ & $0.1235 \pm 0.0021$ \\
        0.565-0.588 & $0.1434 \pm 0.0012$ & $0.1455 \pm 0.0016$ & 0.555-0.565 & $0.1239 \pm 0.0013$ & $0.1238 \pm 0.0016$ & 0.555-0.565 & $0.1283 \pm 0.0019$ & $0.1291 \pm 0.0019$ & 0.57-0.589 & $0.1236 \pm 0.0020$ & $0.1255 \pm 0.0026$ \\
        0.588-0.591 & $0.1470 \pm 0.0072$ & $0.1482 \pm 0.0071$ & 0.565-0.575 & $0.1214 \pm 0.0010$ & $0.1236 \pm 0.0012$ & 0.565-0.575 & $0.1264 \pm 0.0016$ & $0.1255 \pm 0.0017$ & 0.589-0.59 & $0.1371 \pm 0.0053$ & $0.1348 \pm 0.0041$ \\
        0.591-0.607 & $0.1454 \pm 0.0014$ & $0.1457 \pm 0.0014$ & 0.575-0.584 & $0.1203 \pm 0.0010$ & $0.1212 \pm 0.0012$ & 0.575-0.584 & $0.1159 \pm 0.0059$ & $0.1217 \pm 0.0070$ & 0.59-0.635 & $0.1222 \pm 0.0022$ & $0.1237 \pm 0.0022$ \\
        0.607-0.63 & $0.1448 \pm 0.0013$ & $0.1457 \pm 0.0011$ & 0.584-0.594 & $0.1222 \pm 0.0011$ & $0.1212 \pm 0.0013$ & 0.584-0.594 & $0.1316 \pm 0.0063$ & $0.1328 \pm 0.0081$ & 0.635-0.67 & $0.1206 \pm 0.0019$ & $0.1237 \pm 0.0017$ \\
        0.63-0.645 & $0.1465 \pm 0.0014$ & $0.1466 \pm 0.0015$ & 0.594-0.604 & $0.1227 \pm 0.0011$ & $0.1221 \pm 0.0015$ & 0.594-0.604 & $0.1219 \pm 0.0027$ & $0.1239 \pm 0.0047$ & 0.67-0.71 & $0.1220 \pm 0.0017$ & $0.1245 \pm 0.0015$ \\
        0.645-0.66 & $0.1432 \pm 0.0017$ & $0.1434 \pm 0.0017$ & 0.604-0.614 & $0.1219 \pm 0.0011$ & $0.1206 \pm 0.0012$ & 0.604-0.614 & $0.1139 \pm 0.0051$ & $0.1195 \pm 0.0064$ & 0.71-0.766 & $0.1215 \pm 0.0023$ & $0.1247 \pm 0.0023$ \\
        0.66-0.68 & $0.1420 \pm 0.0011$ & $0.1424 \pm 0.0014$ & 0.614-0.623 & $0.1208 \pm 0.0010$ & $0.1189 \pm 0.0012$ & 0.614-0.623 & $0.1230 \pm 0.0037$ & $0.1264 \pm 0.0041$ & 0.766-0.771 & $0.1226 \pm 0.0033$ & $0.1260 \pm 0.0041$ \\
        0.68-0.71 & $0.1438 \pm 0.0009$ & $0.1440 \pm 0.0010$ & 0.623-0.633 & $0.1236 \pm 0.0010$ & $0.1238 \pm 0.0012$ & 0.623-0.633 & $0.1216 \pm 0.0032$ & $0.1244 \pm 0.0035$ & 0.771-0.82 & $0.1194 \pm 0.0021$ & $0.1238 \pm 0.0024$ \\
        0.71-0.766 & $0.1456 \pm 0.0011$ & $0.1457 \pm 0.0012$ & 0.633-0.643 & $0.1216 \pm 0.0010$ & $0.1213 \pm 0.0013$ & 0.633-0.643 & $0.1157 \pm 0.0038$ & $0.1195 \pm 0.0035$ & 0.82-0.88 & $0.1173 \pm 0.0024$ & $0.1211 \pm 0.0025$ \\
        0.766-0.771 & $0.1451 \pm 0.0030$ & $0.1461 \pm 0.0030$ & 0.643-0.653 & $0.1197 \pm 0.0011$ & $0.1191 \pm 0.0012$ & 0.643-0.653 & $0.0921 \pm 0.0139$ & $0.1128 \pm 0.0089$ & 0.88-0.94 & $0.1160 \pm 0.0030$ & $0.1196 \pm 0.0024$ \\
        0.771-0.81 & $0.1449 \pm 0.0013$ & $0.1453 \pm 0.0014$ & 0.653-0.662 & $0.1195 \pm 0.0013$ & $0.1182 \pm 0.0017$ & 0.653-0.662 & $0.1200 \pm 0.0034$ & $0.1213 \pm 0.0041$ & 0.94-1.02 & $0.1123 \pm 0.0042$ & $0.1190 \pm 0.0052$ \\
        0.81-0.85 & $0.1449 \pm 0.0020$ & $0.1457 \pm 0.0020$ & 0.662-0.672 & $0.1226 \pm 0.0008$ & $0.1224 \pm 0.0009$ & 0.662-0.672 & $0.1130 \pm 0.0039$ & $0.1210 \pm 0.0054$ & - & - & -\\
        0.85-0.9 & $0.1450 \pm 0.0019$ & $0.1452 \pm 0.0020$ & 0.672-0.682 & $0.1200 \pm 0.0010$ & $0.1195 \pm 0.0011$ & 0.672-0.682 & $0.1136 \pm 0.0037$ & $0.1207 \pm 0.0048$ & - & - & -\\
        - & - & - & 0.682-0.692 & $0.1200 \pm 0.0015$ & $0.1188 \pm 0.0015$ & 0.682-0.692 & $0.1219 \pm 0.0024$ & $0.1242 \pm 0.0030$ & - & - & -\\
        - & - & - & 0.692-0.701 & $0.1218 \pm 0.0011$ & $0.1206 \pm 0.0011$ & 0.692-0.701 & $0.1252 \pm 0.0014$ & $0.1268 \pm 0.0013$ & - & - & -\\
        - & - & - & 0.701-0.711 & $0.1209 \pm 0.0012$ & $0.1189 \pm 0.0014$ & 0.701-0.711 & $0.1276 \pm 0.0009$ & $0.1273 \pm 0.0009$ & - & - & -\\
        - & - & - & 0.711-0.721 & $0.1211 \pm 0.0013$ & $0.1200 \pm 0.0014$ & 0.711-0.721 & $0.1266 \pm 0.0020$ & $0.1266 \pm 0.0029$ & - & - & -\\
        - & - & - & 0.721-0.731 & $0.1186 \pm 0.0015$ & $0.1158 \pm 0.0014$ & 0.721-0.731 & $0.1118 \pm 0.0058$ & $0.1202 \pm 0.0043$ & - & - & -\\
        - & - & - & 0.731-0.74 & $0.1201 \pm 0.0014$ & $0.1190 \pm 0.0017$ & 0.731-0.74 & $0.1217 \pm 0.0030$ & $0.1228 \pm 0.0031$ & - & - & -\\
        - & - & - & 0.74-0.75 & $0.1214 \pm 0.0011$ & $0.1203 \pm 0.0014$ & 0.74-0.75 & $0.1295 \pm 0.0020$ & $0.1302 \pm 0.0018$ & - & - & -\\
        - & - & - & 0.75-0.76 & $0.1209 \pm 0.0014$ & $0.1190 \pm 0.0017$ & 0.75-0.76 & $0.1231 \pm 0.0025$ & $0.1248 \pm 0.0025$ & - & - & -\\
        - & - & - & 0.76-0.77 & $0.1190 \pm 0.0017$ & $0.1184 \pm 0.0021$ & 0.76-0.77 & $0.1259 \pm 0.0030$ & $0.1260 \pm 0.0023$ & - & - & -\\
        - & - & - & 0.77-0.78 & $0.1198 \pm 0.0017$ & $0.1178 \pm 0.0018$ & 0.77-0.78 & $0.1290 \pm 0.0016$ & $0.1294 \pm 0.0012$ & - & - & -\\
        - & - & - & 0.78-0.799 & $0.1192 \pm 0.0013$ & $0.1166 \pm 0.0019$ & 0.78-0.799 & $0.1270 \pm 0.0011$ & $0.1275 \pm 0.0009$ & - & - & -\\
        - & - & - & 0.799-0.819 & $0.1187 \pm 0.0015$ & $0.1176 \pm 0.0019$ & 0.799-0.819 & $0.1258 \pm 0.0012$ & $0.1264 \pm 0.0010$ & - & - & -\\
        - & - & - & 0.819-0.838 & $0.1187 \pm 0.0016$ & $0.1165 \pm 0.0020$ & 0.819-0.838 & $0.1268 \pm 0.0012$ & $0.1266 \pm 0.0011$ & - & - & -\\
        - & - & - & 0.838-0.884 & $0.1200 \pm 0.0011$ & $0.1193 \pm 0.0012$ & 0.838-0.884 & $0.1271 \pm 0.0010$ & $0.1268 \pm 0.0008$ & - & - & -\\
        - & - & - & 0.884-0.93 & $0.1164 \pm 0.0015$ & $0.1133 \pm 0.0022$ & 0.884-0.93 & $0.1268 \pm 0.0011$ & $0.1270 \pm 0.0010$ & - & - & - \\
    \enddata
\end{deluxetable*}
\end{longrotatetable}

\section{Final Spectra}\label{sec:w69b}
\subsection{Results for WASP-39\,b, WASP-17\,b, and WASP-121\,b}
We show our final spectra compared to past published spectra of the same data in \autoref{fig:final_spec} for comparison. Our analysis of WASP-39\,b and WASP-17\,b broadly agree with the results presented in past publications, and thus we refer the reader to the conclusions of those works for the atmospheric composition of said planets. A first look at \autoref{fig:final_spec}, might make it seem like there is a significant difference in the result for WASP-121\,b, but if we account for a vertical absolute transit depth offset due to differences in the system parameters and detrending method used, and fix the limb darkening to the values from \citet{Evans_2018}, our results are actually in agreement with the past values (this is discussed further in \autoref{sec:ld}). Overall, the resulting errors on the spectral points for our analysis is sometimes larger, which is likely due to the use of consistent methods for all of the planets, which may not be the optimal methods for each dataset individually (for example, the choice of spectral extraction aperture, see \autoref{sec:aper} and \autoref{app:aper}), and is necessary for the goal of our uniform archival project.  %While we are unsure what the exact cause of this difference in spectra is, we note that \citet{Estrela_2021} uses an exponential ramp to fit for systematics in both the WFC3/IR and STIS data, although the exponential systematics behavior is believed to be due to charge trapping behavior that does not affect the STIS detectors in the same way, and thus this difference in systematics treatment may be causing the difference. %Even offsetting vertically by the absolute transit depth, a difference in the spectral shape remains (see \autoref{sec:w69_offset}). As the \citet{Estrela_2021} analysis uses the four parameter nonlinear limb darkening law from \citet{Claret_2000} rather than a quadratic limb darkening scheme as we use, it is possible that a difference in limb darkening treatment is the cause of the difference in spectra. However, since \citet{Estrela_2021} do not report their fit limb darkening values, we are unable to verify that this is the cause definitively. \\

\subsection{Results for WASP-69\,b}
\begin{figure*}
    \centering
\gridline{\fig{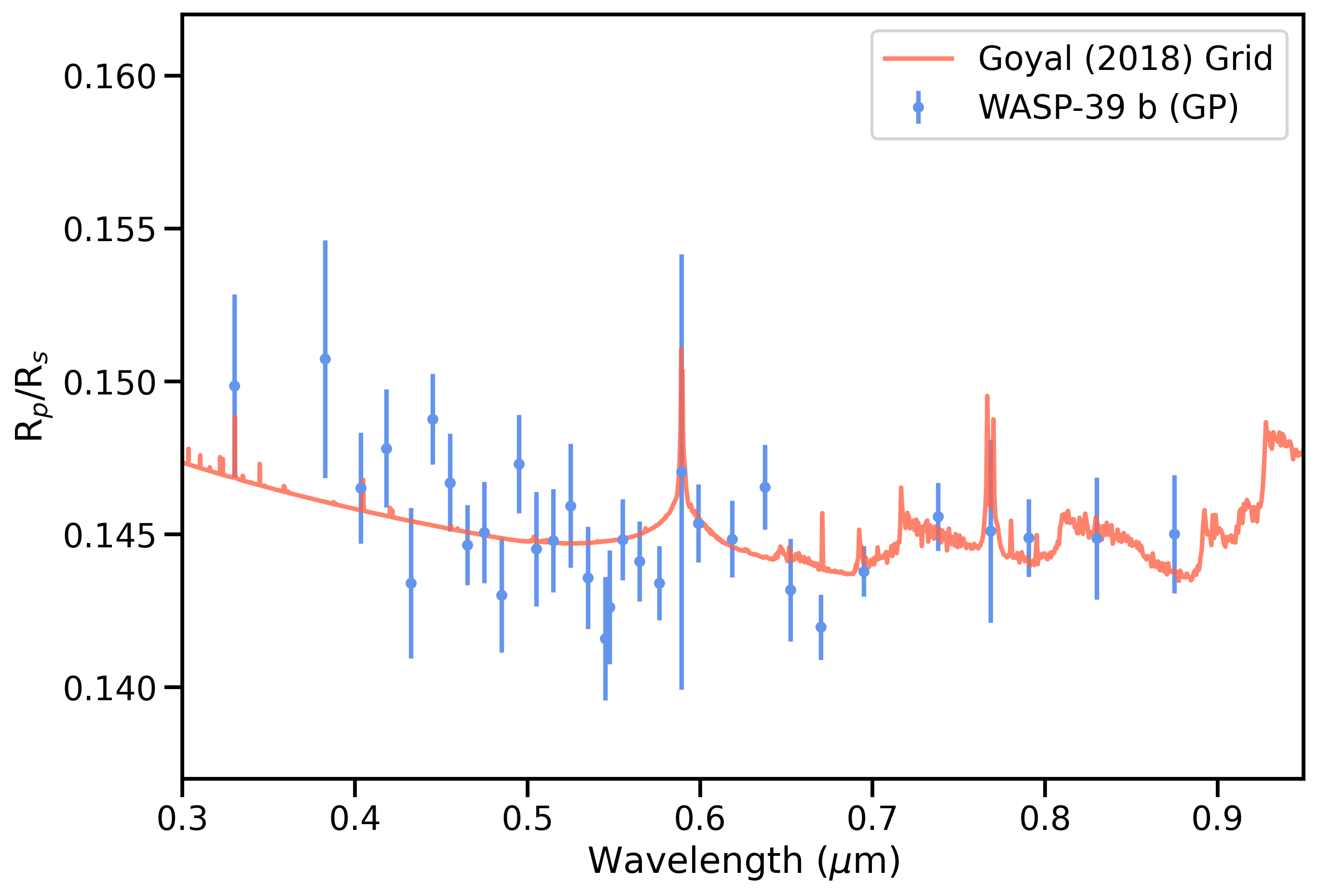}{0.49\textwidth}{}
          \fig{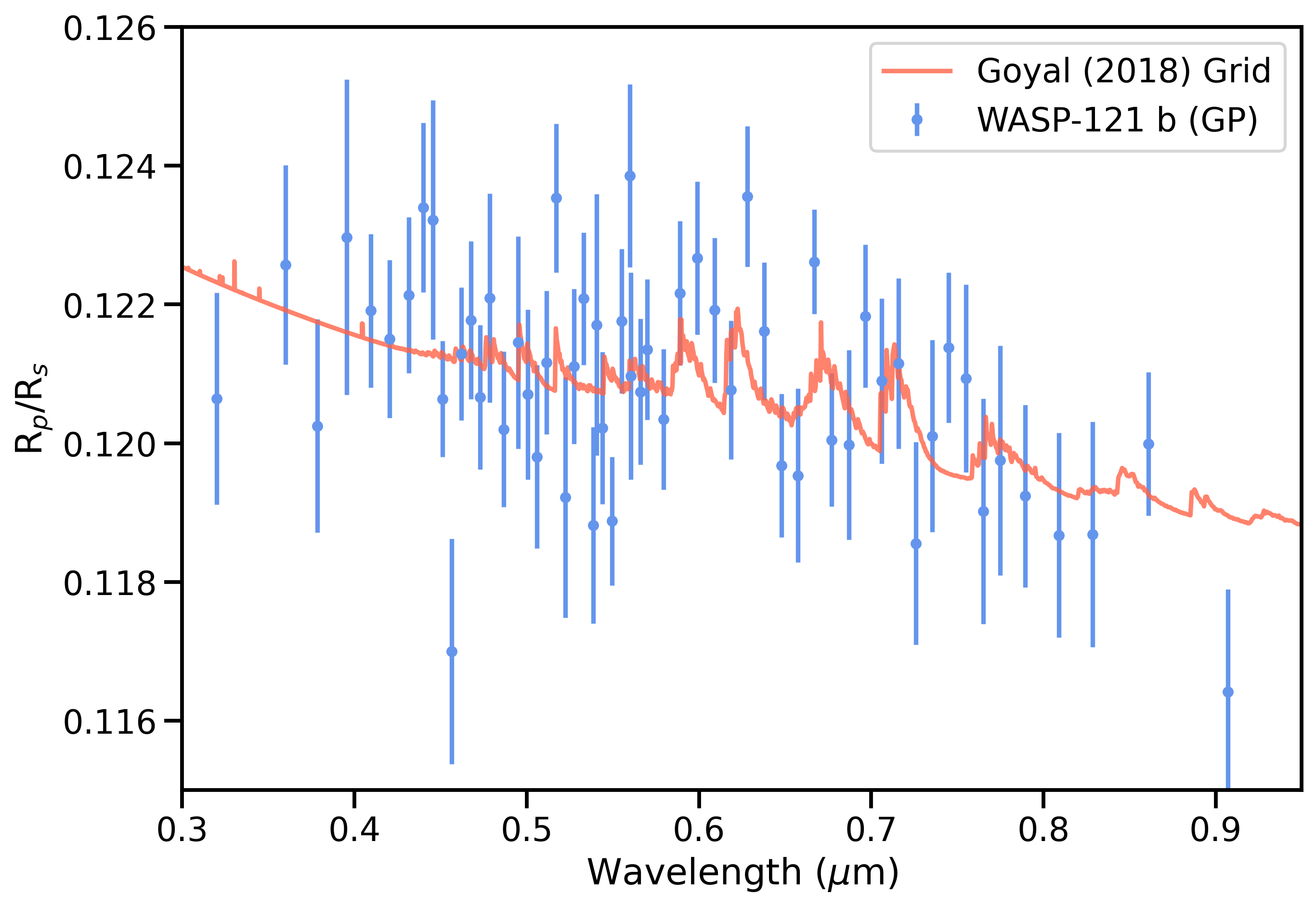}{0.49\textwidth}{}}
\gridline{\fig{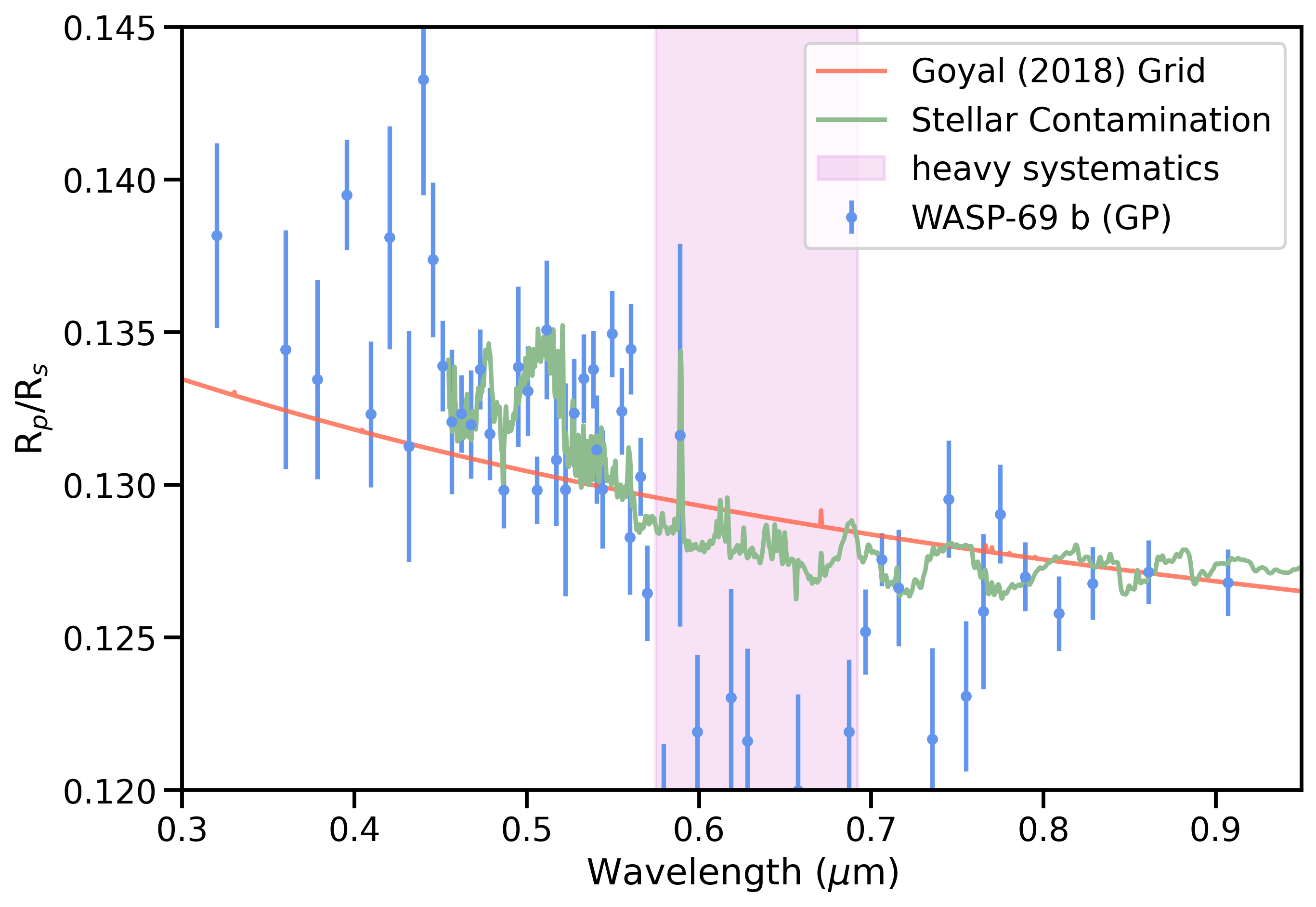}{0.49\textwidth}{}
          \fig{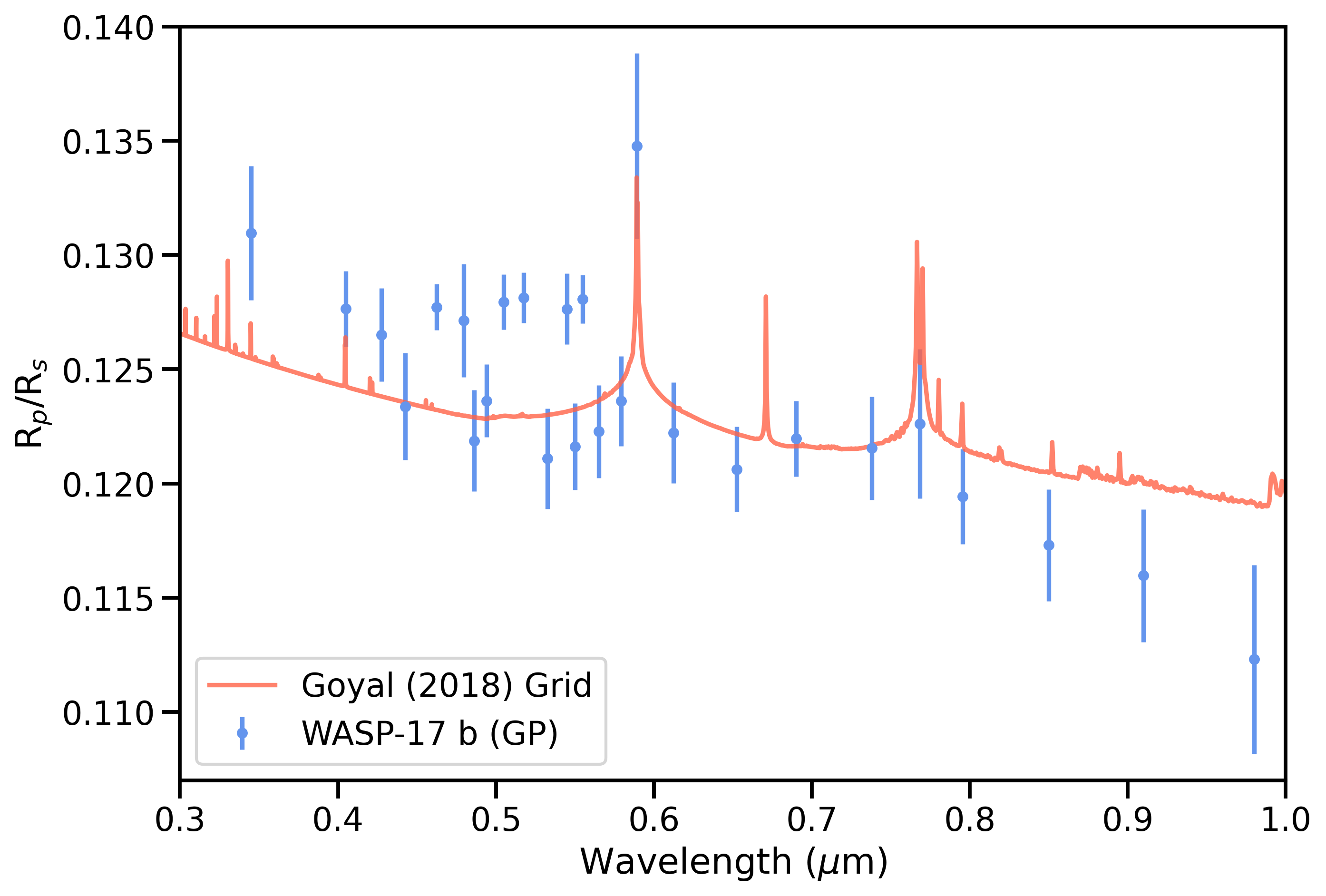}{0.49\textwidth}{}}
    \caption{Our final best spectrum for each of our four test planets, shown along with models from \citet{goyal_2019}. Note that due to the general lack of features in this wavelength range, there are multiple models equally consistent with the data, but one model is shown as demonstrative of such a family of models. For WASP-39\,b, WASP-121\,b, and WASP-17\,b, due to the similarities in our final spectra, we refer the reader to the planet specific papers and the modeling results therein for an in-depth discussion of the specific properties of the planets' atmospheres. For WASP-69\,b, for which our final spectrum differs significantly from that in past results, we present our best-fit models specifically: a potential haze scattering slope forward model \citep{goyal_2019} and stellar contamination model. It is impossible to confidently distinguish between these models with the quality of the data. The points in the shaded region are affected by large systematics (see \autoref{fig:slc_750}) and so are left out of the analysis.}
    \label{fig:models}
\end{figure*}

Our analysis of WASP-69\,b differs significantly from that of \citet{Estrela_2021}, especially regarding the handling of systematics. We consider the GP detrended spectra to be our final result (see \autoref{app:systematics} for the changes associated with the use of different detrending methods), and show these spectra along with corresponding best-fit models from the \citet{goyal_2019} grid in \autoref{fig:models}. Due to our spectrum's final consistency with previously published results, we simply show a representative model from a family of models consistent with the data, and defer to the more in-depth modeling work completed in past publications for WASP-39\,b \citep{Wakeford_2018}, WASP-121\,b \citep{Evans_2018}, and WASP-17\,b \citep{Alderson_2022}. However, for WASP-69\,b, due to the significant difference in our result in comparison to past work \citep{Estrela_2021}, we decide to perform a new modeling exploration effort below. \\

We carry out a new model comparison analysis to determine the atmospheric composition suggested by our spectrum of WASP-69\,b. To do this, we compare our final spectrum to the \citet{goyal_2019} forward model grid computed for WASP-69\,b. This grid contains models across 22 equilibrium temperatures ($400-2600$ K), four planetary gravities ($5-50$ m/s$^2$), five atmospheric metallicities (1x-200x), four C/O ratios (0.35-1.0), four scattering haze parameters, four uniform cloud parameters, and we consider the rainout condensation grid. Our best-fit model using a basic chi-squared minimization (which is the same for the linear and the GP detrending) is 1263K, [Fe/H] = -2.3, C/O = 0.75, haze parameter of 1100, and cloud parameter of 0 (no clouds). We note that this low [Fe/H] seems to be entirely driven by the non-detection of Na and K (both of which lie in the contaminated, masked section), and that the C/O is fully unconstrained. Essentially, this forward model fit just suggests a significant haze scattering slope. We believe that the fit temperature, which is significantly higher than the $T_{eq}=988$ K, is also due to the strength of this large scattering slope signal. This model is shown with the data in the WASP-69\,b panel in \autoref{fig:models} (bottom left).\\

This optical slope can be indicative of aerosols, as suggested by our forward model, but an upwards trend in the optical is also a signature of potential stellar activity. Therefore, we additionally run a stellar contamination retrieval using the \texttt{Exoretrievals} \citep{Espinoza_2019} framework, which follows \citet{Rackham_2018} and calculates a stellar contaminated spectrum allowing for some temperature starspot's covering fraction, both in and out of the transit chord (occulted and unocculted, respectively). The best-fit retrieved model is also shown in the bottom left of \autoref{fig:models}, and corresponds to contamination by unocculted spots colder than the photosphere. The strength of this signature is highly degenerate with the strength of the haze scattering slope, and so it is impossible to confidently distinguish between the two with the precision of the STIS data. Additional observations at higher resolution and precision with JWST NIRISS/SOSS (as this target is too bright for NIRSpec/PRISM) may be able to add extra evidence in favor of one or the other, especially if these observations in a different epoch show a different stellar contamination signal. 

\section{Lessons Learned from HST Data}\label{sec:lessons}
\subsection{Linear models with more degrees of freedom can lead to high variation in spectra}
\begin{figure}
    \centering
    \includegraphics[width=0.45\textwidth]{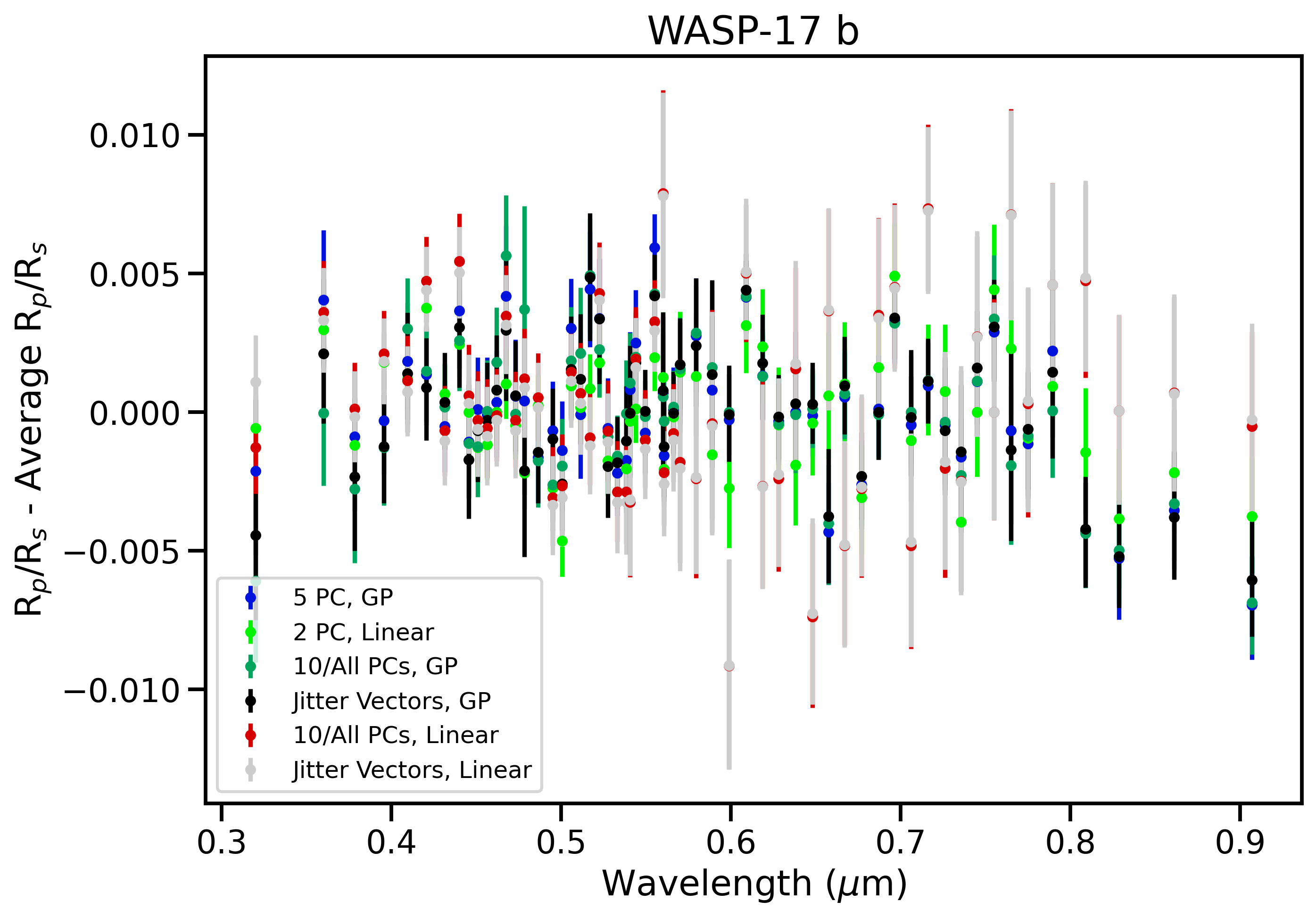}
    \caption{The spectrum of WASP-17\,b fit using different detrending methods, each with its average $R_p/R_s$ removed. It can be seen that the silver and red points, corresponding to linear detrending fits with 10 detrending vectors (either PCs or original vectors, respectively) have a larger variance than the rest of the points, which encompass the best-fit linear and GP detrending models (2 PCs and 5 PCs, respectively), as well as the two GP fits using those same 10 detrending vectors. To see these spectra in context with the rest of the tested methods, see \autoref{fig:sys_test_sub}.}
    \label{fig:variation}
\end{figure}

When one provides extra degrees of freedom to a model, it increases the flexibility at the risk of decreasing the predictive ability of the model. When detrending systematics in HST spectra, this flexibility can lead to a large variability in the spectra, which is not seen in results with less flexible models. With linear detrending models, which assume the form of the systematics generating function, this behavior can clearly be seen in the difference in spectral fits between those using low numbers of PCs and those with large numbers of PCs (see \autoref{app:systematics}). On the other hand, we don't see this behavior materialize in GP detrending models, likely because of the inherent flexibility of those models combined with our exponential priors on the GP parameters which penalize overfitting. We show this effect in our spectral fits for WASP-17\,b in  \autoref{fig:variation}, where it can be seen that the silver and red spectra, which use 10 detrending vectors (either in PC or original vector form, respectively), have significantly more spread than the rest of the points, which consist of the best-fit linear and GP models (2 PCs and 5 PCs, respectively) and also the two GP 10 detrending vectors fits. We note that the average $R_p/R_s$ for each spectrum has been removed here to show the relative size of this effect, as there is an inherent shift in absolute $R_p/R_s$ when correcting for systematics in different ways (see \autoref{app:systematics}).

\subsection{Testing error estimation through comparing single transits} \label{sec:error_test}
The fact that two transits are observed with the G430L grating for these test planets gives us a valuable opportunity to test which systematics detrending method's estimated errors are more representative of the true error in the data. We can treat each of the G430L transits as a separate noise instance of the same data. If we fit each transit's spectroscopic light curves separately, the estimated errors should capture the spread between the two resulting spectra if they are truly indicative of our uncertainty in the results. We carry out this test for the best-fit systematics model for the linear and GP methods (using the typical 2 PCs and 5 PCs, respectively) for WASP-17\,b, as 1) the data are phased almost identically, getting rid of the inevitable increase in transit fit precision from the increased phase coverage, and 2) the data lacks post-transit baseline, and thus is a larger challenge for detrending methods and again, does not gain any additional information that would be more helpful in a joint fit (see \autoref{fig:wlc}). \\

\begin{figure}[htb!]
    \centering
    \includegraphics[width=0.45\textwidth]{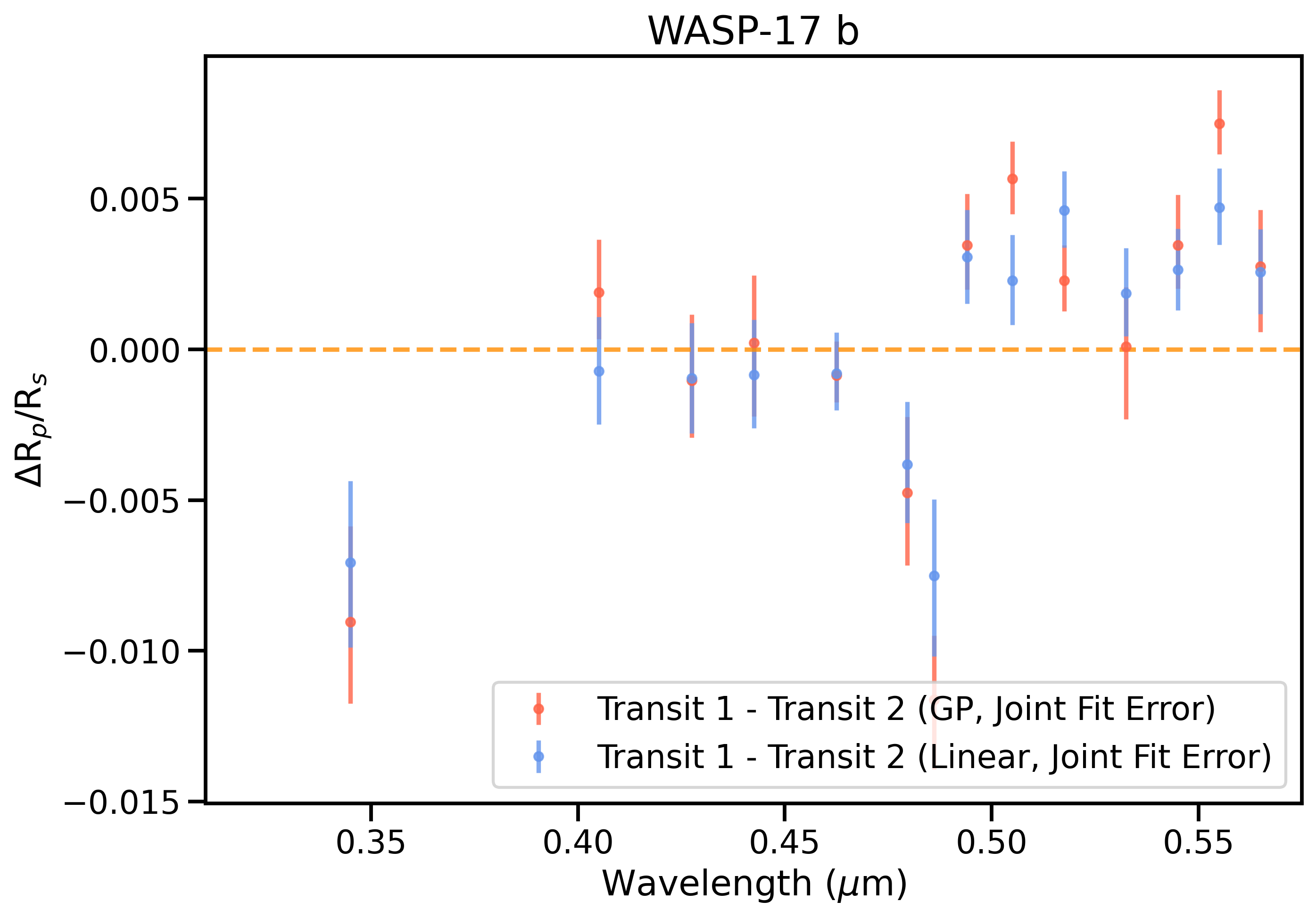}
    \caption{The difference in $R_p/R_s$ between the individually fit WASP-17\,b G430L transits, plotted with the error bars from the corresponding method's joint fit error. If the joint fit is correctly estimating the scatter in predicted outcomes given the state of the data, the errors should encompass the spread in values from each transit's individual fit, and be consistent with zero. It can be seen that, besides the first spectral point, the GP's errors are consistent with zero throughout the spectrum, but that there are multiple linear points that are inconsistent with zero. This suggests caution for using linear detrending methods to estimate the error with problematic data sets.}
    \label{fig:err_test}
\end{figure}

We fit each of the G430L transits of WASP-17\,b individually, and then subtract the average transit depth for each transit (to get rid of any absolute transit depth difference, which can be caused by a variety of effects, see \autoref{app:systematics}), and then subtract the resulting normalized second transit from the first transit. We do this for both linear and GP detrending methods, and then compare the resulting scatter in the data to the size of the error bars estimated by the corresponding method's G430L joint fit. We show the result of this test in \autoref{fig:err_test}. Most points from both methods are consistent at $3 \sigma$ with zero, with only two GP and linear points instead consistent at $5 \sigma$. Given the unfavorable phasing and transit coverage of the transit observations used in this example, we believe our methods are correctly accounting for the uncertainty in the data over repeated observations.

\subsection{Limb darkening choice has a significant effect on final spectrum}\label{sec:ld}
\begin{figure}[htb!]
    \centering
    \includegraphics[width = 0.45\textwidth]{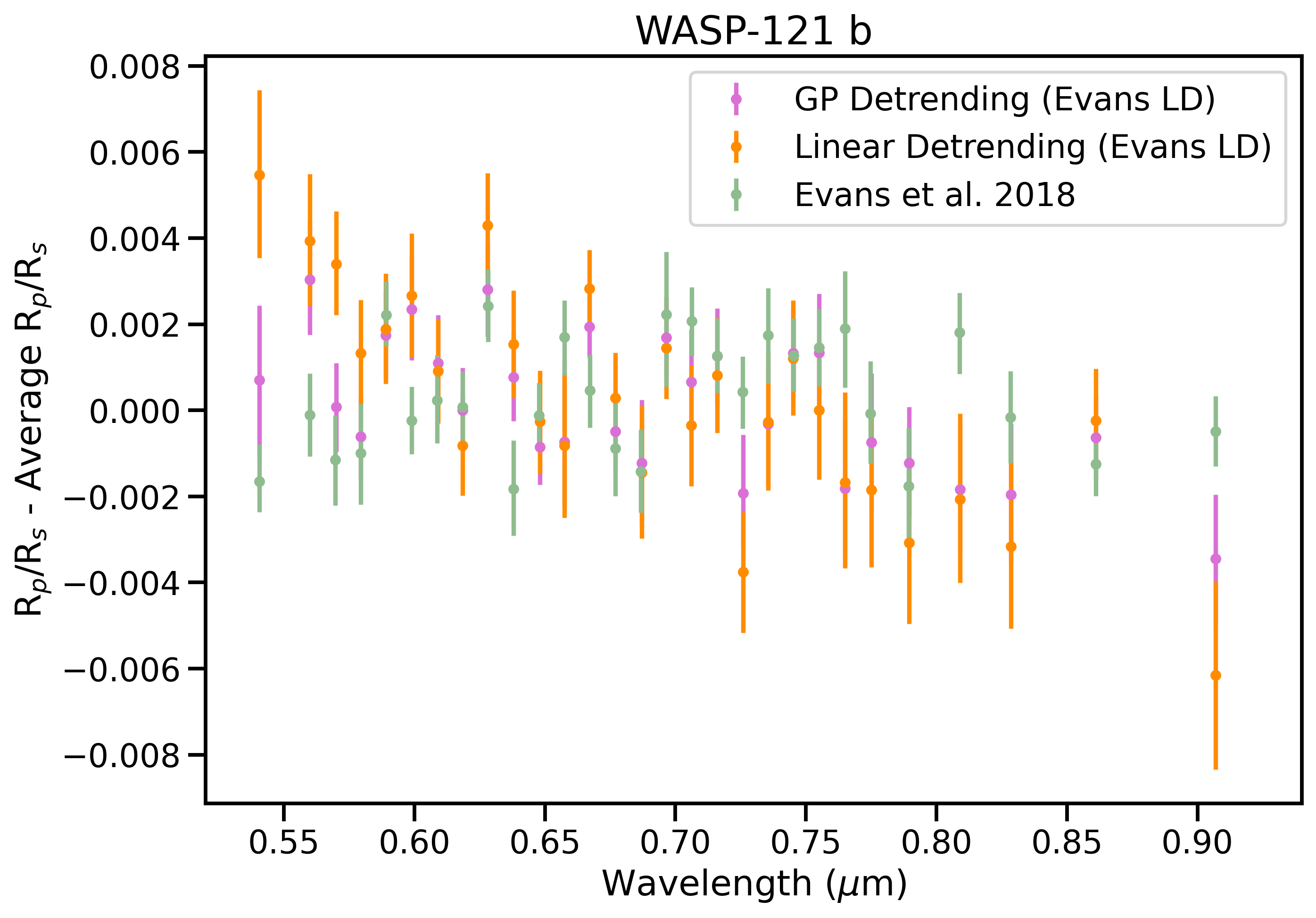}
    \caption{A refit of the G750L spectrum of WASP-121\,b, with the limb darkening fixed to that fit for in \citet{Evans_2018}, compared to the results from said work. Note that the absolute $R_p/R_s$ has been removed from all data sets, to account for any vertical offset due to differences in detrending method or system parameters. The difference in the spectra seen in \autoref{fig:final_spec} has mostly disappeared after changing the limb darkening, showing that the choice of whether to fix to theory or fit to data can have an effect on the end result when analyzing HST data.}
    \label{fig:w121_offset}
\end{figure}

We fix the limb darkening coefficients to those predicted from theory to ensure consistency across the datasets. The inconsistent coverage of ingress and egress and behavior of systematics between observations of different planets makes the reliability of fitting for these values unpredictable. However, some other works choose to fit for these values using the data, as there are likely imperfections in our theory of the behavior of limb darkening. For example, \citet{Evans_2018} fit for the limb darkening coefficients in their analysis of the WASP-121\,b STIS data. As can be seen in \autoref{fig:final_spec}, there is a noticeable difference between our analysis of this same WASP-121\,b STIS G750L data and those of \citet{Evans_2018}. To test if this could be due to the differences in our treatment of limb darkening, we repeat our G750L fit of WASP-121\,b, but with the limb darkening coefficients fixed to those reported in \citet{Evans_2018}. The result of this is shown in \autoref{fig:w121_offset}, where it can be seen that much of the difference in the spectra has disappeared after accounting for the difference in limb darkening. We note that the average $R_p/R_s$ has also been removed from each final spectrum in order to account for any vertical offset due to a difference in detrending methods and different system parameter usage, but the shape of our reduction changed between \autoref{fig:final_spec} and \autoref{fig:w121_offset}. Since we are uncertain of what the best approach to limb darkening is (and as a full deep-dive into the intricacies of this issue are beyond the scope of this paper), we simply urge caution to relying heavily on one technique or the other, and suggest attempting both and comparing results in in-depth analyses of specific planet spectra. 

\subsection{Choice of aperture radius for spectral extraction does not significantly affect results}\label{sec:aper}
We test varying the spectral extraction aperture radius to assess if this could affect our final spectra. We typically use a wide aperture radius of 15 pixels, as optimal extraction's method of weighing pixels should stop any significant unnecessary noise from being extracted with the true flux. However, past results have shown that using an aperture radius of 6.5 pixels is ideal for spectral extraction with STIS \citep[e.g.][]{Sing_2011, Nikolov_2014}, and so we compare the use of these two aperture sizes for spectral extraction with our typical spectroscopic light curve fitting method. We find that the use of these two aperture sizes result in final spectra that are consistent to within $1 \sigma$ for both GP and linear detrending techniques, though the resulting errors on the spectroscopic light curves do vary (see \autoref{app:aper}). The choice of which aperture is ideal to use varies for each data set, and the strength of said data's systematics, but as long as the resulting spectroscopic light curve's errors are correctly estimated, it should not affect the results of the spectrum.

\section{Conclusions} \label{sec:conclusion}
In this work, we present our analysis framework for the HST/STIS portion of the SHEL program. This begins with our joint-fit analysis, in which we use all available high quality data from the literature along with all TESS observations of a target together to constrain precise system parameters by fitting all high quality transits and RV observations simultaneously. This enables us to have well defined and uniformly derived priors with which to use for our HST transit fits. We then describe our data reduction and cleaning methods, which were thoroughly tested with our four test planets until a set of uniform parameters were found that could clean all of the data equally well (see \autoref{fig:extracted_spec}). Our uniform analysis continues with our light curve fitting, in which we describe our in-depth analysis and testing of systematics detrending methods, with which we are able to find best-fitting methods for both linear and GP detrending models that agree well between all of the test planets. We find that the choice of detrending method does not seem to significantly impact the final spectrum's shape, but does affect the uncertainties. We then compare our final fit spectra with that already present in the literature, and find that all but one agree well with past results. We find that our analysis of WASP-69\,b results in a spectrum significantly different than previously published work, and thus fit both atmospheric and stellar contamination models to our final spectrum of WASP-69\,b. All of our best GP detrended spectra and corresponding models from the \citet{goyal_2019} grid are shown in \autoref{fig:models}, though for WASP-39\,b, WASP-121\,b, and WASP-17\,b we note that the shown models are representative of a family of models consistent with the data, and we refer the reader to the original planet focused papers (\citet{Wakeford_2018} for WASP-39\,b, \citet{Evans_2018} for WASP-121\,b, and \citet{Alderson_2022} for WASP-17\,b) for more in-depth modeling conclusions as our spectra are consistent with those presented in previous work. Finally, we discuss a few interesting lessons that were gleaned through this testing process that may help others in their analysis of transmission spectroscopy data in the future. \\

In this work, we described the first stage of the SHEL project. In future work, we will carry out this analysis for all planets with high quality HST/STIS transmission spectroscopy data. In addition, we will perform the same type of uniform analysis for HST/WFC3 data, such that all archival HST data are available in a standard form for any analysis that can utilize it. However, we note that as planets continue to be observed and additional high quality data becomes available, especially transits using JWST, it will almost always be worthwhile to re-run the joint fits as we do with the new data included. This would also require then running the spectroscopic light curve fitting again with those joint fit results as priors, but this is the best way to ensure that there are not differences in the data due to differences in the system parameters used in analysis. \\

\begin{acknowledgments}
We thank our reviewer Drake Deming for a thoughtful and helpful review process. This research is based on observations made with the NASA/ESA Hubble Space Telescope obtained from the Space Telescope Science Institute (STScI), which is operated by the Association of Universities for Research in Astronomy, Inc., under NASA contract NAS 5–26555. These observations are associated with programs 12473 and 14767. The specific observations analyzed can be accessed via this \dataset[DOI]{https://doi.org/10.17909/2pd5-pg65}. Support for this work was provided by NASA through grants under the HST-AR-16634 program from STScI. This paper includes data collected with the TESS mission, obtained from the Mikulski Archive for Space Telescopes (MAST) data archive at STScI. Funding for the TESS mission is provided by the NASA Explorer Program. STScI is operated by the Association of Universities for Research in Astronomy, Inc., under NASA contract NAS 5–26555. This work also makes use of observations made with the NASA/ESA/CSA James Webb Space Telescope. The data were obtained from MAST at STScI, which is operated by the Association of Universities for Research in Astronomy, Inc., under NASA contract NAS 5-03127 for JWST. These observations are associated with programs 1201 and 1366. N.H.A. acknowledges support by the National Science Foundation Graduate Research Fellowship under Grant No. DGE1746891. M.T.P. acknowledges the support of the Fondecyt-ANID Post-doctoral fellowship 3210253.
\end{acknowledgments}


\vspace{5mm}
\facilities{HST (STIS), JWST (NIRSpec, NIRCam, NIRISS), TESS, NGTS, OHP:1.93m (SOPHIE), Euler1.2m (CORALIE, EulerCam), ESO:3.6m (HARPS), GTC (OSIRIS), TRAPPIST, Danish 1.54m Telescope, Magellan:Clay (MIKE)}


\software{astropy \citep{2013A&A...558A..33A,2018AJ....156..123A}, barycorrpy \citep{Kanodia_2018}, batman \citep{batman}, corner \citep{corner}, dynesty \citep{dynesty}, george \citep{hodlr}, juliet \citep{juliet}, jupyter \citep{Kluyver2016jupyter}, matplotlib \citep{Hunter:2007}, multinest \citep{Feroz_2008, Feroz_2009, Feroz_2019}, NumPy \citep{harris2020array}, pandas \citep{reback2020pandas, mckinney-proc-scipy-2010}, pymultinest \citep{buchner_2014}, SciPy \citep{2020SciPy-NMeth}, seaborn \citep{seaborn}, transitspectroscopy \citep{transitspectroscopy}}
\pagebreak
\appendix
\section{Systematics model testing}\label{app:systematics}
\begin{figure}[h!]
\gridline{\fig{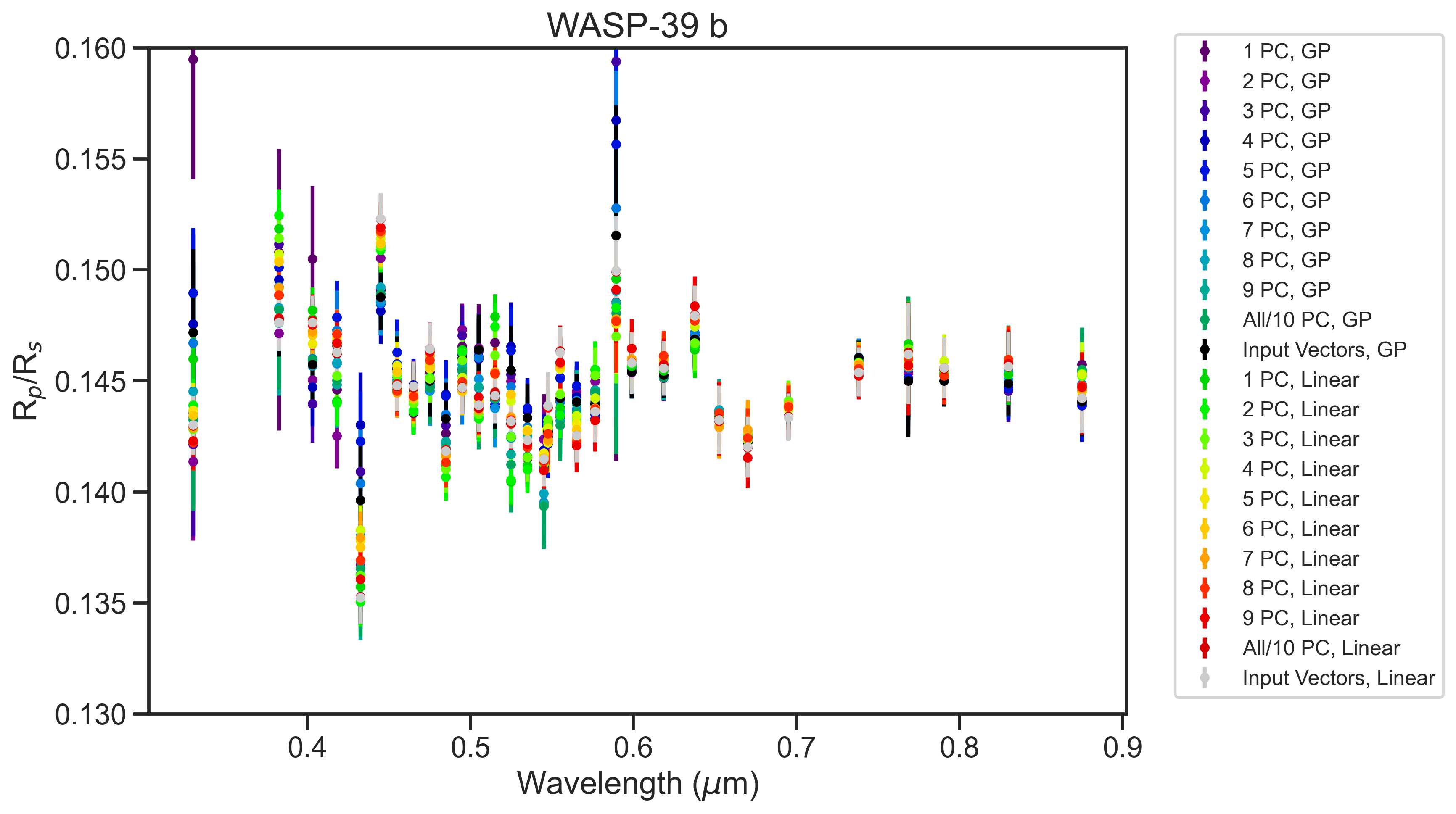}{0.49\textwidth}{}
          \fig{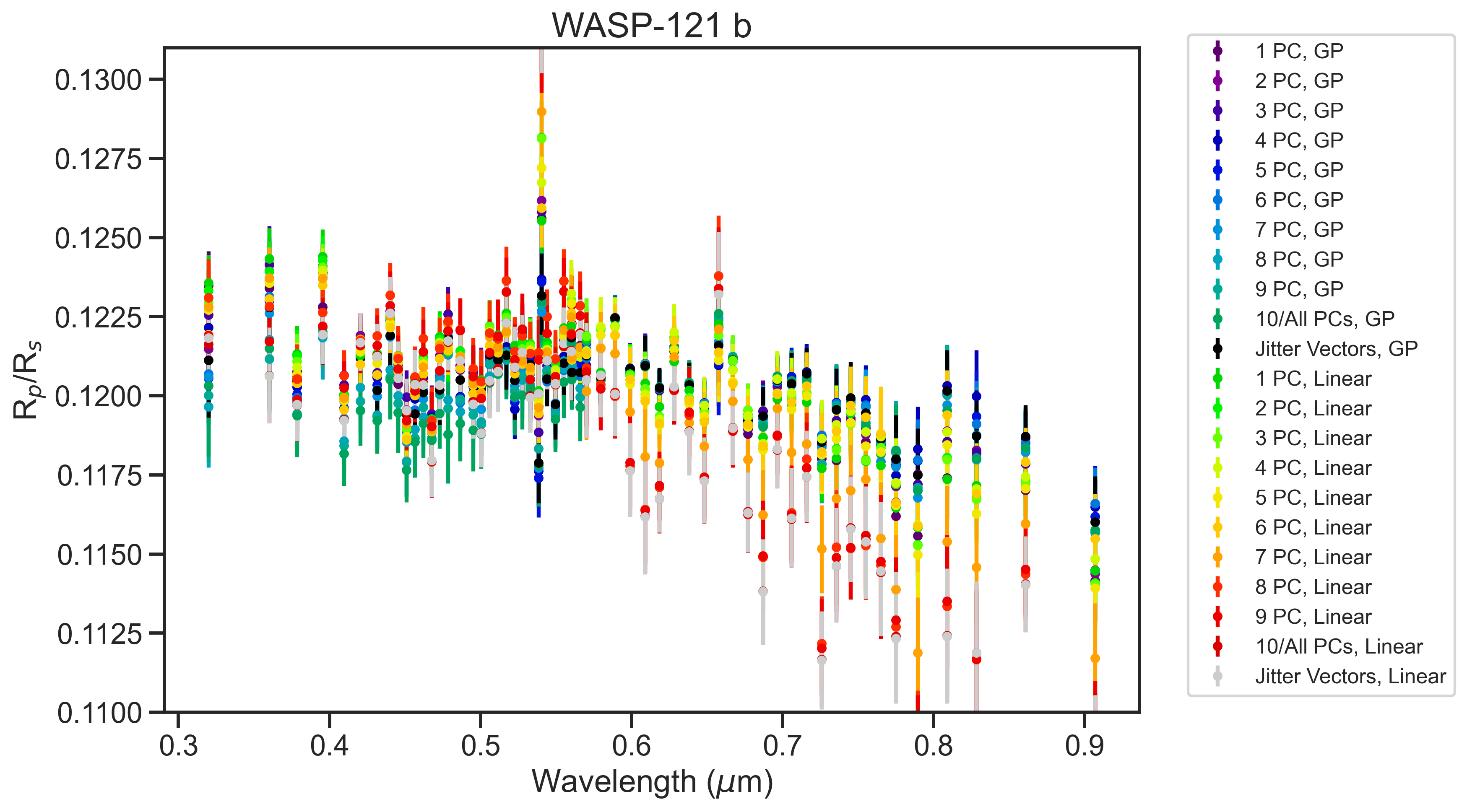}{0.49\textwidth}{}}
\gridline{\fig{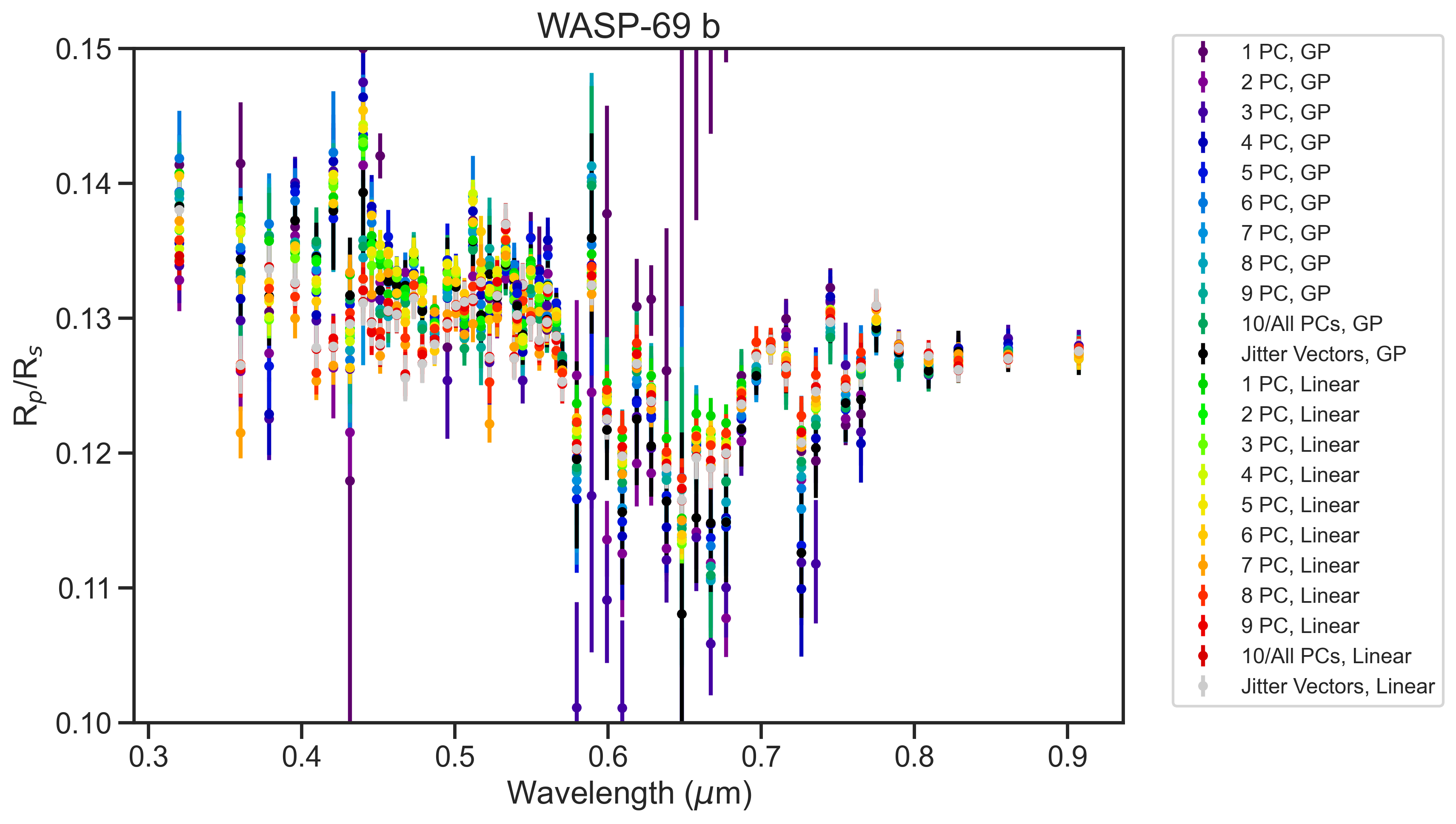}{0.49\textwidth}{}
          \fig{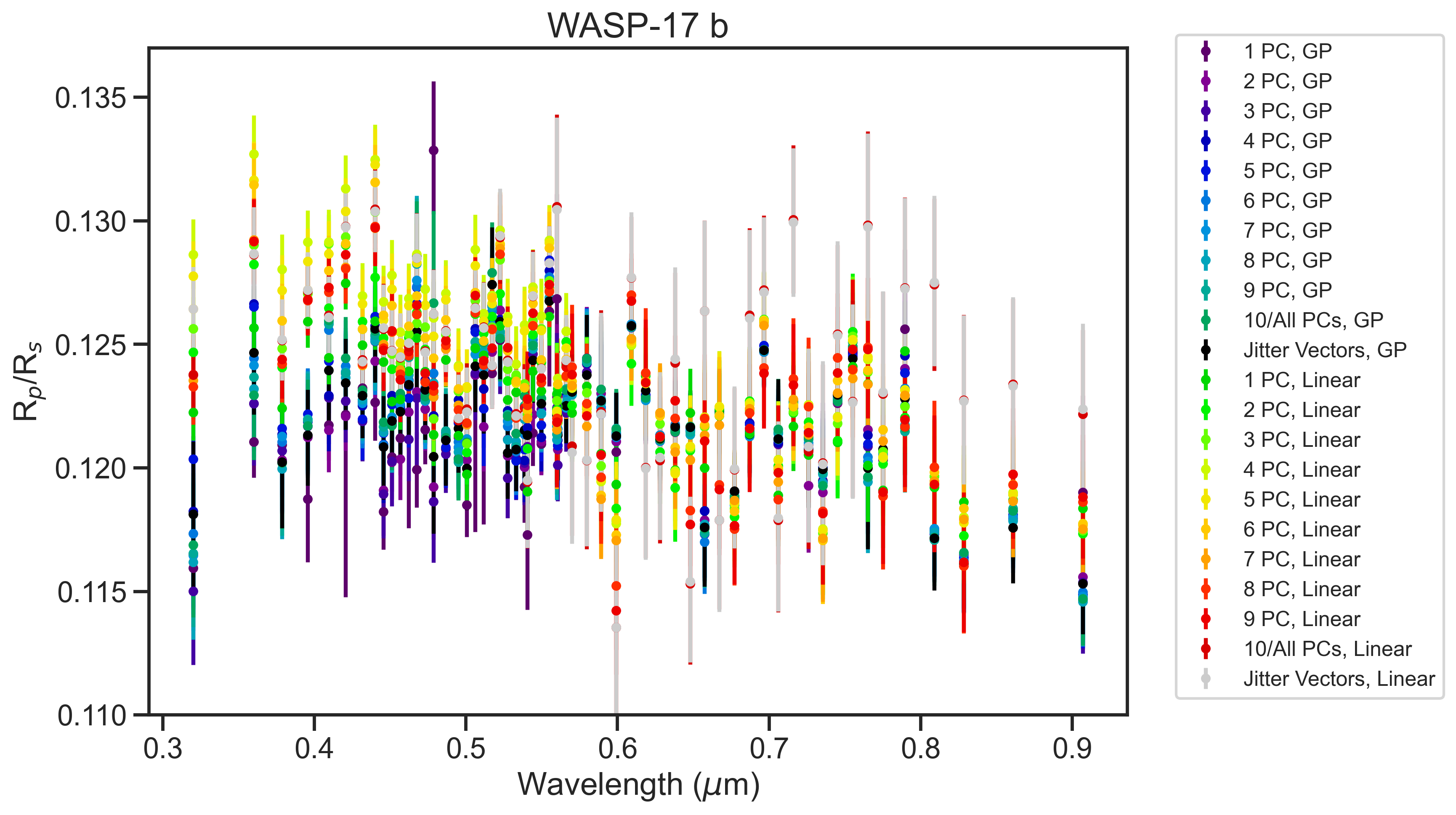}{0.49\textwidth}{}}
\caption{The combined STIS G430L and G750L spectrum of WASP-39\,b (top left), WASP-121\,b (top right), WASP-69\,b (bottom left), and WASP-17\,b (bottom right). Each color (common between planets) corresponds to a different systematics detrending technique, either using linear or GP methods and either between 1-10 PCs or the 10 initial input vectors.}\label{fig:sys_test}
\end{figure}

We performed tests on our four test planets (WASP-17\,b, WASP-121\,b, WASP-69\,b, and WASP-39\,b) to determine the optimal systematics model. With our 10 selected detrending vectors -- 6 engineering jitter vectors (RA, DEC, Latitude, Longitude, V2\_roll, V3\_roll), 3 trace movement vectors (one for the position on the detector, and two associated with the shape of the trace) and the HST orbital phase to the fourth power -- we fit each dataset's spectroscopic light curves with those input directly, as well as from 1 to 10 principal components obtained from using PCA on these detrending vectors, using both a linear and a gaussian process (GP) detrending method. For the linear detrending method, each vector's coefficient is given a uniform prior from -100 to 100. For the GP, we use a Matern-3/2 kernel\footnote{Note that we did a similar test using the exponential-squared kernel from \textit{george} and saw negligible differences in the fit.} as implemented in \textit{george} \citep{hodlr} with an exponential prior of 1 for each coefficient. Each spectroscopic light curve is fit using nested sampling, specifically using the dynamic nested sampling method from \textit{dynesty} (CITE). This results in 22 spectra per grating, for which we have 2 gratings for each planet (G430L and G750L), for a total of 88 spectra. Note that the G430L spectra are composed of two separate transits that are phased and fit simultaneously, while the G750L spectra are only one transit. We note that WASP-121\,b, WASP-69\,b, and WASP-17\,b are all fit using the spectral bin widths from \citet{Evans_2018}, but that we instead use the generally wider bin widths from \citet{Wakeford_2018} as a comparison in case the binning scheme had any effect on the result of this test, which we do not see any evidence for. The 22 spectra for each dataset are shown overlaid in \autoref{fig:sys_test}. The Jupyter notebooks used to generate these tests are available on Github\footnote{\url{https://github.com/natalieallen/stis_pipeline}}. \\

\begin{figure}[h!]
\gridline{\fig{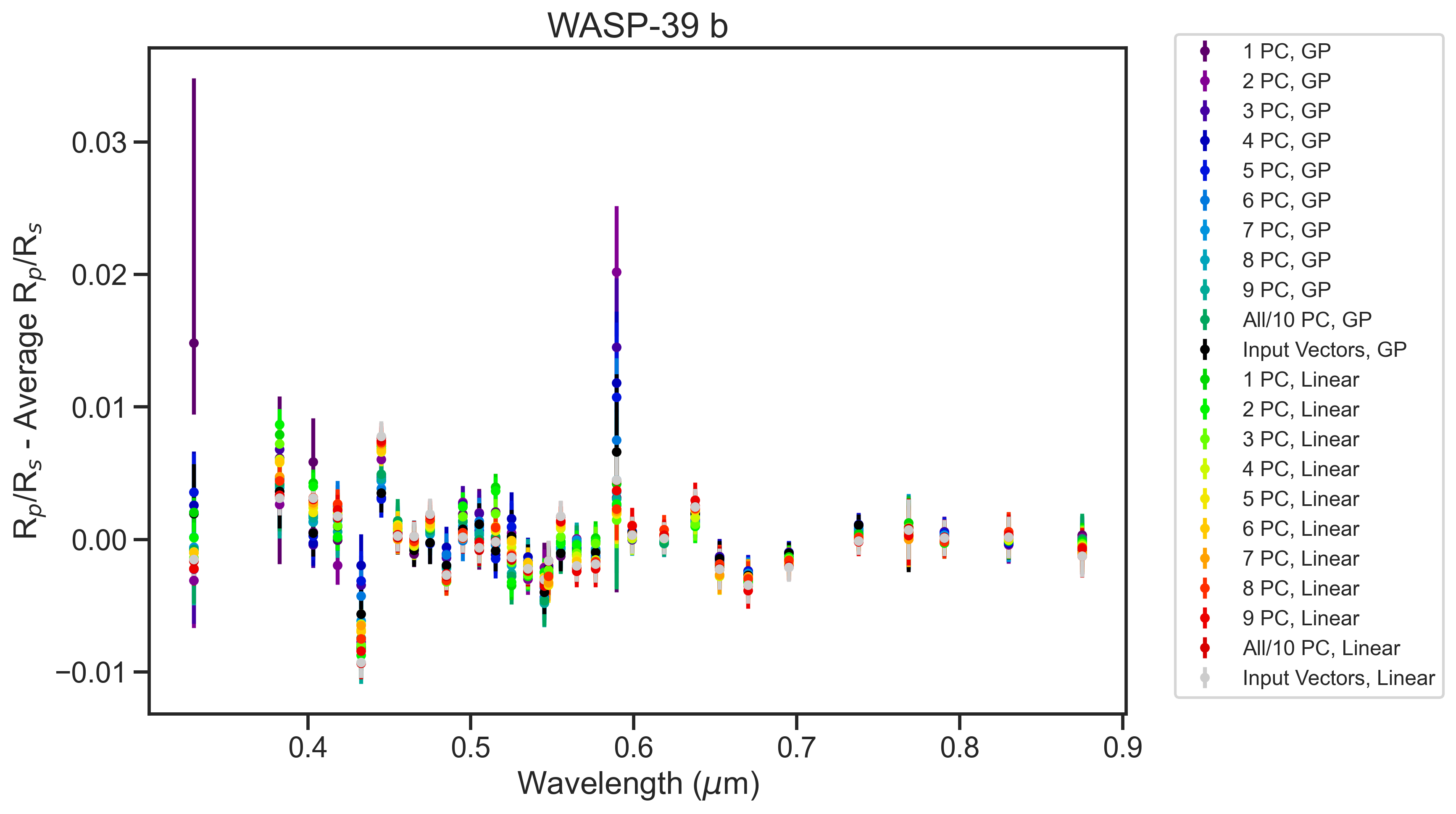}{0.49\textwidth}{}
          \fig{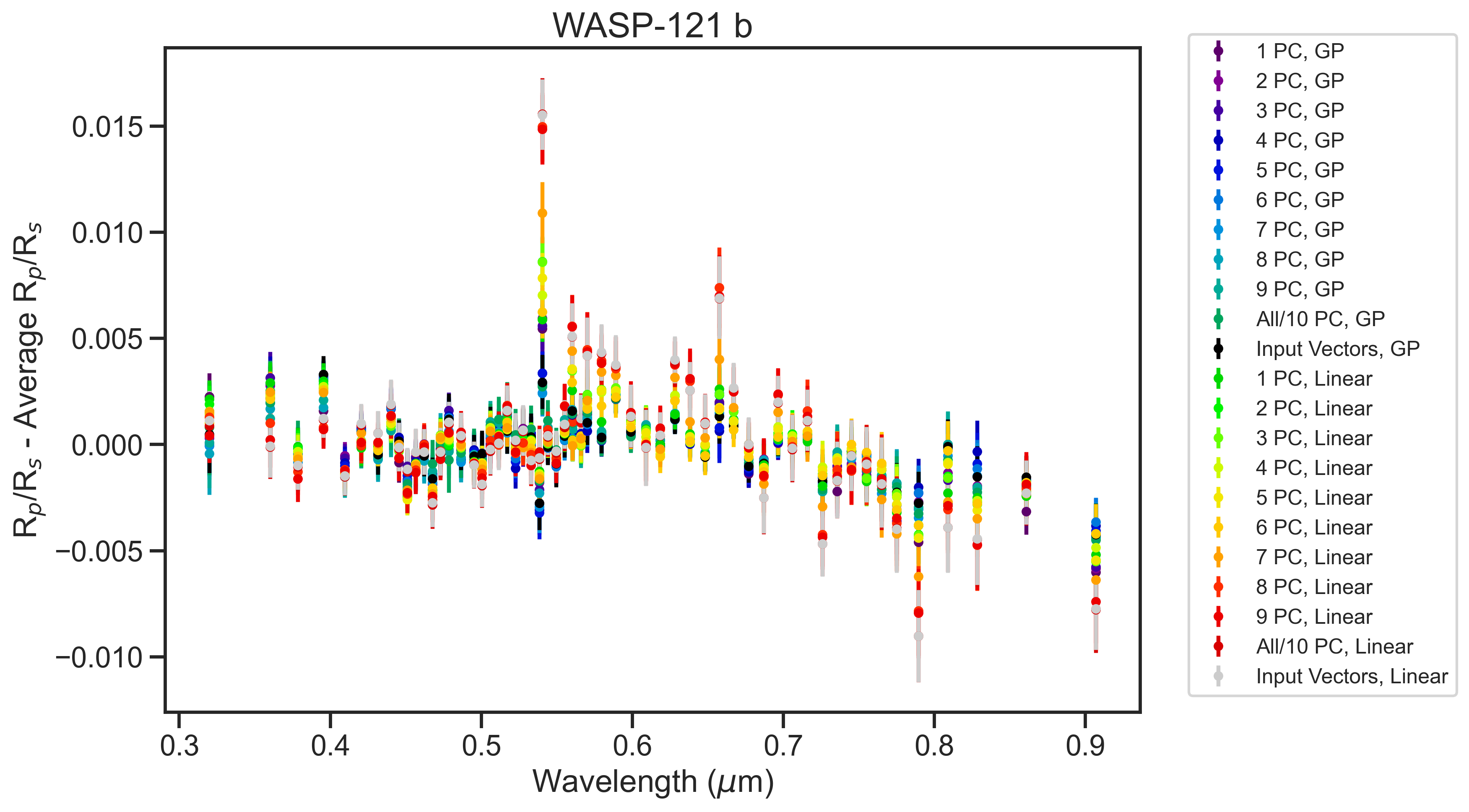}{0.49\textwidth}{}}
\gridline{\fig{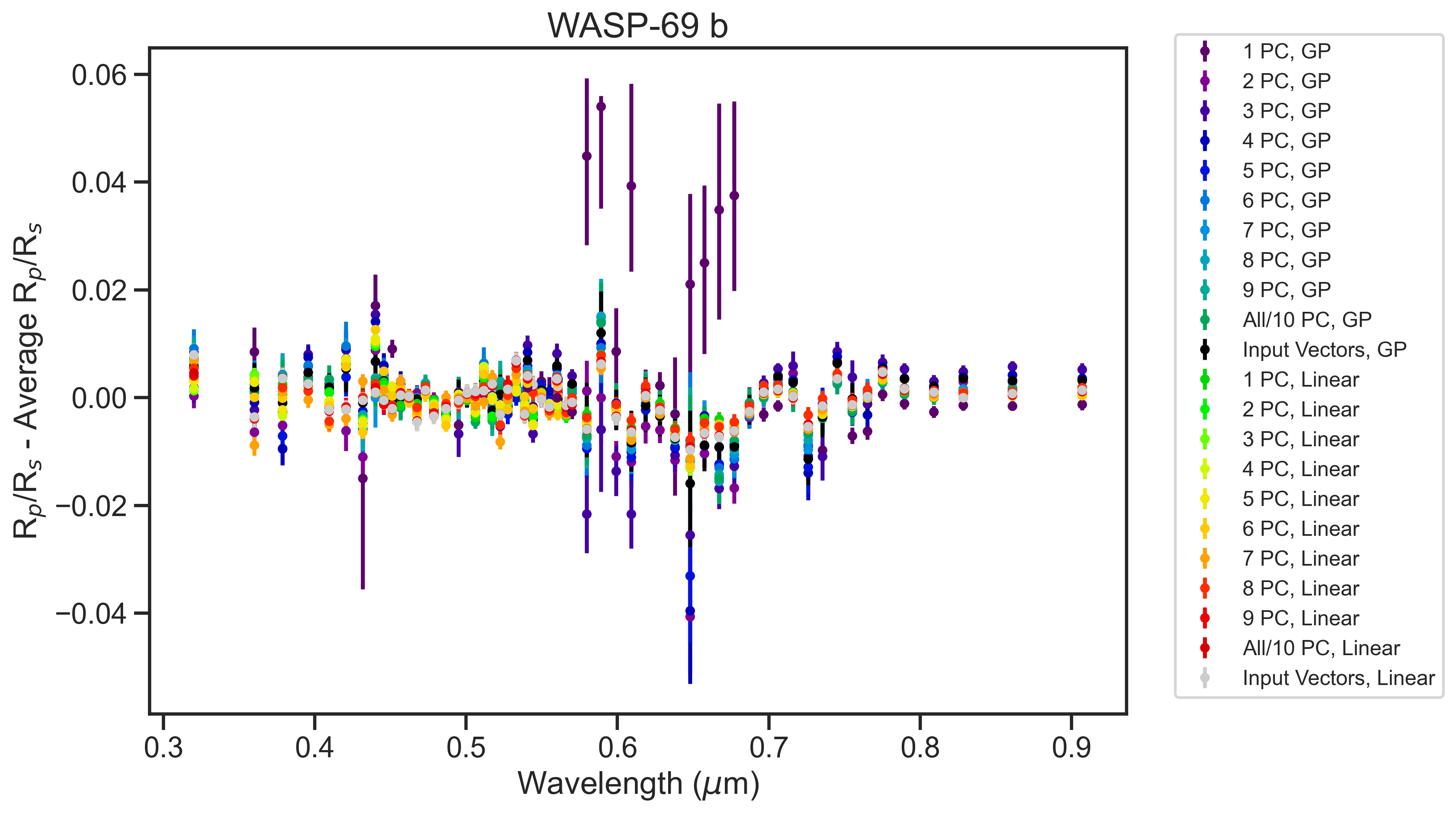}{0.49\textwidth}{}
          \fig{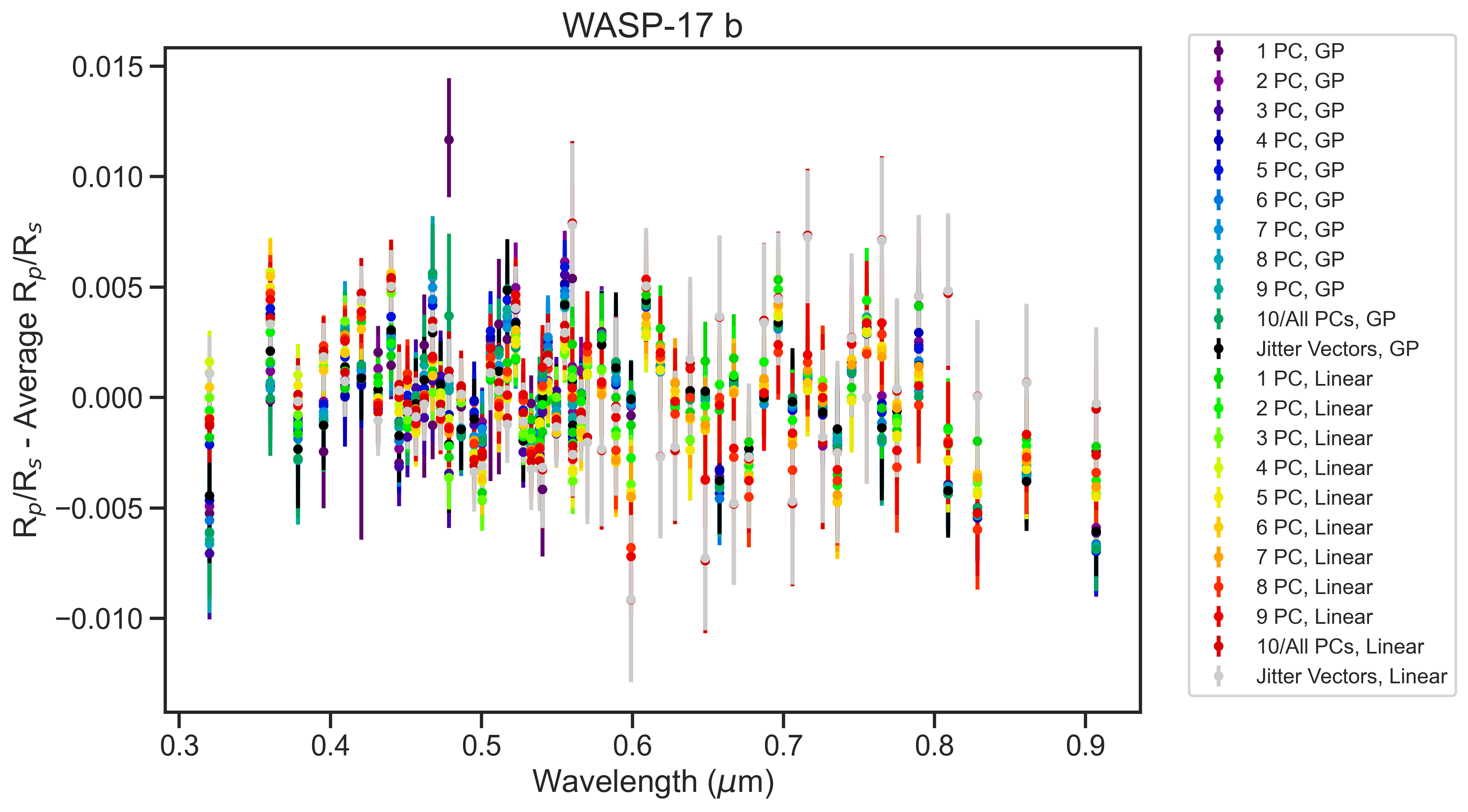}{0.49\textwidth}{}}
\caption{The same as \autoref{fig:sys_test}, but each spectrum has its average $R_p/R_s$ subtracted from it to remove any absolute $R_p/R_s$ offsets. This compares spectral shape rather than true $R_p/R_s$.}\label{fig:sys_test_sub}
\end{figure}

It can be seen in \autoref{fig:sys_test} that the choice of systematics model and number of detrending vectors affect the absolute $R_p/R_s$. Additionally, the errors estimated by the linear model can be significantly smaller than those obtained by the GP \citep[which is expected, see ][for a recent review on GPs and why the suggested errors are larger than for polynomial methods]{aigrain_2022}. This is discussed further in \autoref{sec:error_test}. While the level of derived error on $R_p/R_s$ varies strongly depending on the level of systematics in the light curves and choice of detrending method and associated detrending vectors, we note that this error varies from an inflation on the HST pipeline derived errors (which is mostly photon noise dominated, though it also considers read-out noise) from approximately 100 to 1500 ppm. The spread in the obtained spectra differs between planets, but is typically overall greater for the G430L data than it is for the G750L data, which is due to the fact that the G430L's two transits have double the systematics to deal with simultaneously. However, it seems as though the overall spectrum shape for each planet remains similar regardless of detrending method. In \autoref{fig:sys_test_sub}, we show each of the spectra shown in \autoref{fig:sys_test}, but with each spectrum's average $R_p/R_s$ subtracted from it. From this, it can be seen that the shape of many of the spectra match relatively well, and thus the detrending method is not likely to cause spurious detections or hide atmospheric features that would be present with another detrending method unless the extremes are used. \\

\begin{figure}[h!]
\gridline{\fig{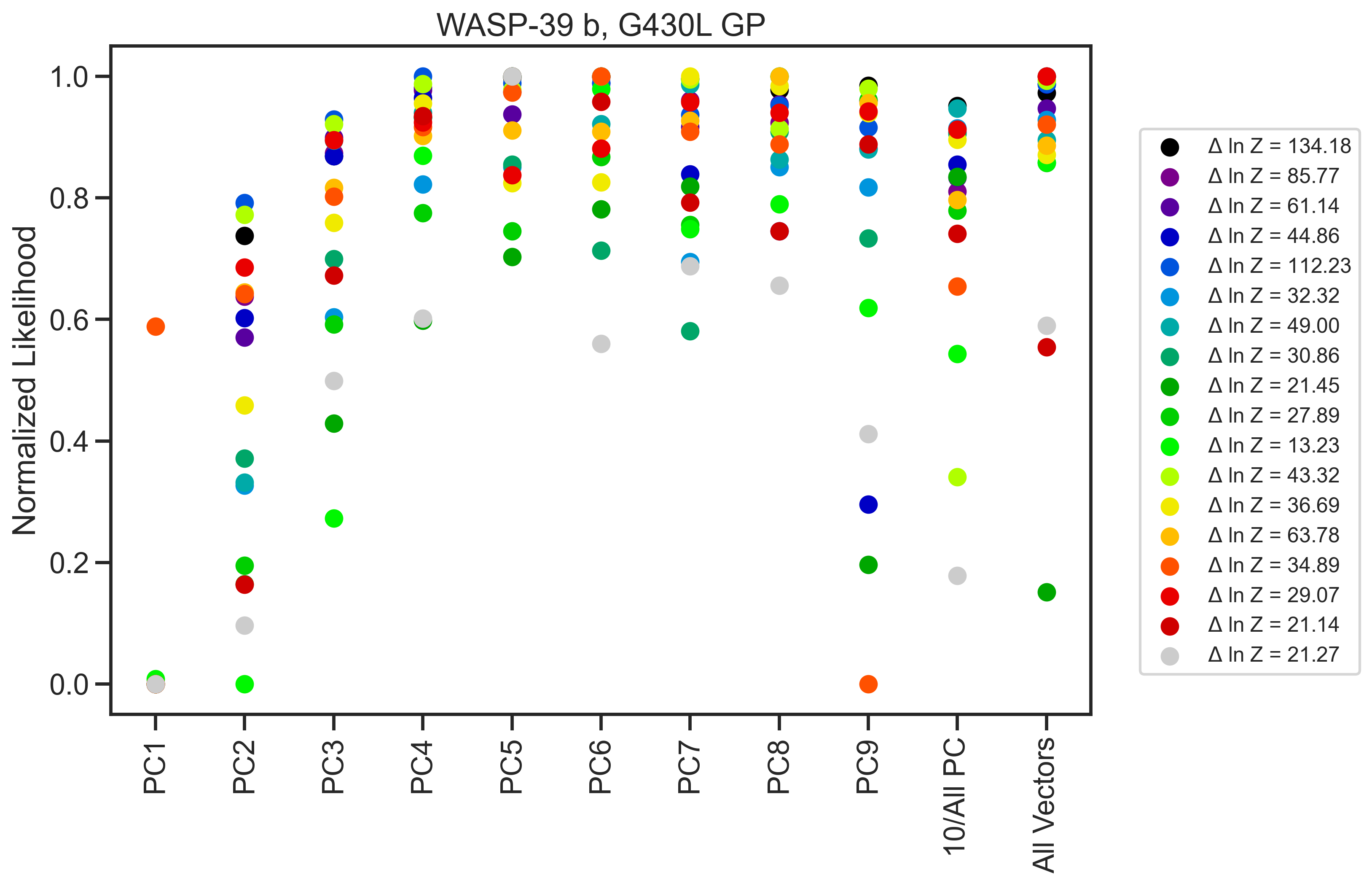}{0.49\textwidth}{}
          \fig{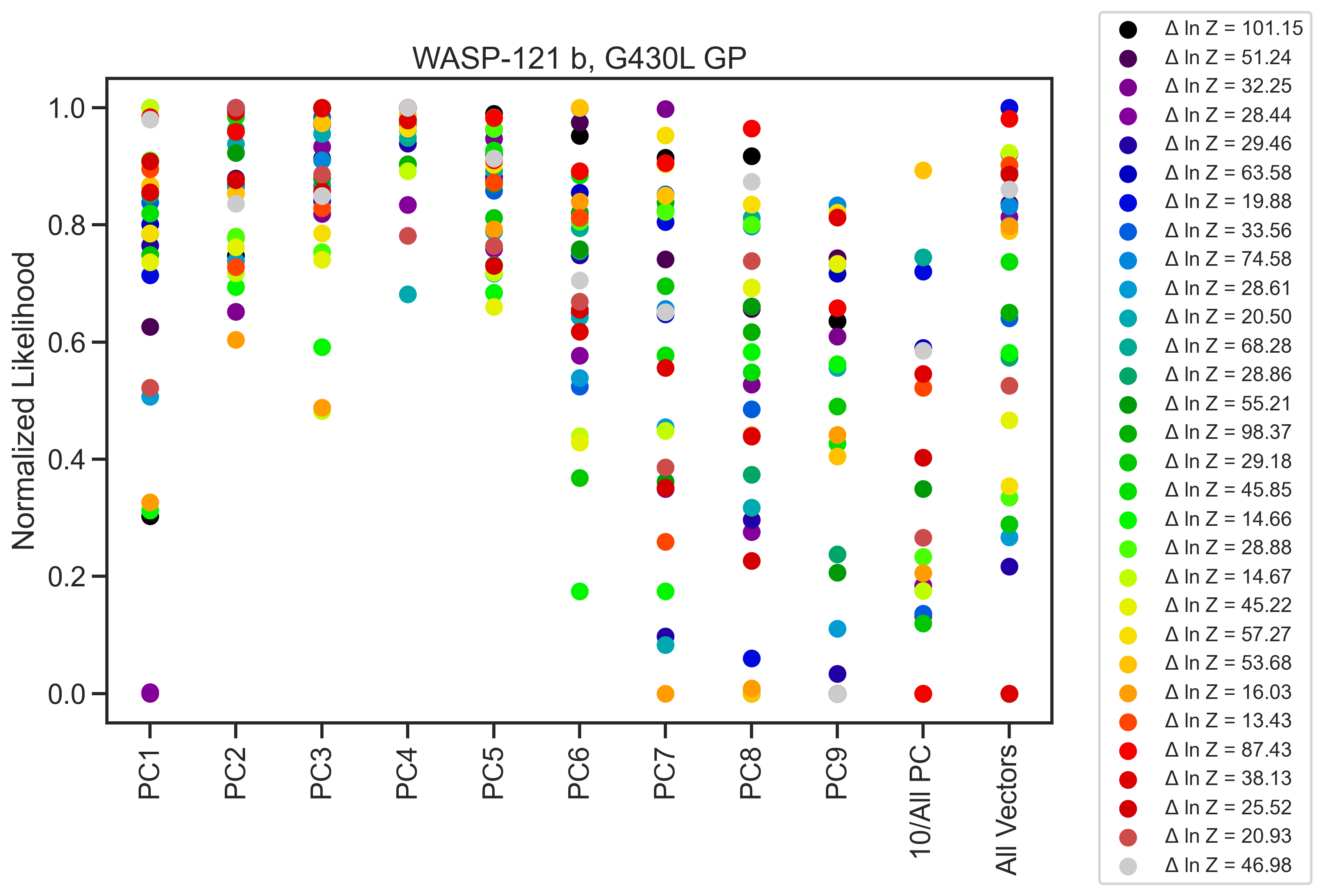}{0.49\textwidth}{}}
\gridline{\fig{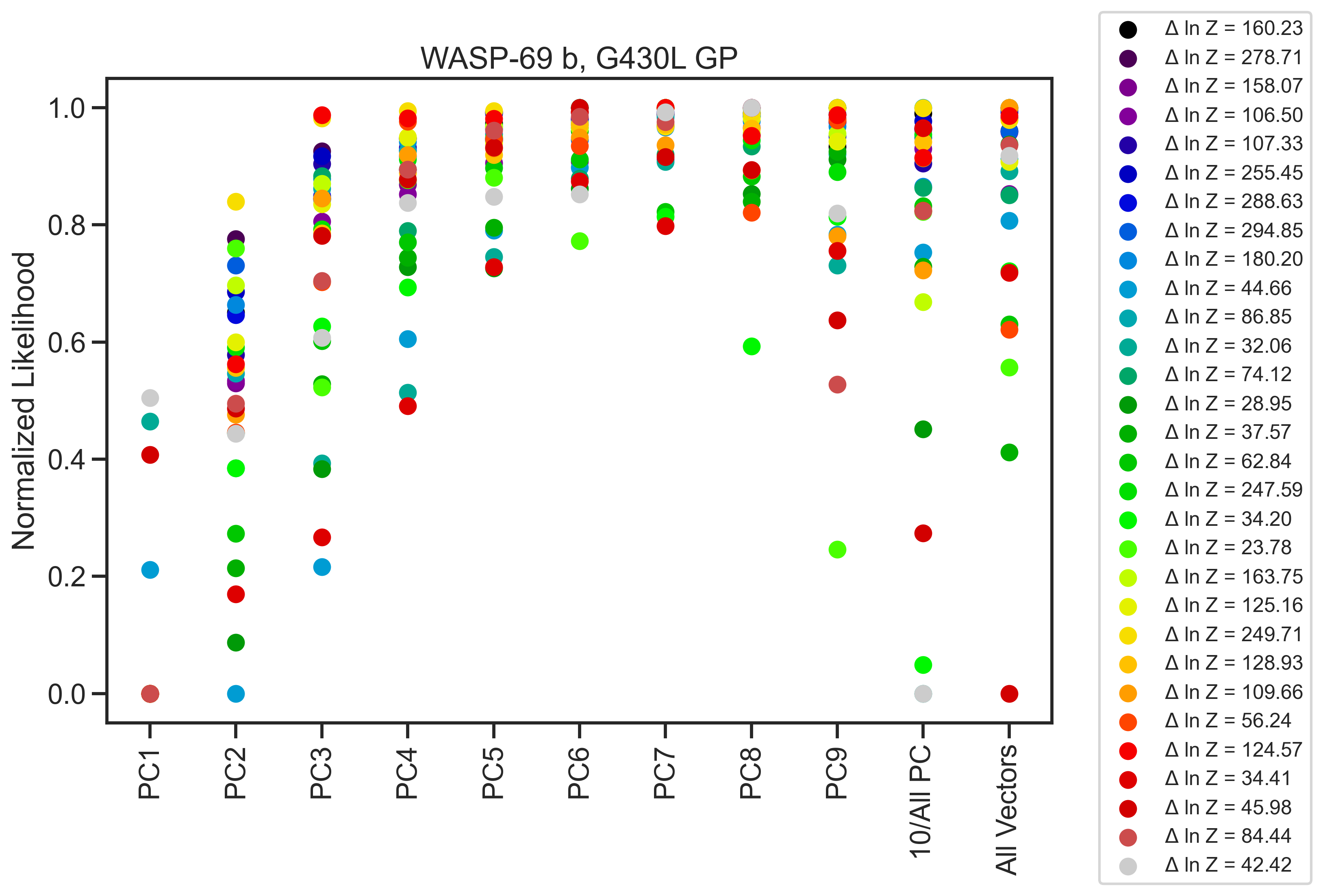}{0.49\textwidth}{}
          \fig{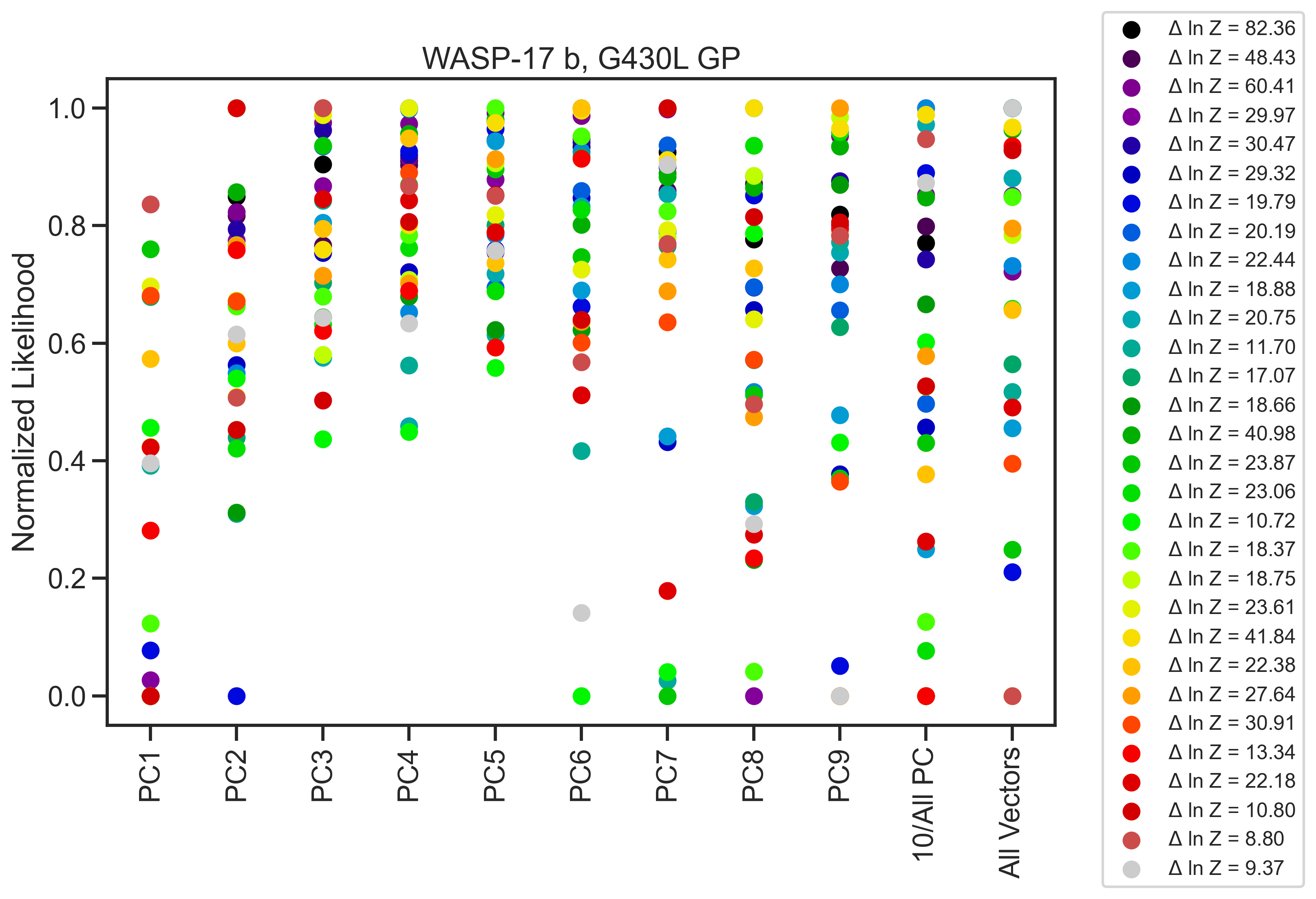}{0.49\textwidth}{}}
    \caption{The normalized 0 to 1 \textit{ln Z} for each set of regressors from 1 to 10 PCs, and also for all original vectors, for G430L GP detrending. Each color corresponds to a wavelength bin. The true \textit{$\Delta$ ln Z} this is detailing is given in the legend. The more bunched towards 1 a regressor set is, the better overall fit it is across all wavelength bins.}
    \label{fig:evidence1}
\end{figure}

\begin{figure}[h!]
\gridline{\fig{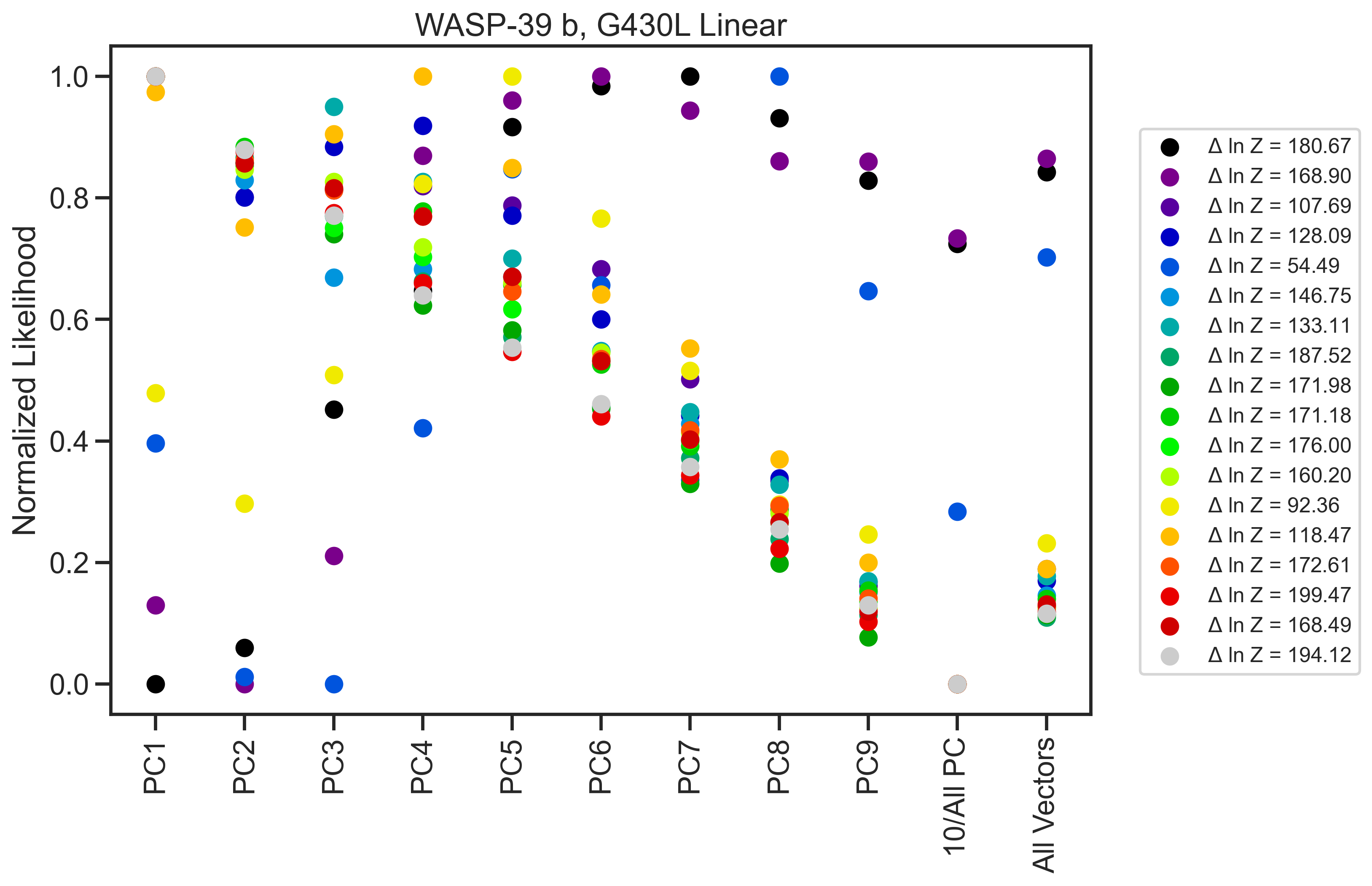}{0.49\textwidth}{}
          \fig{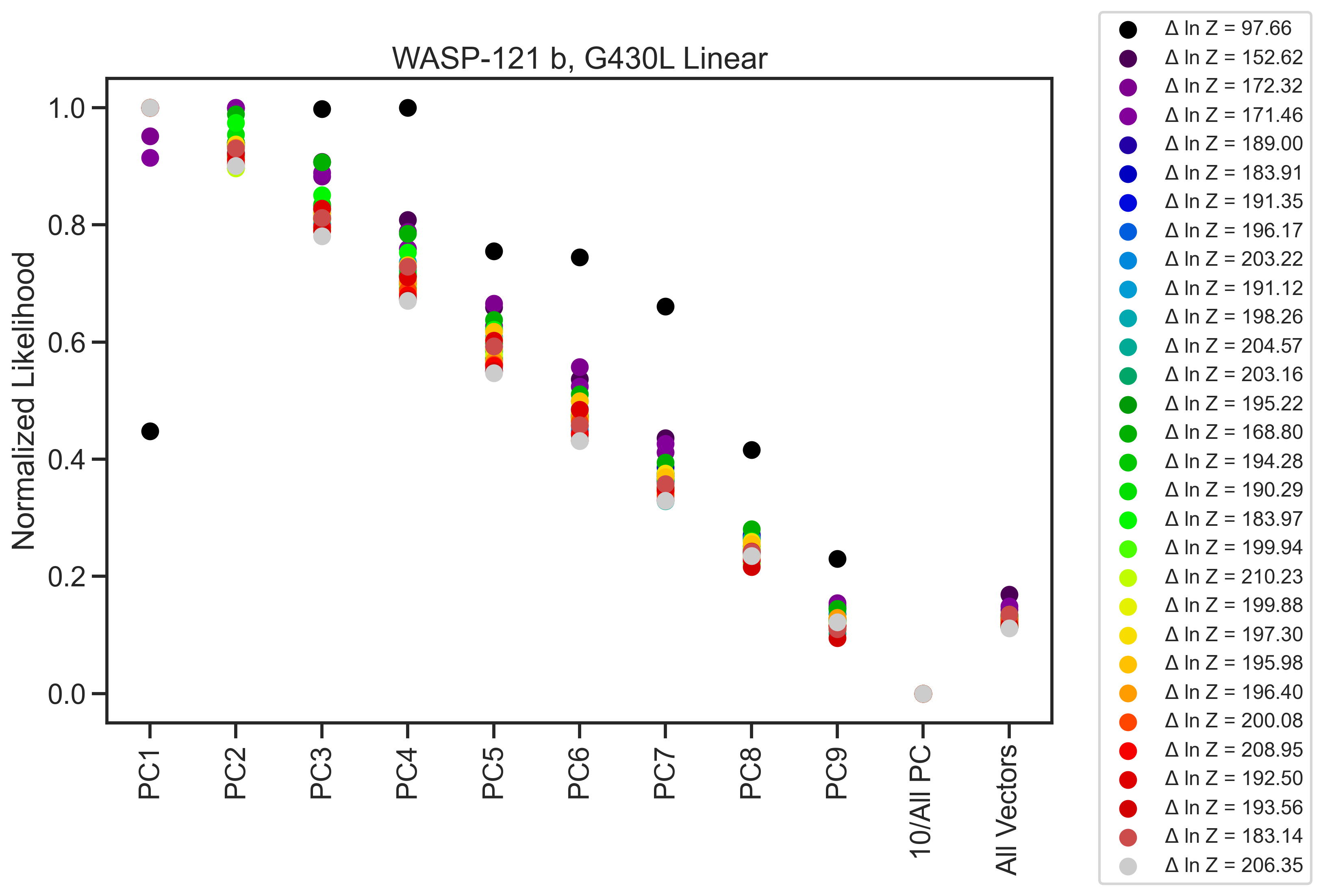}{0.49\textwidth}{}}
\gridline{\fig{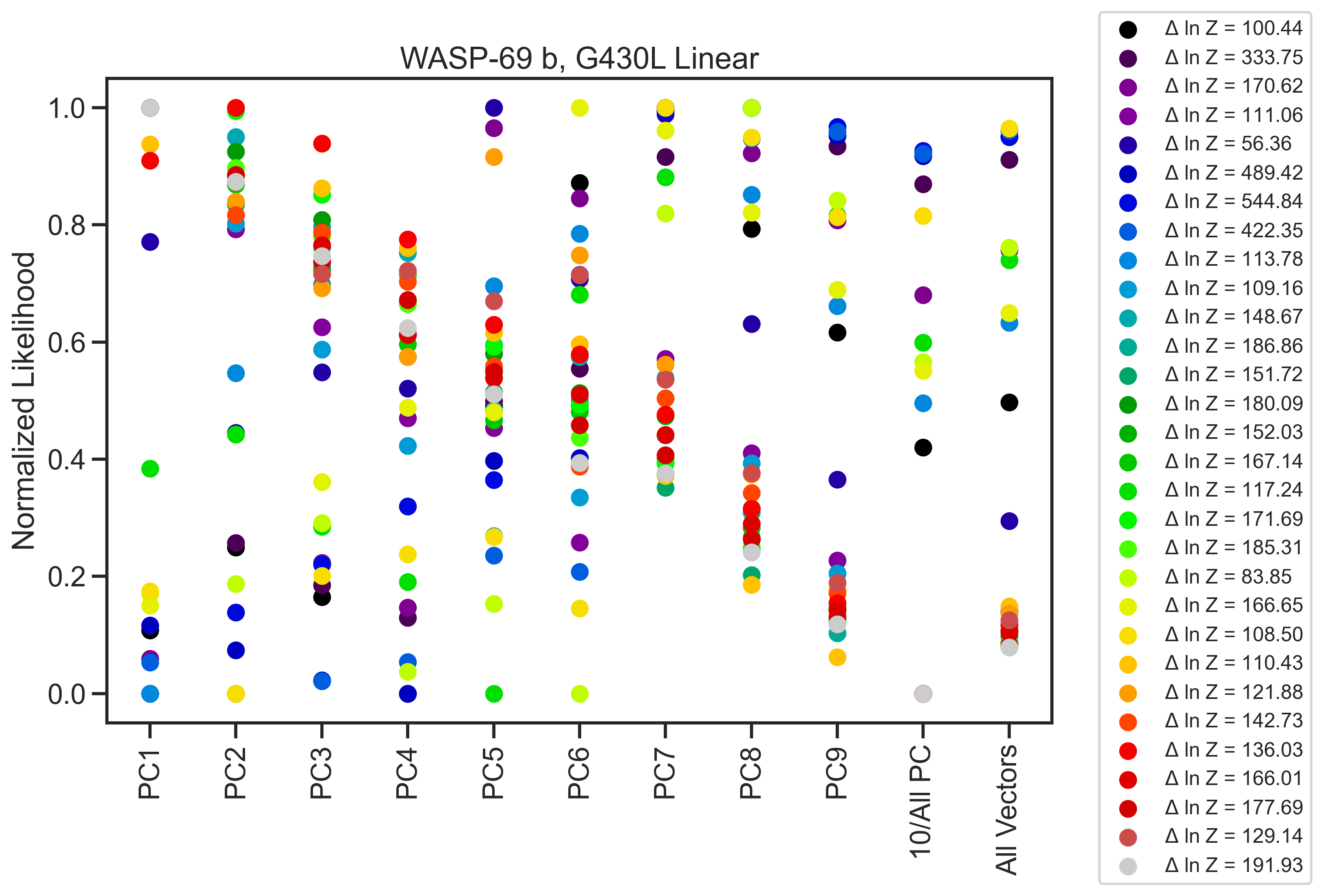}{0.49\textwidth}{}
          \fig{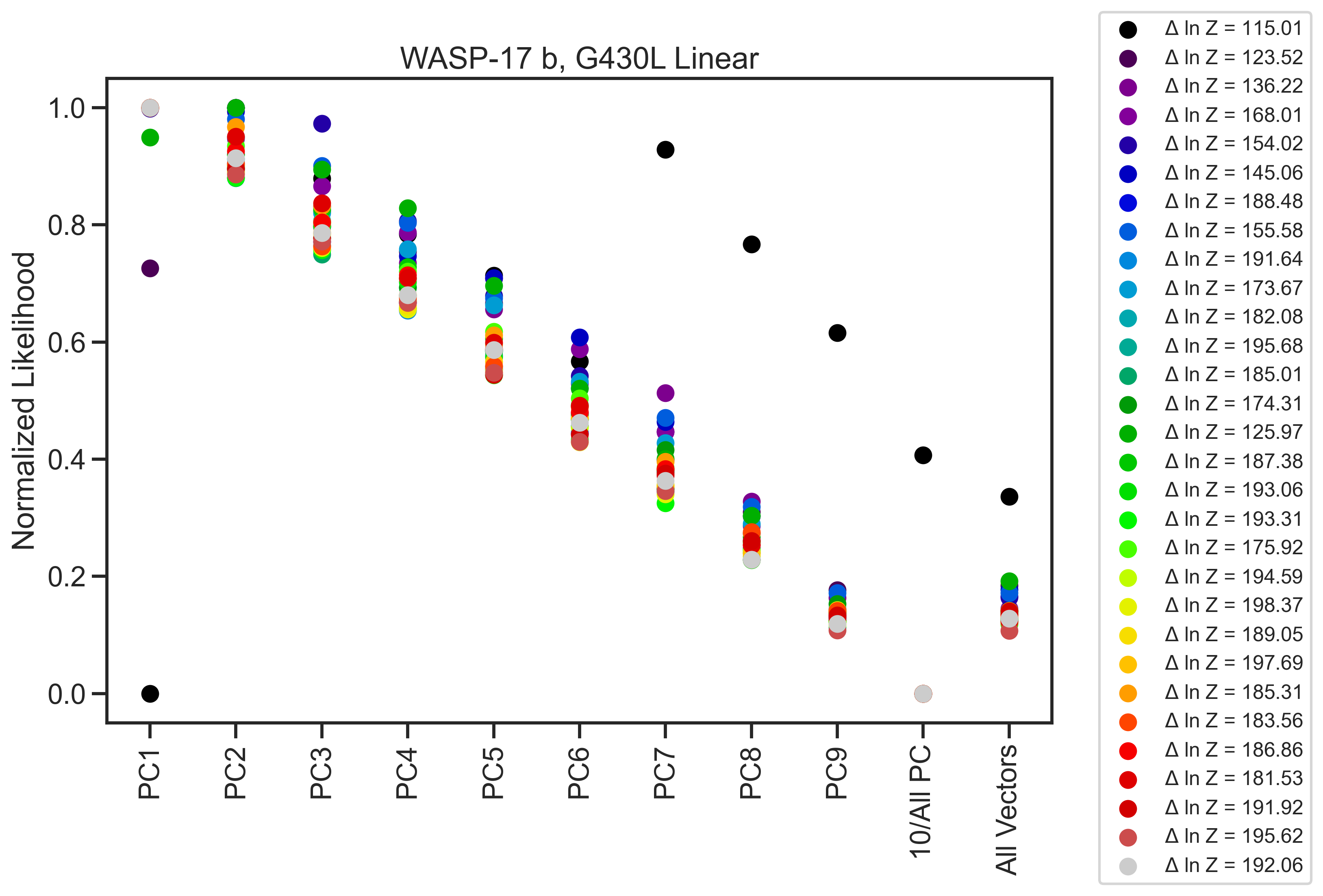}{0.49\textwidth}{}}
    \caption{Same as \autoref{fig:evidence1}, but for G430L linear detrending.}
    \label{fig:evidence2}
\end{figure}

\begin{figure}[h!]
\gridline{\fig{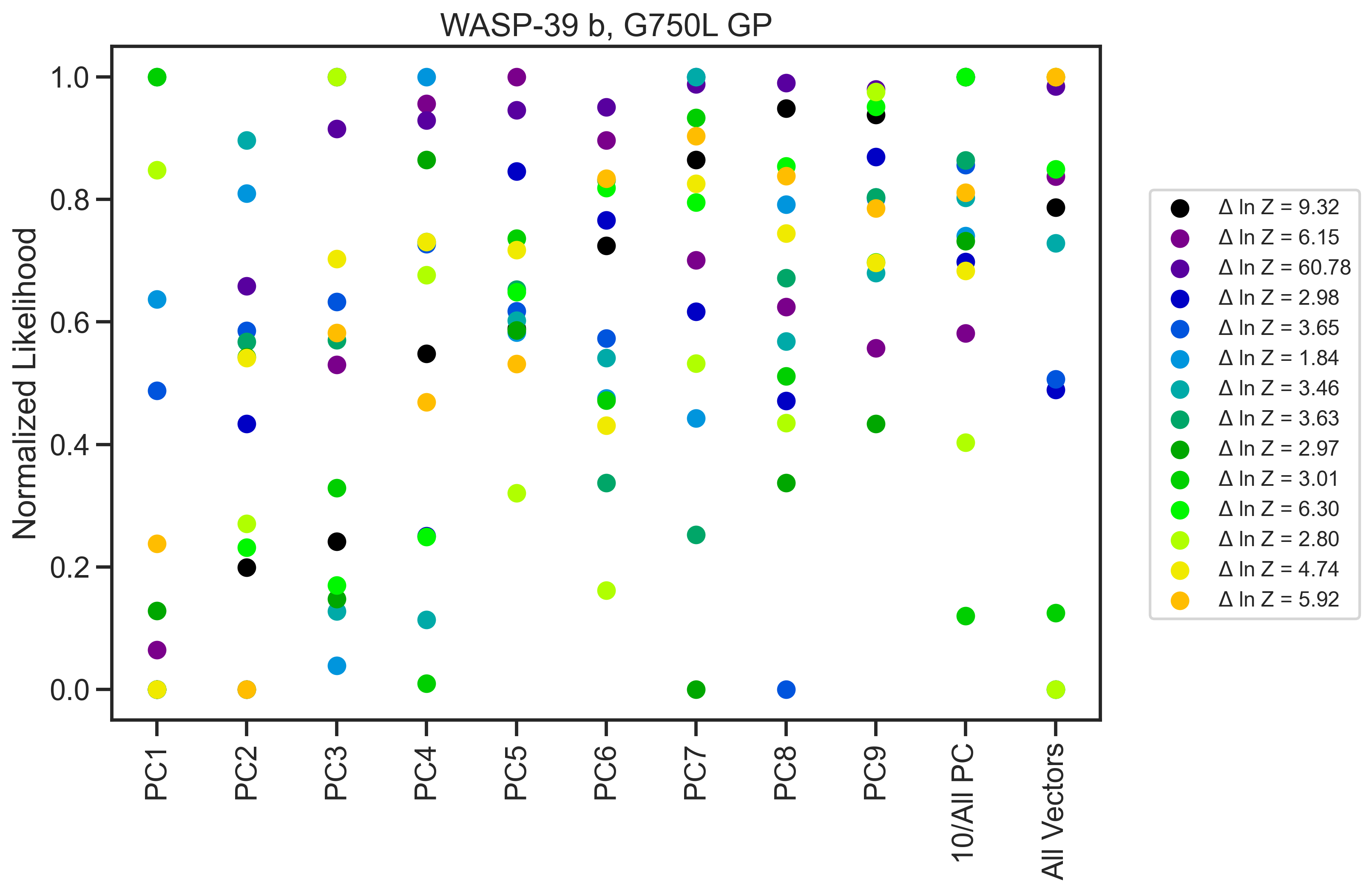}{0.49\textwidth}{}
          \fig{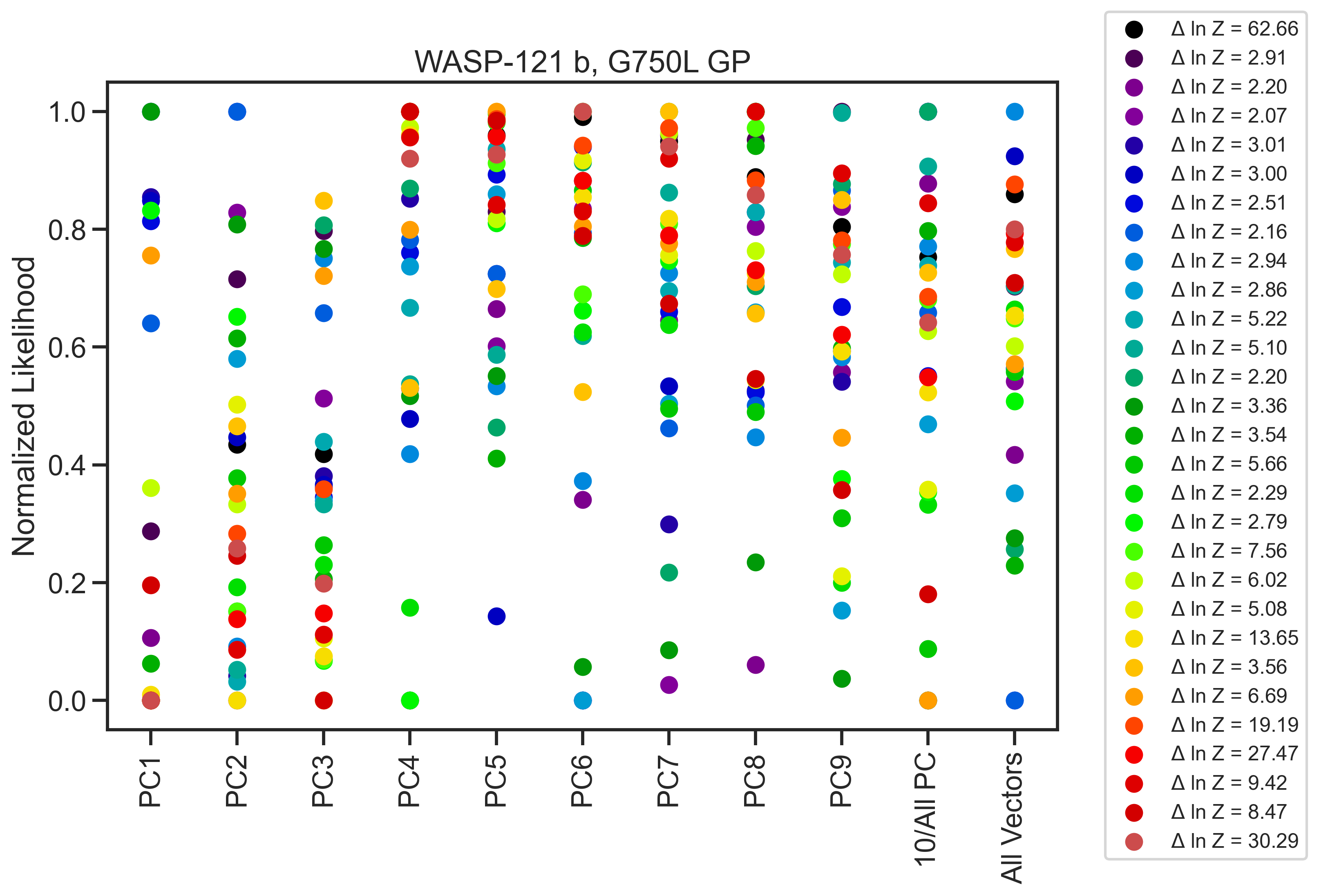}{0.49\textwidth}{}}
\gridline{\fig{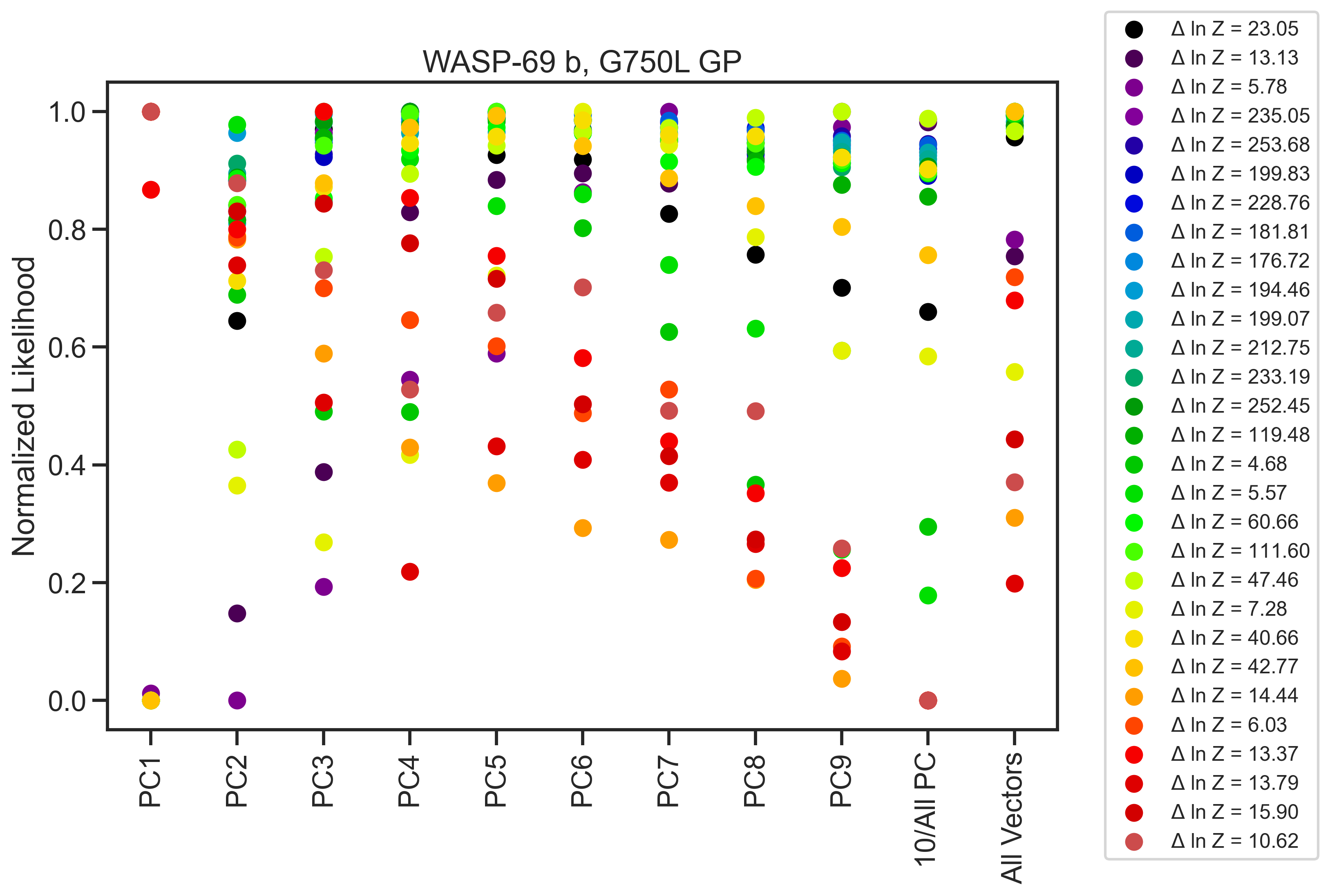}{0.49\textwidth}{}
          \fig{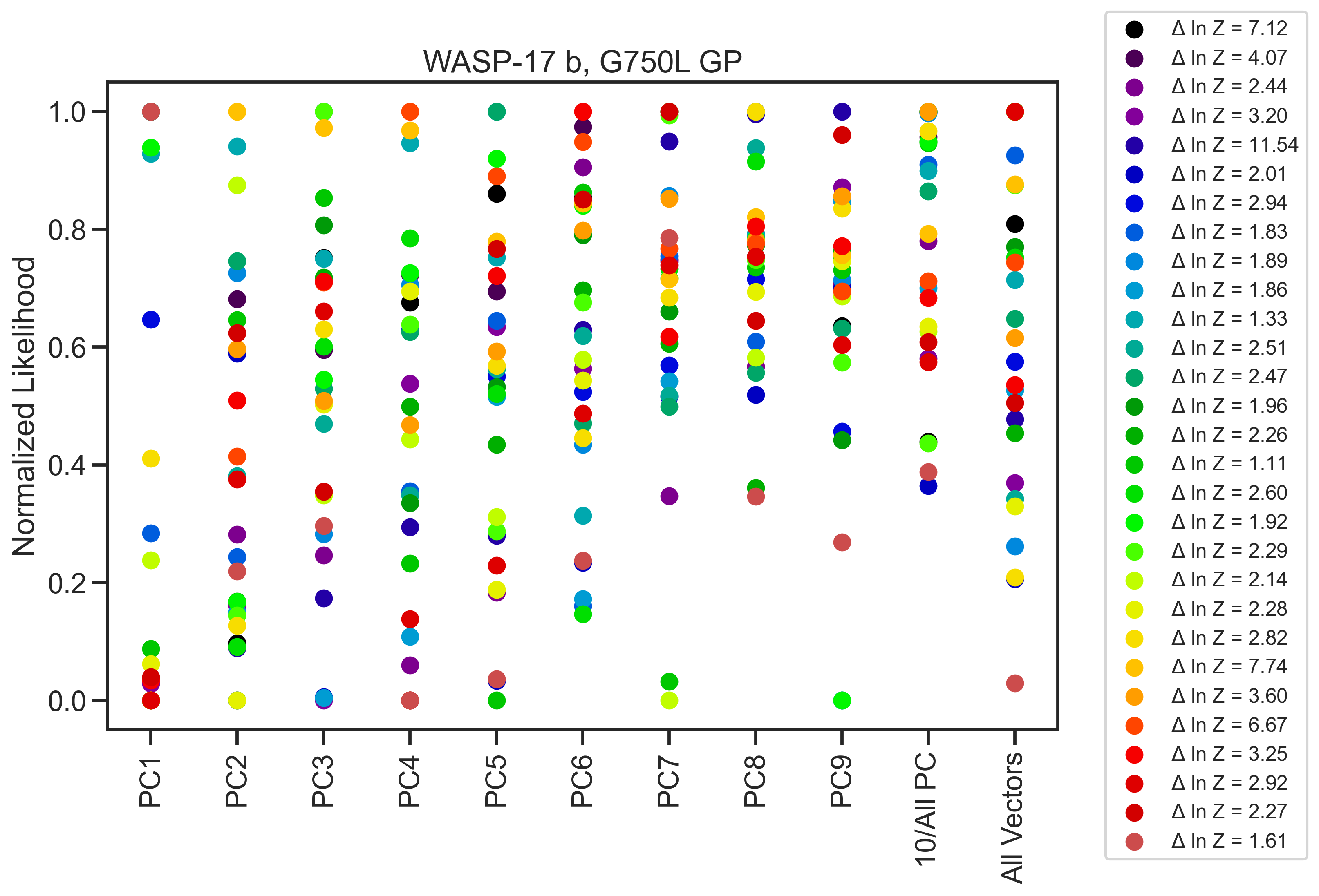}{0.49\textwidth}{}}
    \caption{Same as \autoref{fig:evidence1}, but for G750L GP detrending.}
    \label{fig:evidence3}
\end{figure}

\begin{figure}[h!]
\gridline{\fig{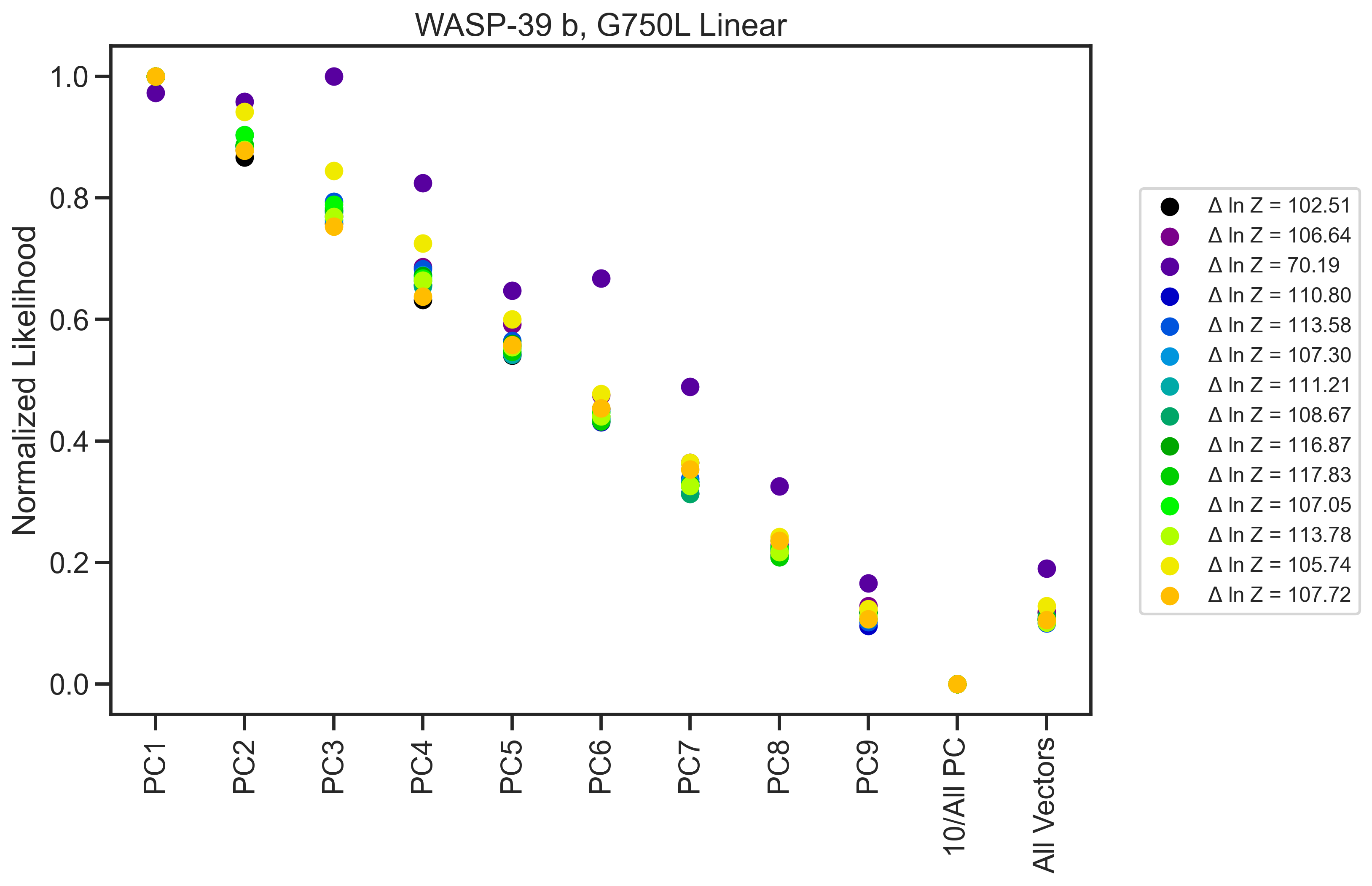}{0.49\textwidth}{}
          \fig{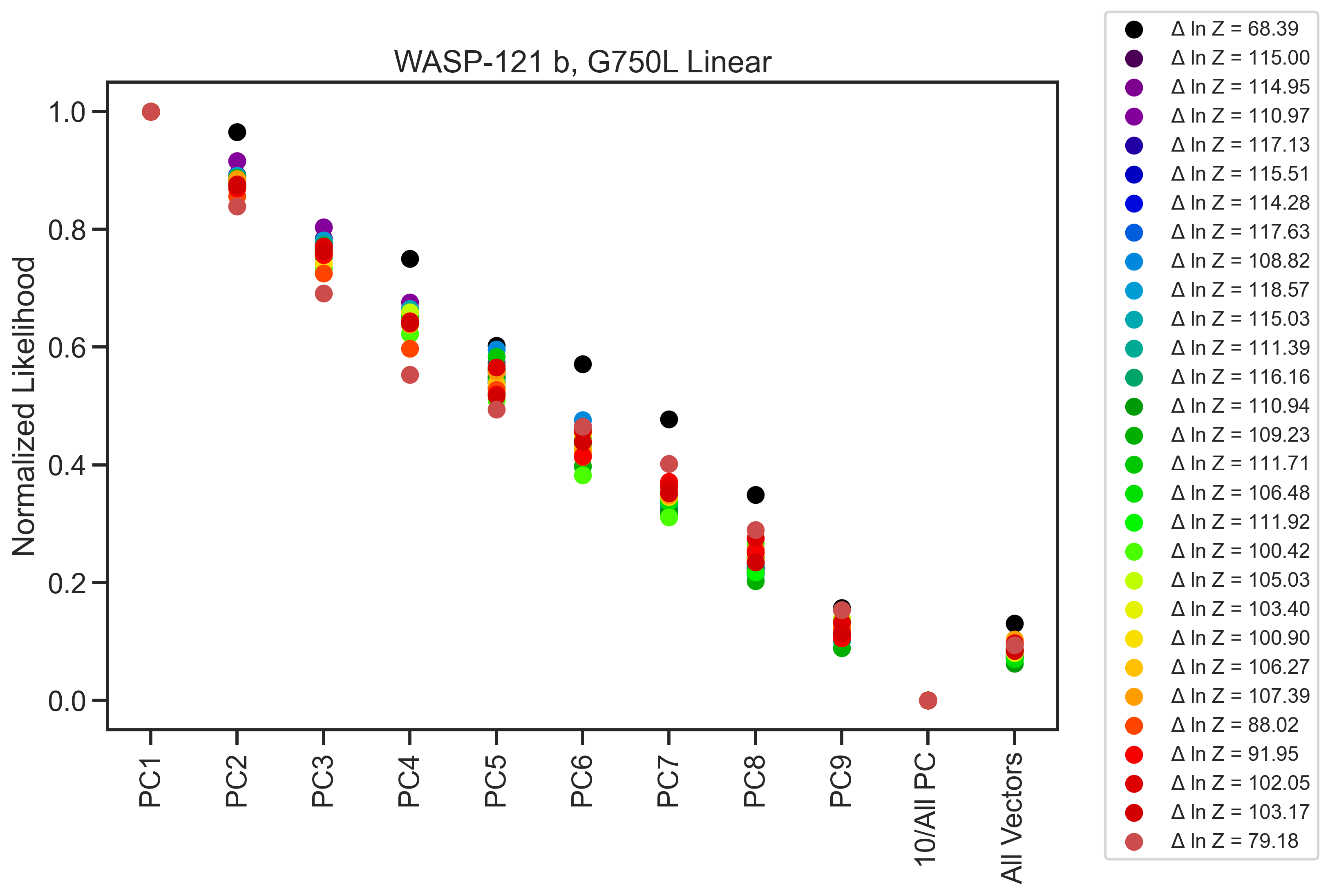}{0.49\textwidth}{}}
\gridline{\fig{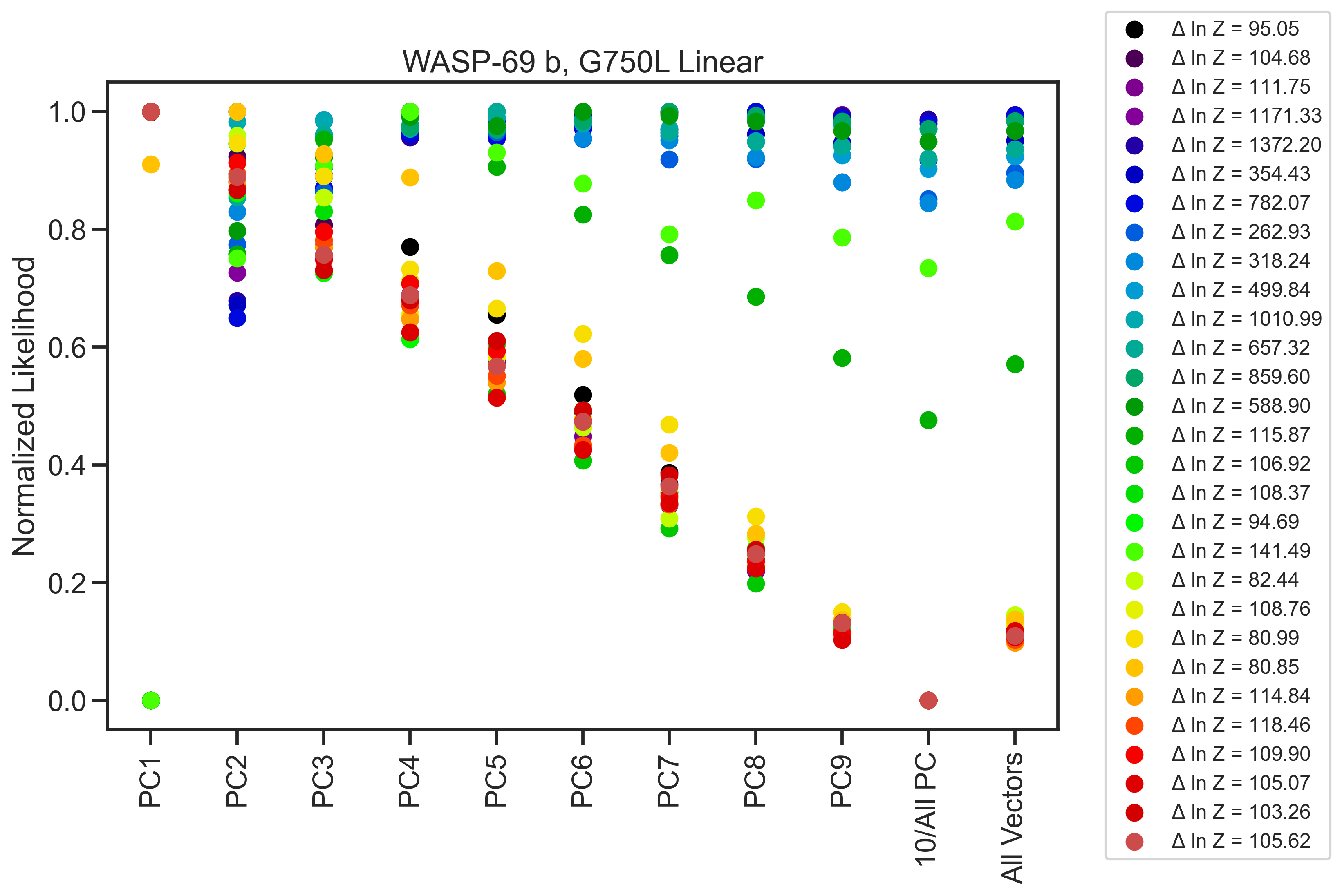}{0.49\textwidth}{}
          \fig{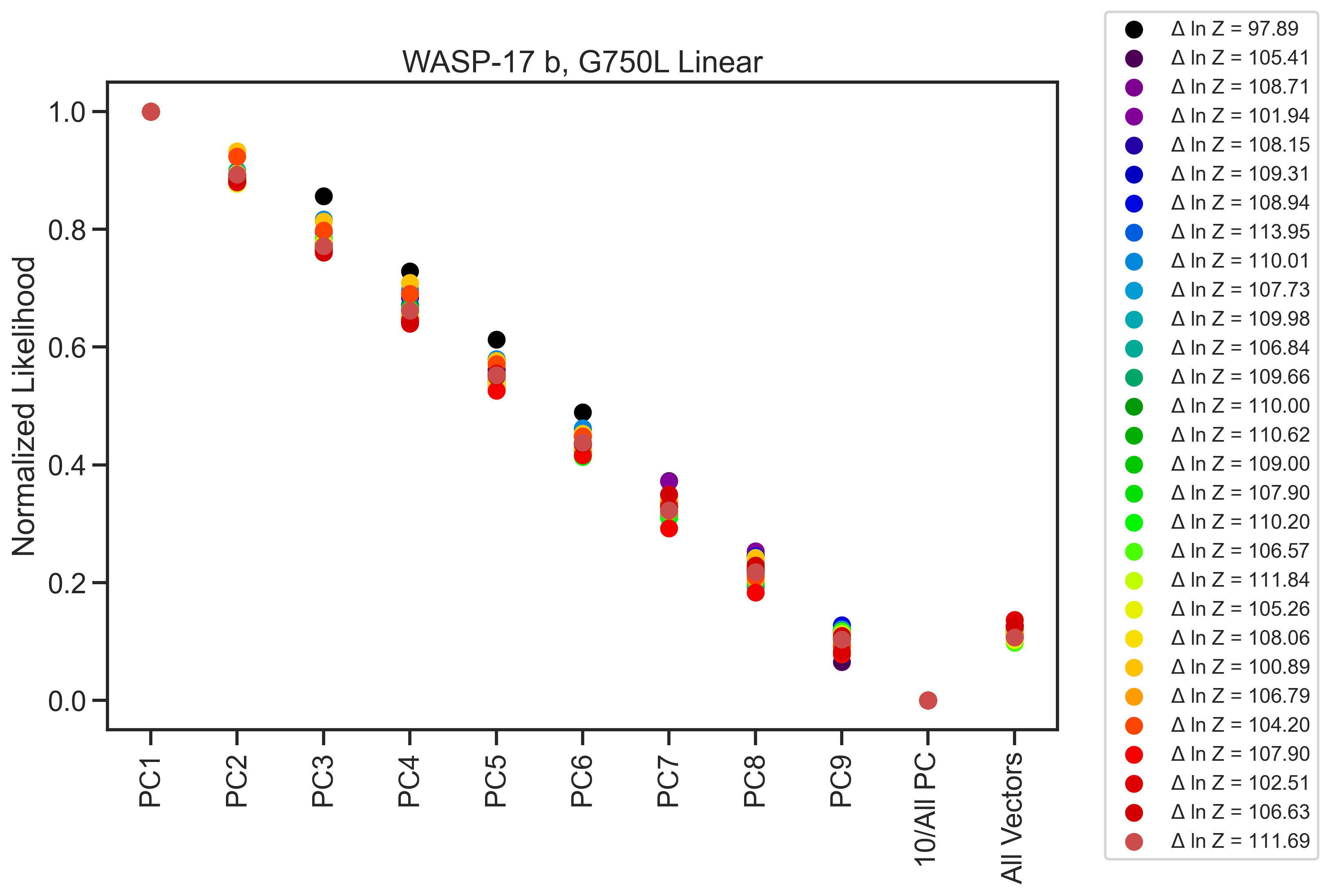}{0.49\textwidth}{}}
    \caption{Same as \autoref{fig:evidence1}, but for G750L linear detrending.}
    \label{fig:evidence4}
\end{figure}

Each spectroscopic light curve fit has an associated Bayesian evidence \textit{ln Z}, which can be, with a prior defined for each model, used to define the probability of each model given the data by taking into account the model fit itself but also the number of free parameters. If the number of free parameters in a model is not supported by the fit, this value will decrease to penalize overfitting. We normalize the evidence from each fit onto a 0 to 1 scale, and plot the resulting normalized evidence values as a function of detrending method per wavelength bin in \autoref{fig:evidence1}, \autoref{fig:evidence2}, \autoref{fig:evidence3}, and \autoref{fig:evidence4}, for the G430L GP, G430L linear, G750L GP, and G750L linear, respectively. The legend shows the absolute $\Delta$ \textit{ln Z}, which gives the scale for each wavelength bin. The idea is that the more bunched towards 1 the evidence values are, the better of a fit that detrending method offers on average. By comparing the results between our test datasets between the linear and GP methods, a clear pattern emerges. The G430L GP fits show a clear turnover in evidence, increasing the overall goodness of fit of the spectrum for a few PCs and then in turn decreasing. This same trend is there for the G750L GP fits, but weaker as, as previously discussed, this spectrum is only one transit and thus is significantly less dependent on detrending method. The exact point of the turnover varies slightly, but we find the most consistent is 5 PCs, and so we use this for our final GP-detrended spectra in \autoref{fig:final_spec}. On the other hand, the linear fits show interesting decreasing behavior after around 1-3 PCs, in which those first detrending methods are about equivalent and then they steeply decrease. This is simply saying that the improvement in fit for additional vectors past the first few are not supported by the data. Thus, we choose 2 PCs for our final linear-detrended spectra in \autoref{fig:final_spec}. \\

We note that while we choose the single best performing systematics model for each method, this choice has an effect on the absolute $R_p/R_s$. If the situation requires knowledge of the absolute $R_p/R_s$, we instead suggest to take a weighted average of the different PC spectra results as the final spectrum, noting that this would increase the errors to encompass the spread in the data from using different systematics models.  \\

\section{Additional Spectroscopic Light Curves}\label{app:slc}
\begin{figure*}[h!]
    \gridline{\fig{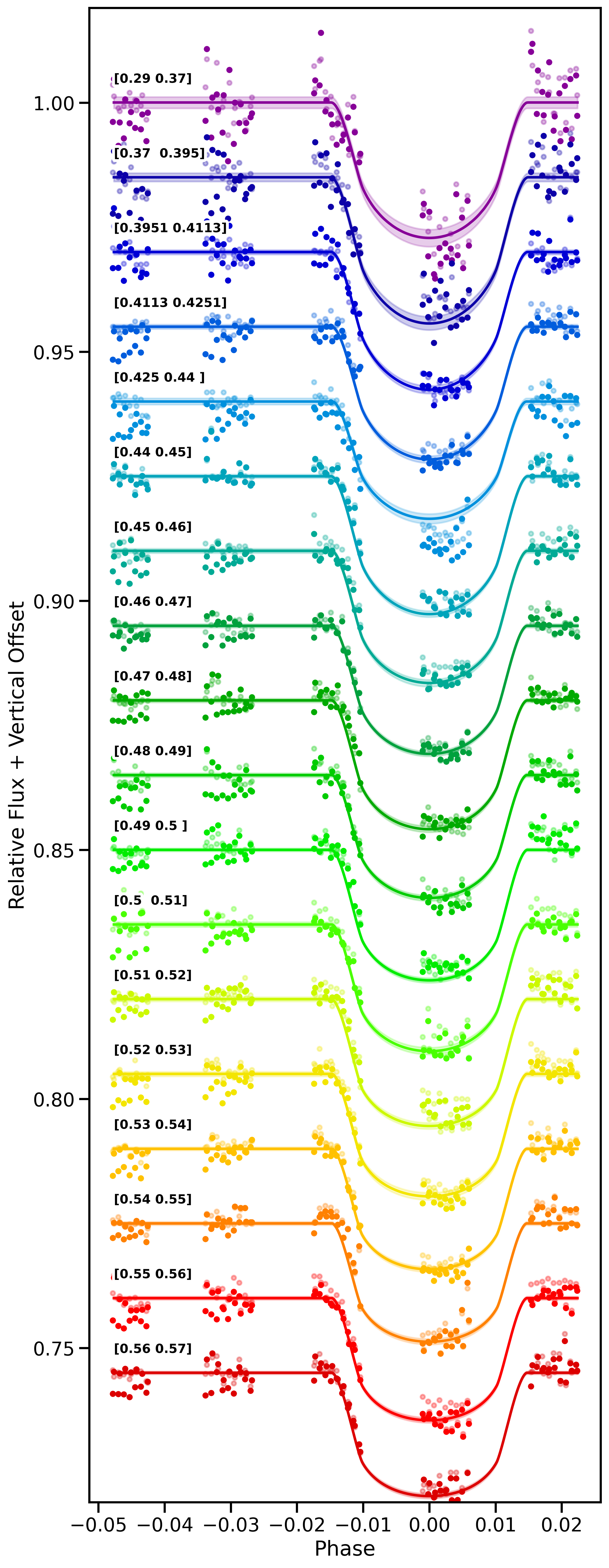}{0.4\textwidth}{}
          \fig{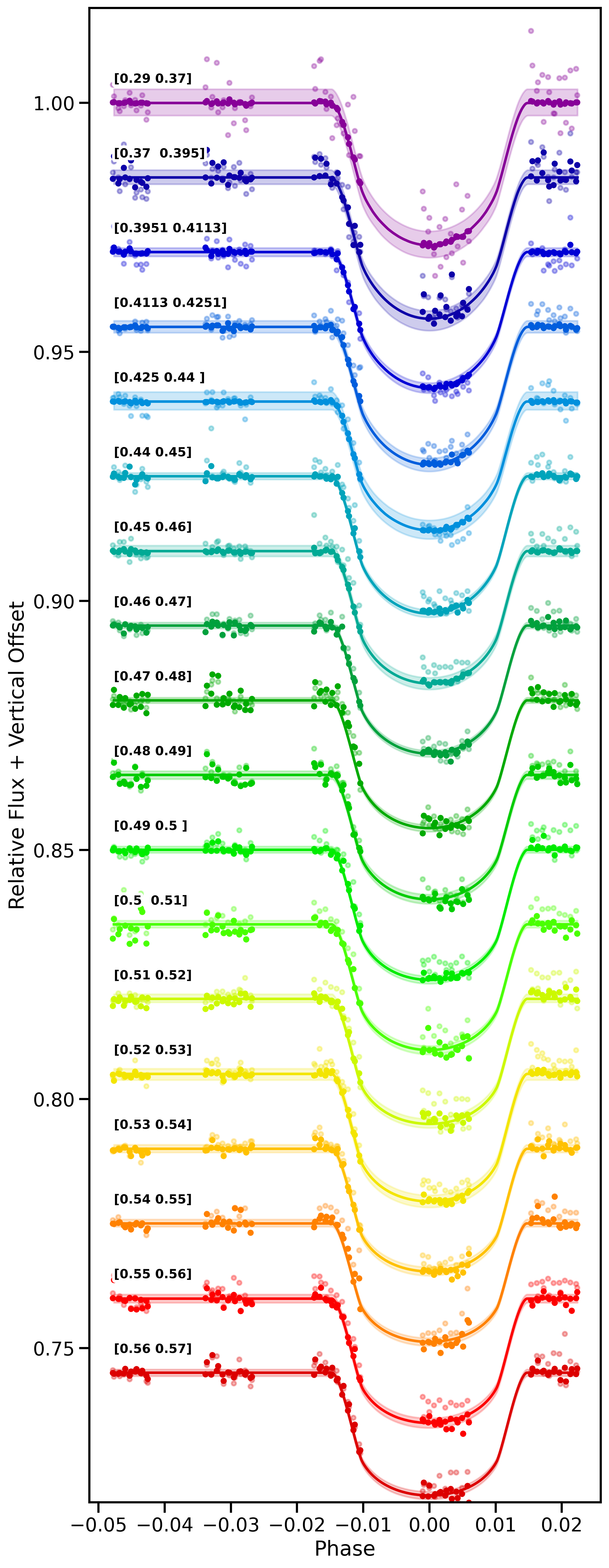}{0.4\textwidth}{}}
    \caption{The same as \autoref{fig:slc_430}, but for WASP-39\,b.}
    \label{fig:slc_w39b_430}
\end{figure*}

\begin{figure*}[h!]
    \gridline{\fig{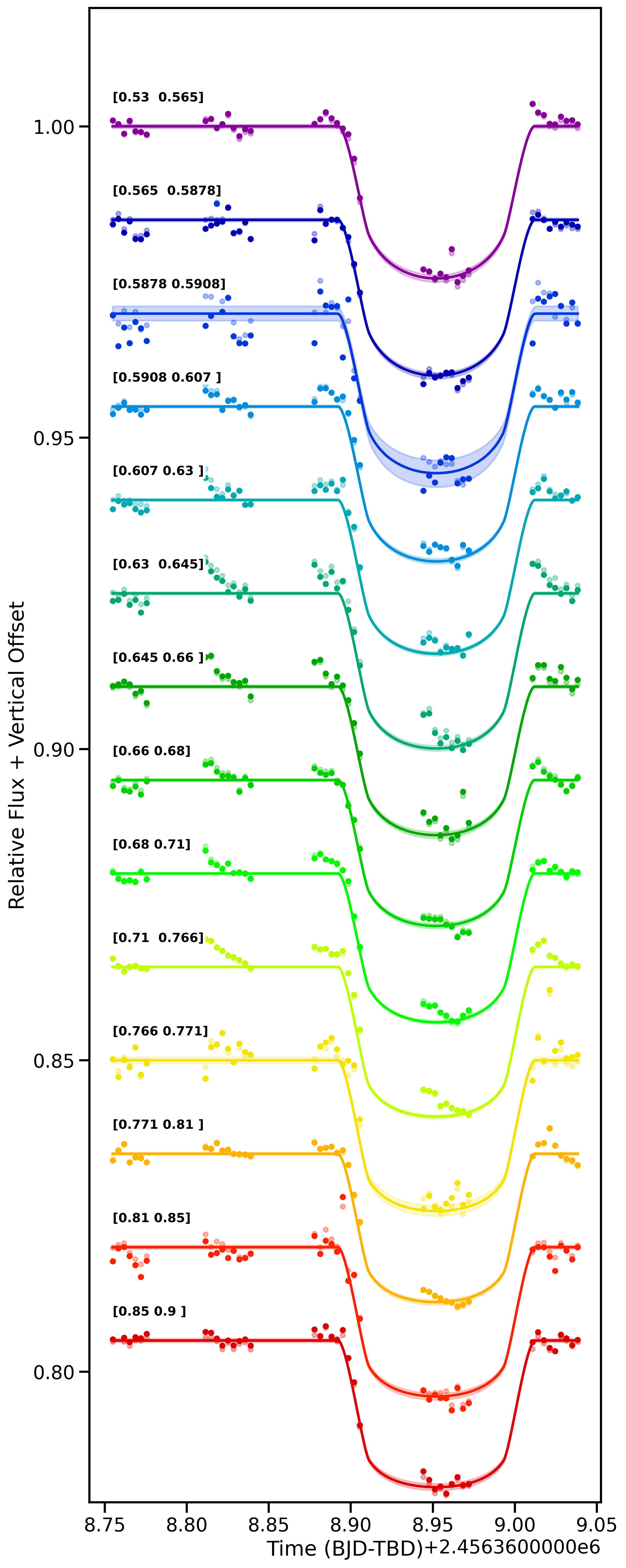}{0.4\textwidth}{}
          \fig{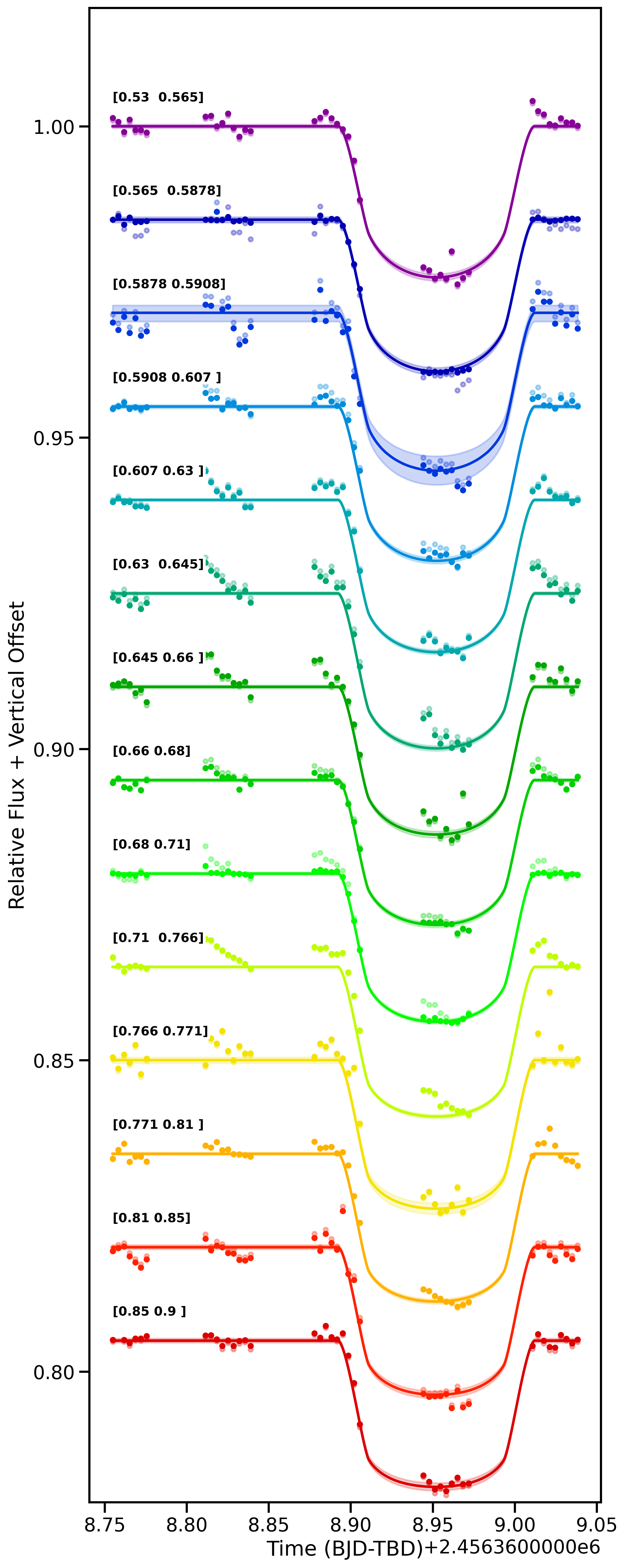}{0.4\textwidth}{}}
    \caption{The same as \autoref{fig:slc_750}, but for WASP-39\,b.}
    \label{fig:slc_w39b_750}
\end{figure*}

\begin{figure*}[h!]
    \gridline{\fig{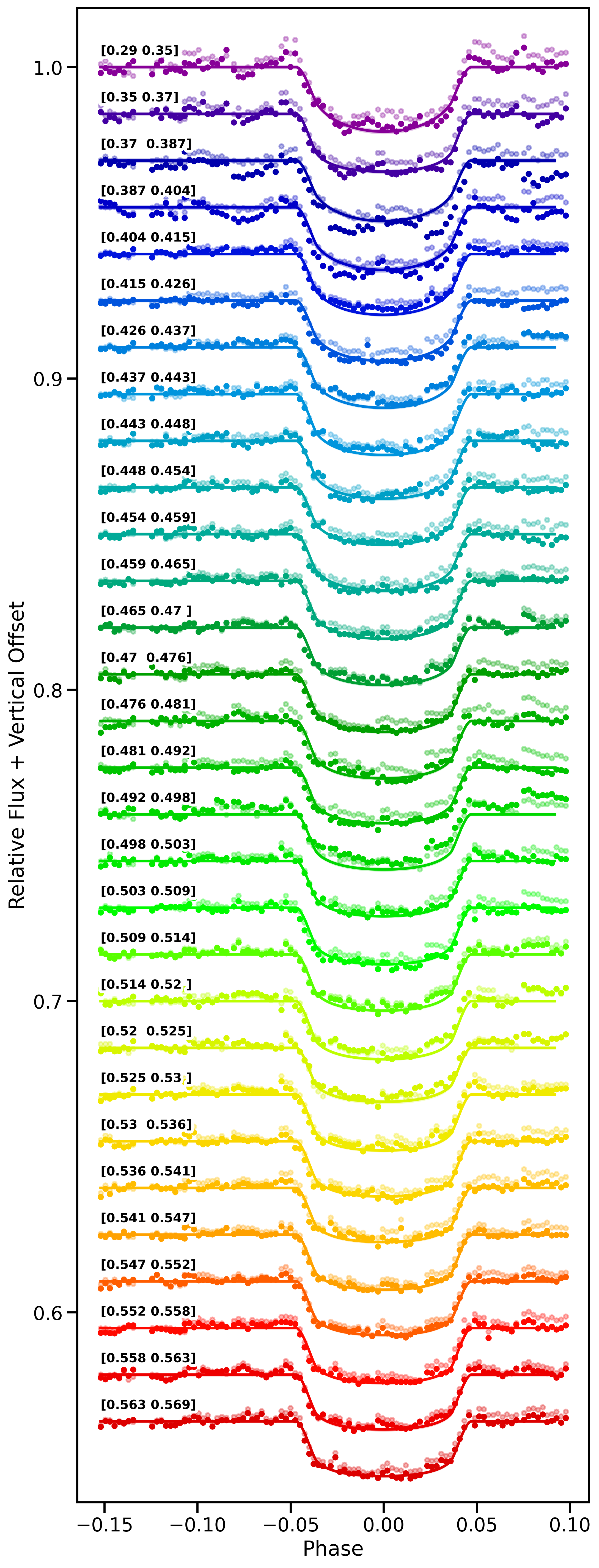}{0.4\textwidth}{}
          \fig{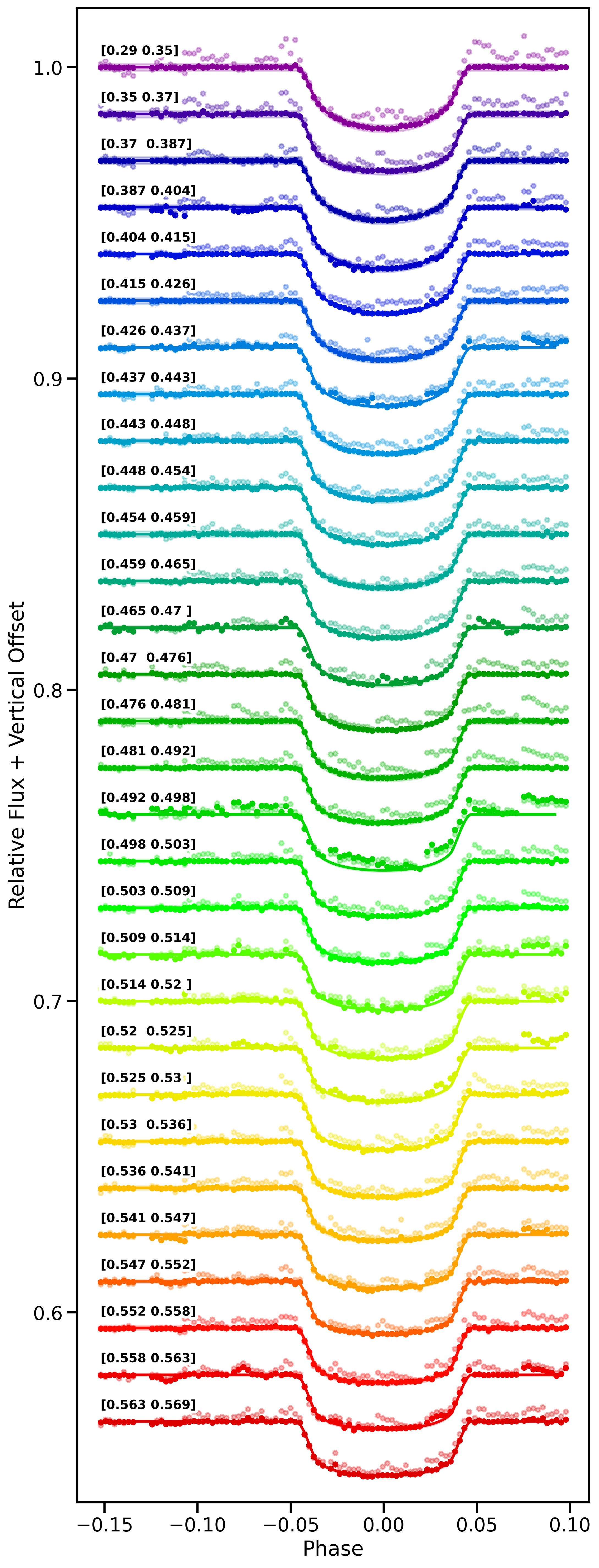}{0.4\textwidth}{}}
    \caption{The same as \autoref{fig:slc_430}, but for WASP-121\,b.}
    \label{fig:slc_w121b_430}
\end{figure*}

\begin{figure*}[h!]
    \gridline{\fig{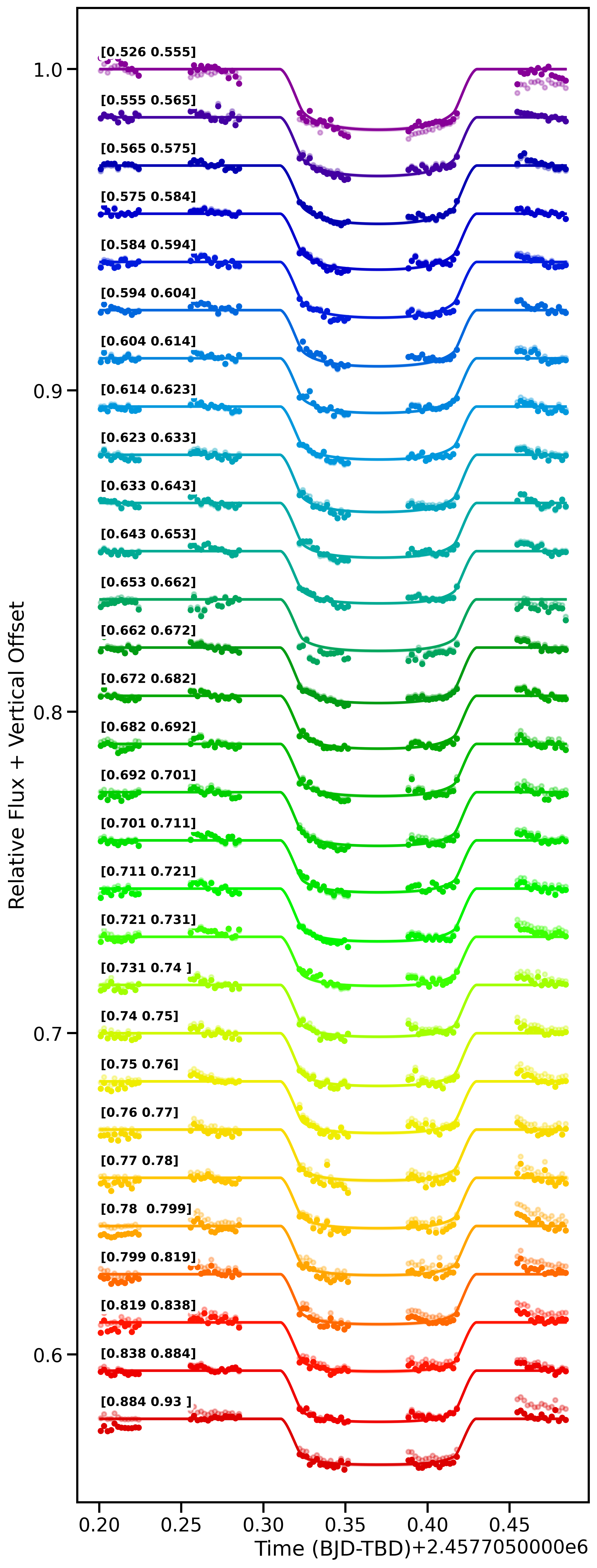}{0.4\textwidth}{}
          \fig{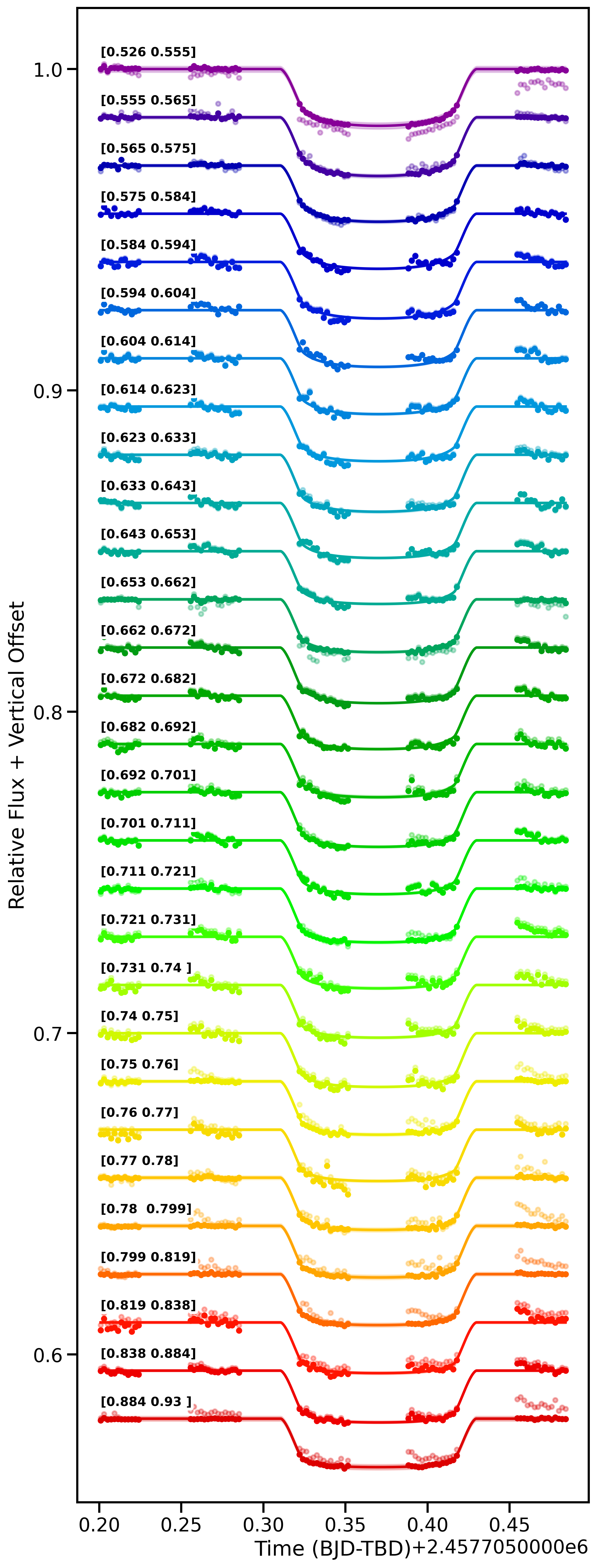}{0.4\textwidth}{}}
    \caption{The same as \autoref{fig:slc_750}, but for WASP-121\,b.}
    \label{fig:slc_w121b_750}
\end{figure*}

\begin{figure*}[h!]
    \gridline{\fig{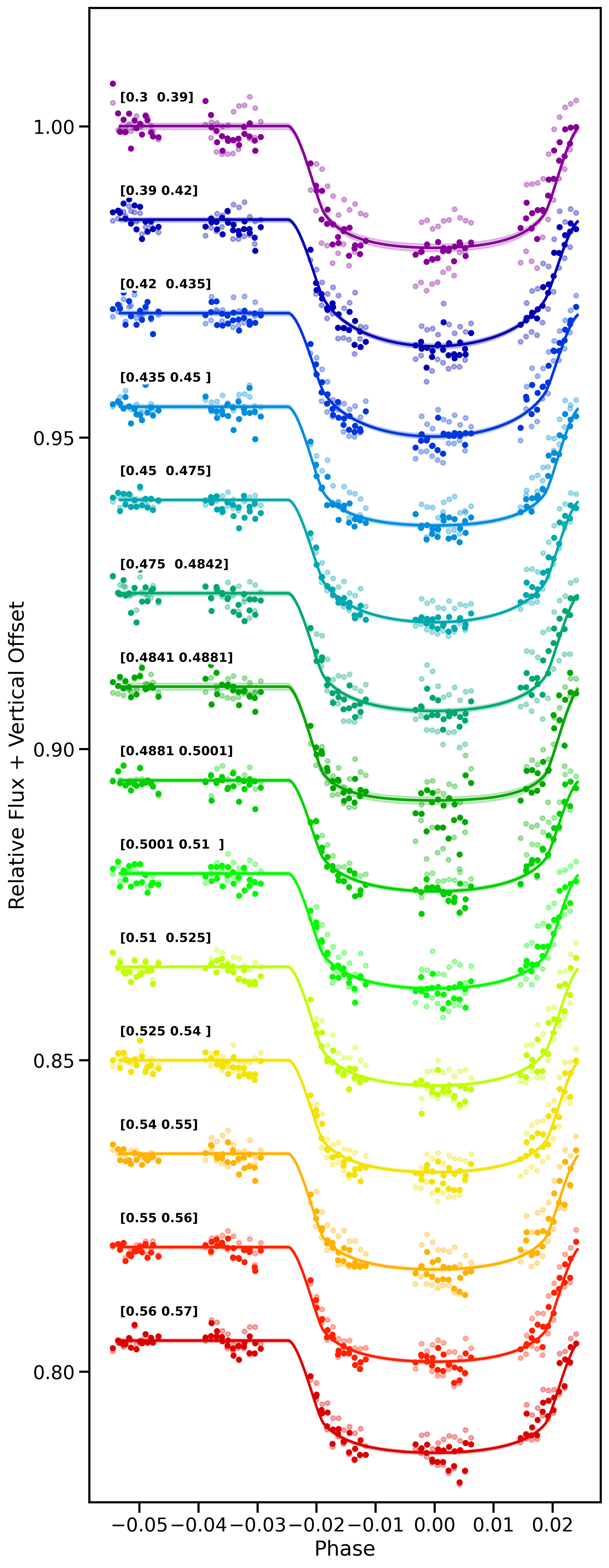}{0.4\textwidth}{}
          \fig{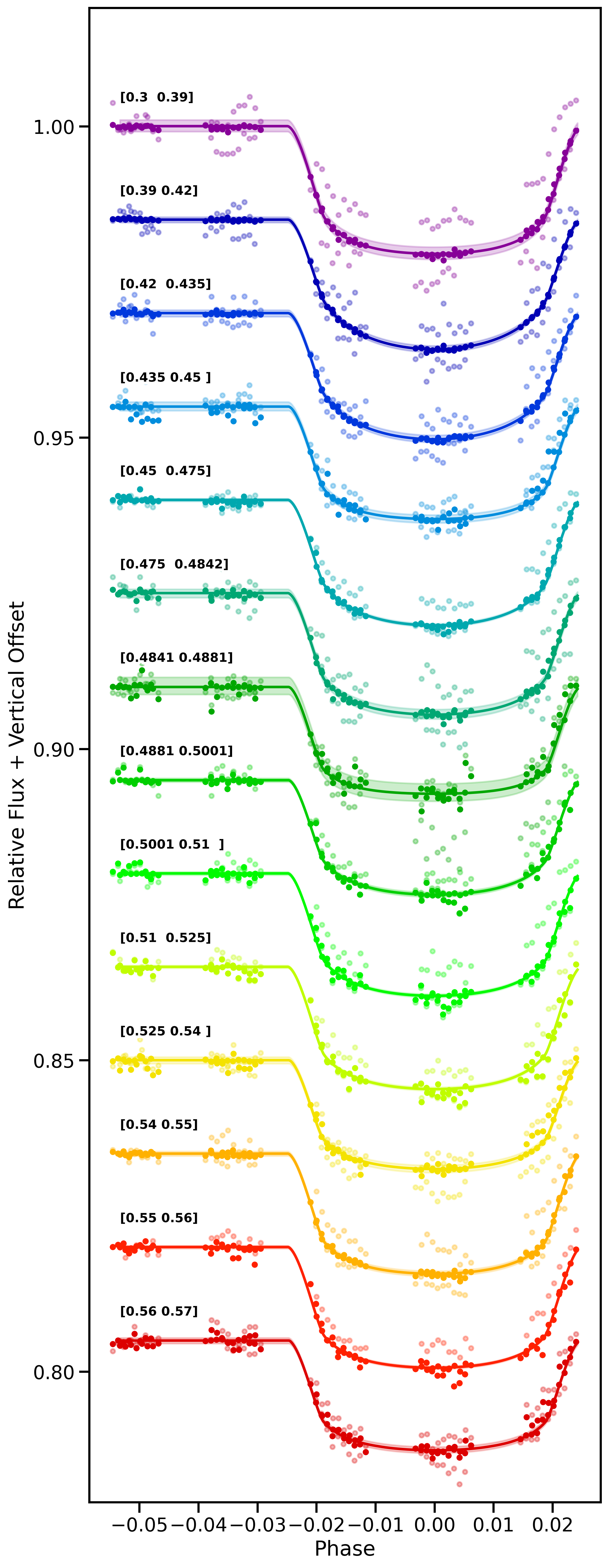}{0.4\textwidth}{}}
    \caption{The same as \autoref{fig:slc_430}, but for WASP-17\,b.}
    \label{fig:slc_w17b_430}
\end{figure*}

\begin{figure*}[h!]
    \gridline{\fig{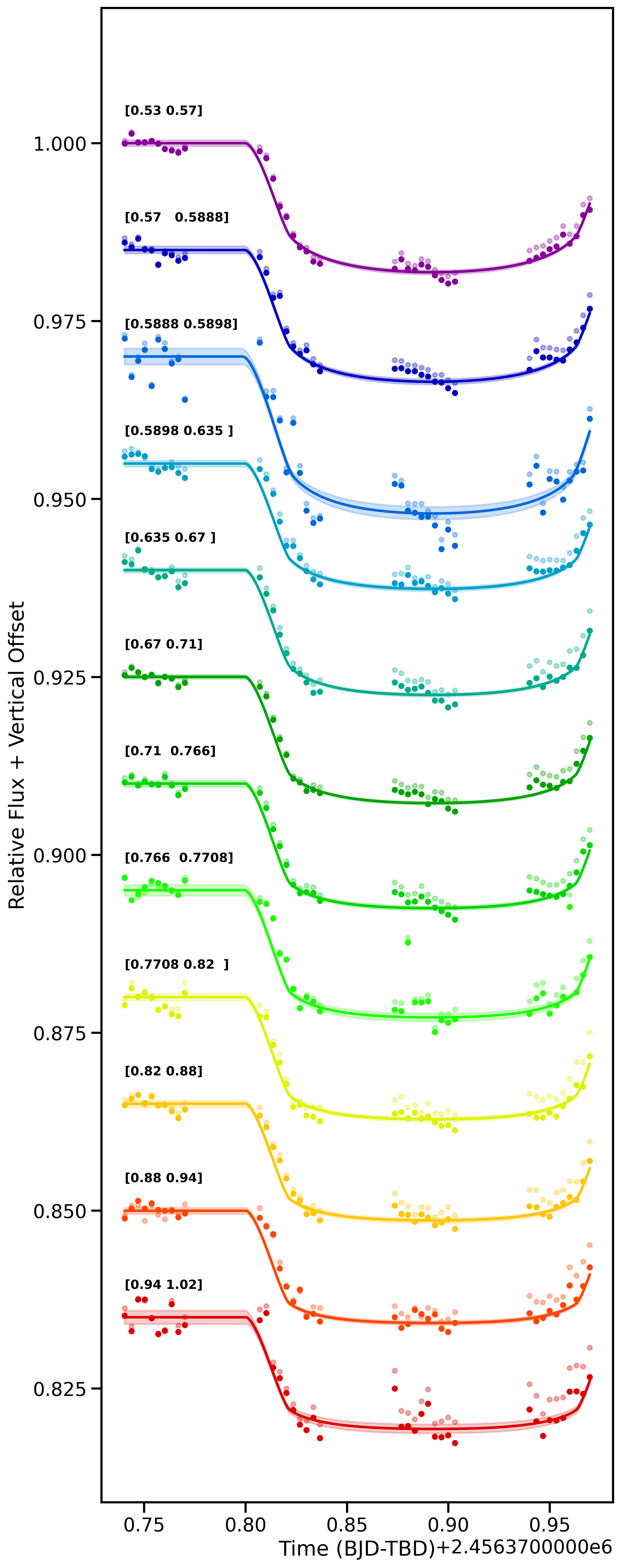}{0.4\textwidth}{}
          \fig{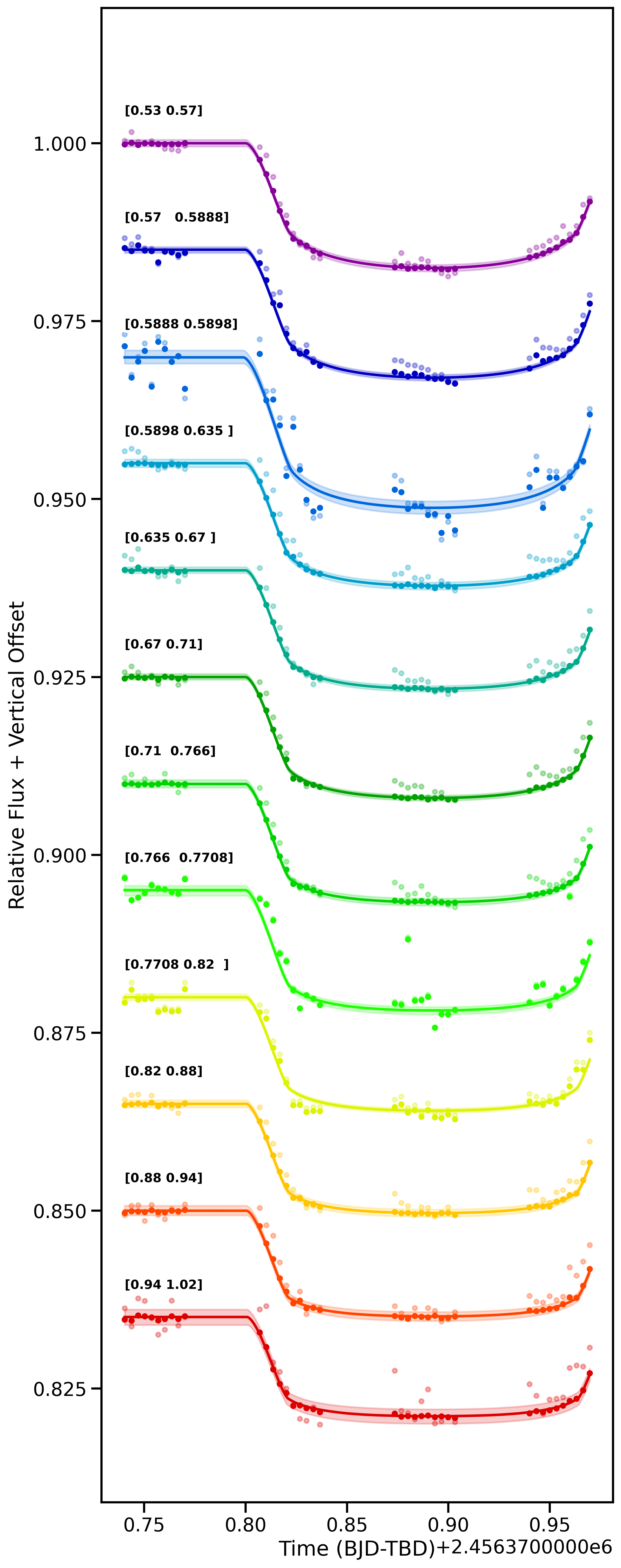}{0.4\textwidth}{}}
    \caption{The same as \autoref{fig:slc_750}, but for WASP-17\,b.}
    \label{fig:slc_w17b_750}
\end{figure*}

This section contains the spectroscopic light curves not included in the main text: WASP-39\,b G430L (\autoref{fig:slc_w39b_430}) and G750L (\autoref{fig:slc_w39b_750}), WASP-121\,b G430L (\autoref{fig:slc_w121b_430}) and G750L (\autoref{fig:slc_w121b_750}), and WASP-17\,b G430L (\autoref{fig:slc_w17b_430}) and G750L (\autoref{fig:slc_w17b_750}).

\section{WASP-69\,b Extra Figures}
\subsection{WASP-69\,b Full Spectrum}\label{app:w69_full}
\begin{figure*}[h!]
    \centering
    \includegraphics[width=0.65\textwidth]{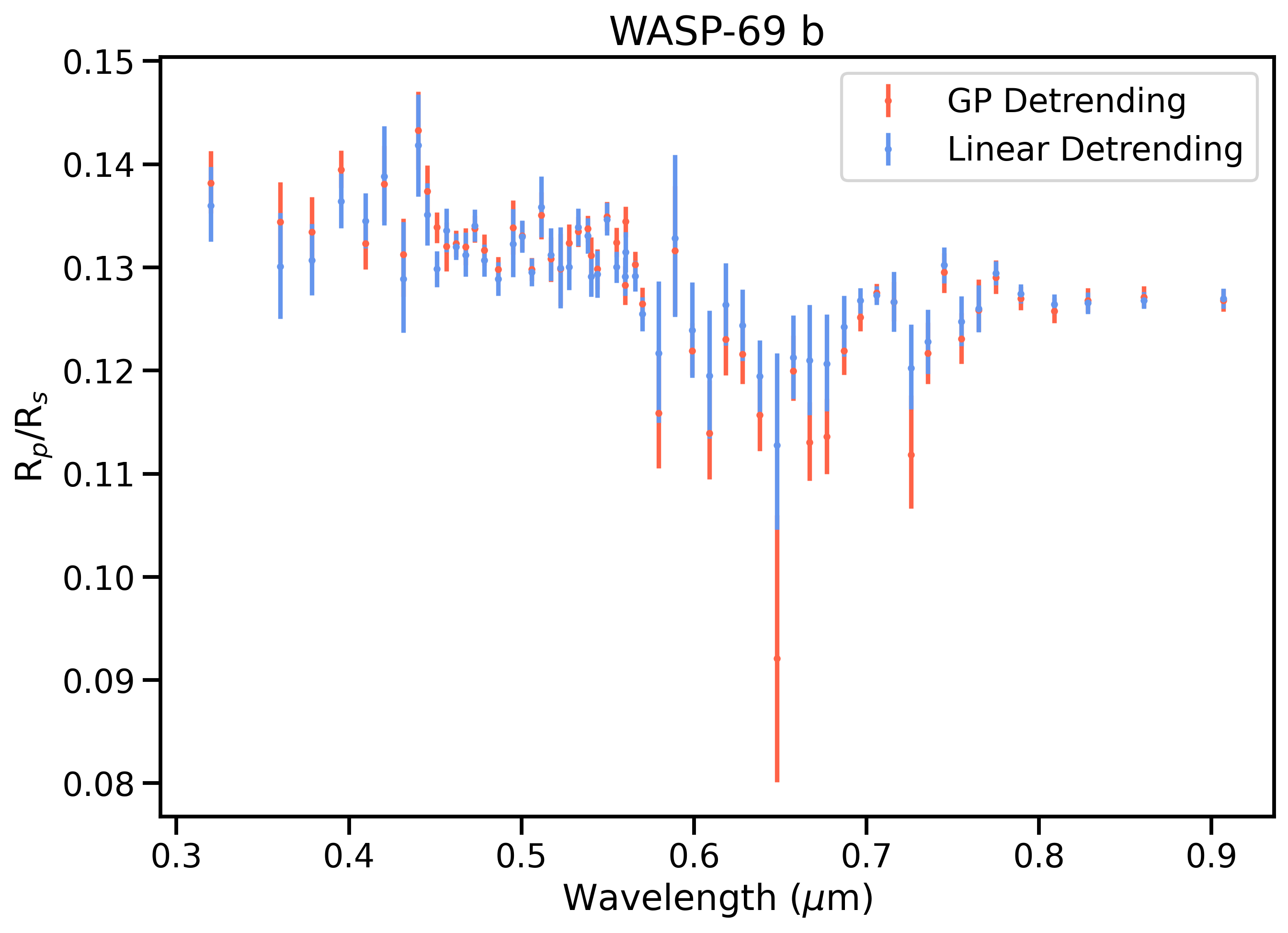}
    \caption{The same as the bottom left panel of \autoref{fig:final_spec}, but without a restricted y-axis scale, showing the full spectrum of WASP-69\,b.}
    \label{fig:w69_full}
\end{figure*}

The observation of WASP-69\,b taken with the G750L mode had especially strong systematics present (see \autoref{fig:wlc}), and thus to show the variation level in the less affected part of the spectrum better, we restrict the y-axis scale in \autoref{fig:final_spec}. \autoref{fig:w69_full} instead shows the full range of the WASP-69\,b spectrum.

\subsection{Offset with previous result}\label{sec:w69_offset}
\begin{figure*}
        \gridline{\fig{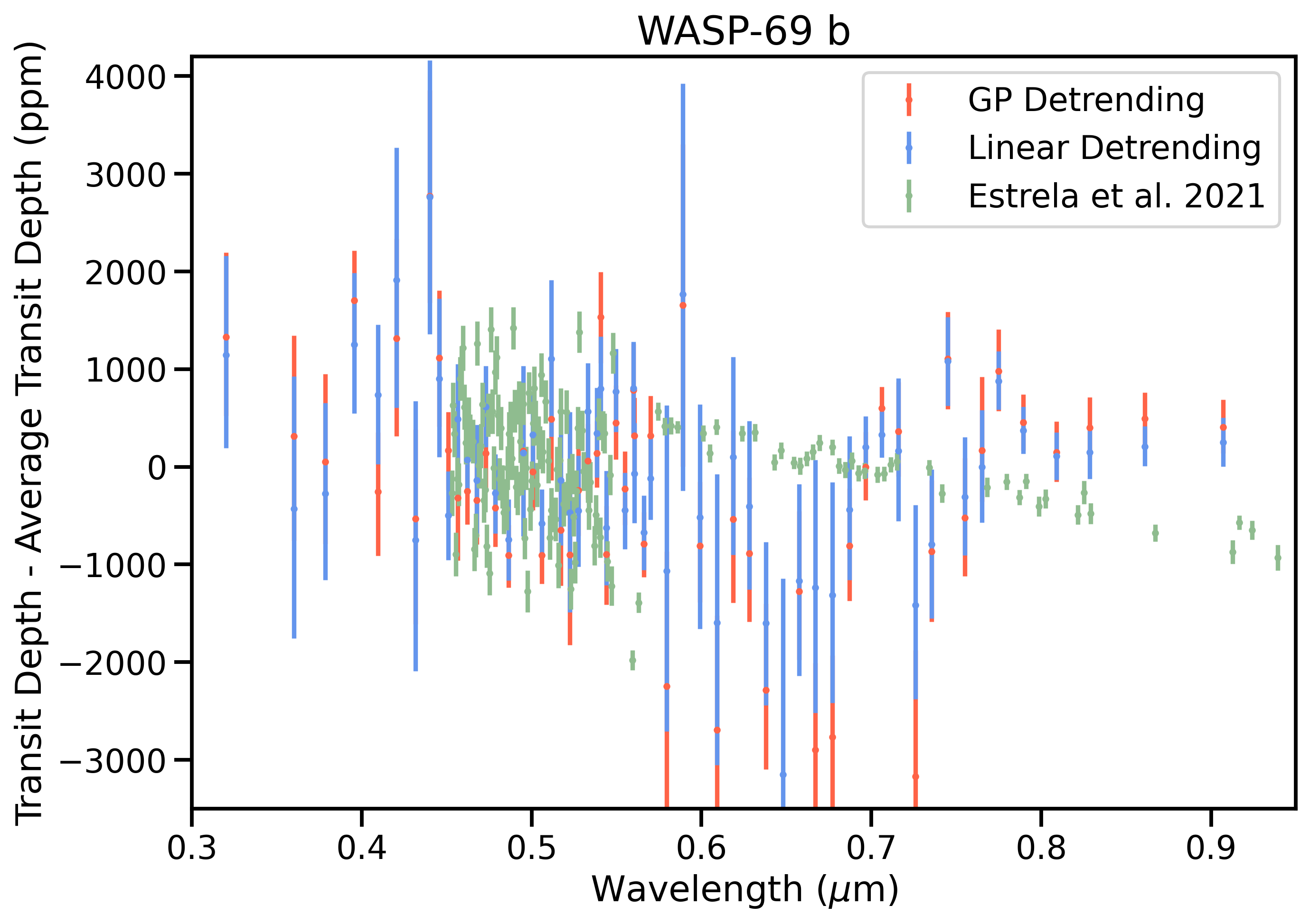}{0.4\textwidth}{}
          \fig{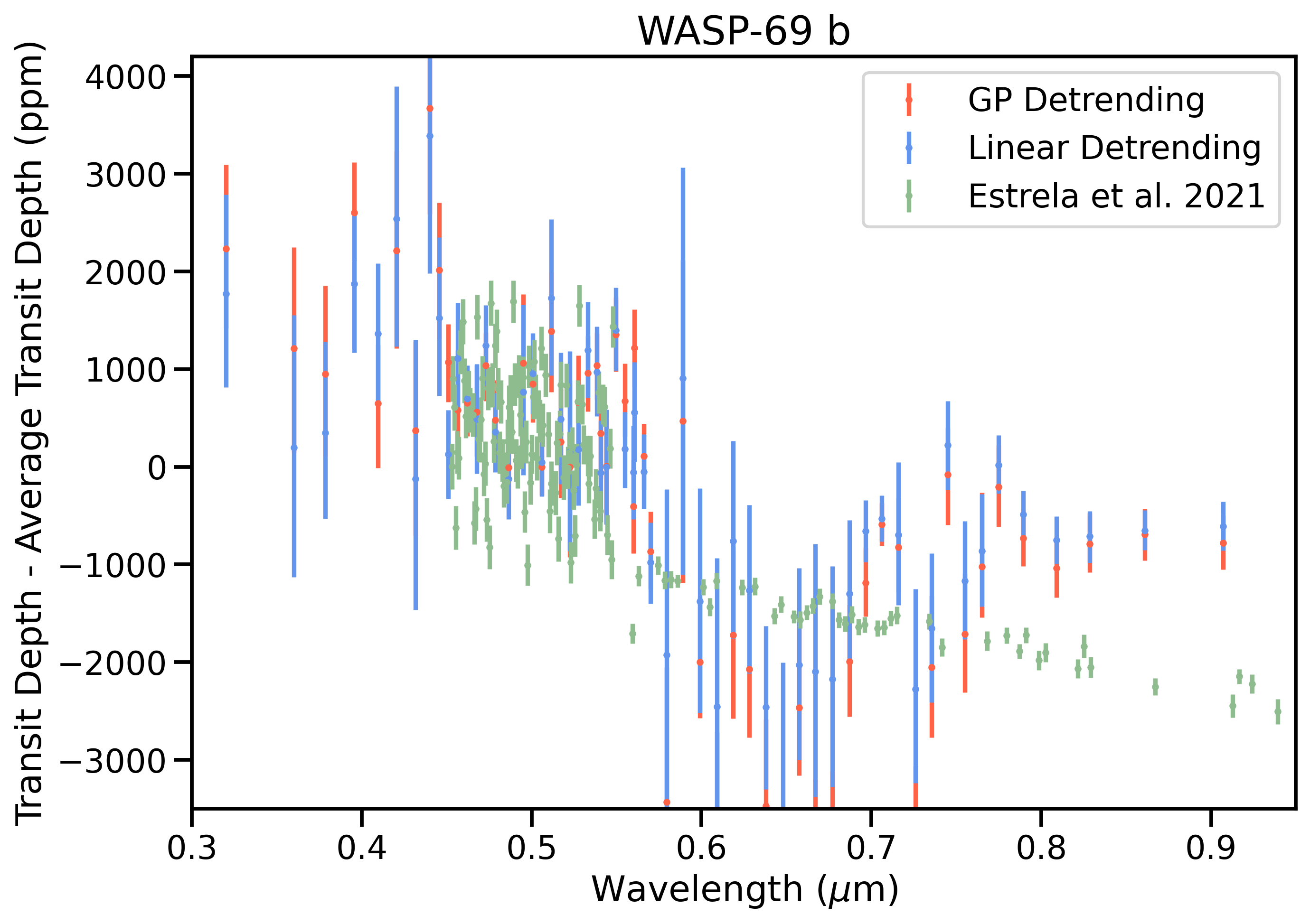}{0.4\textwidth}{}}
    \caption{The transit depth with the average transit depth removed for both of our analyses (GP and linear) and the analysis from \citet{Estrela_2021}. The left has removed the average transit depth on each instrument separately, while the right has removed the overall average transit depth for each reduction. In both cases, the leftover spectral shape is distinctly different within the error bars, meaning that there is more than just a vertical offset between reductions.}
    \label{fig:w69_sub}
\end{figure*}

As can be seen in \autoref{fig:final_spec}, our WASP-69\,b spectrum seems to have both a vertical offset and a spectral shape difference from the previously published spectrum from \citet{Estrela_2021}, but without accounting for this offset it is difficult to say whether there is a spectral shape difference for certain. Therefore, we remove the average transit depth from both our analyses and those from \citet{Estrela_2021} and compare the results in \autoref{fig:w69_sub}. Since \citet{Estrela_2021} fit for an offset between the G430L and G750L data, we try removing the average transit depth both in the overall spectrum and also per instrument, but in both cases the spectral shape between our spectrum and that from \citet{Estrela_2021} remain different.

\section{Spectral Extraction Aperture Size}\label{app:aper}
\begin{figure}[h!]
\gridline{\fig{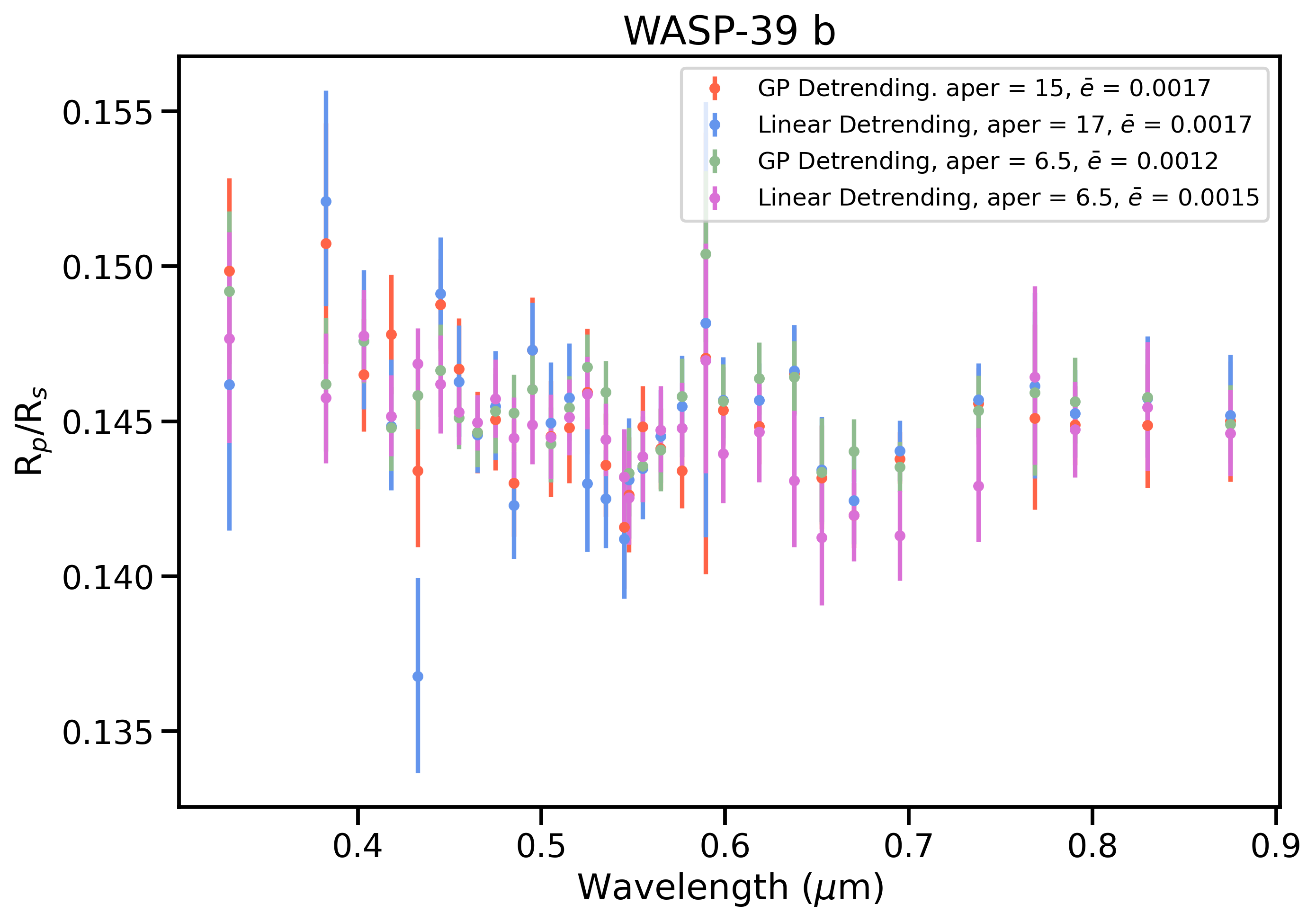}{0.49\textwidth}{}
          \fig{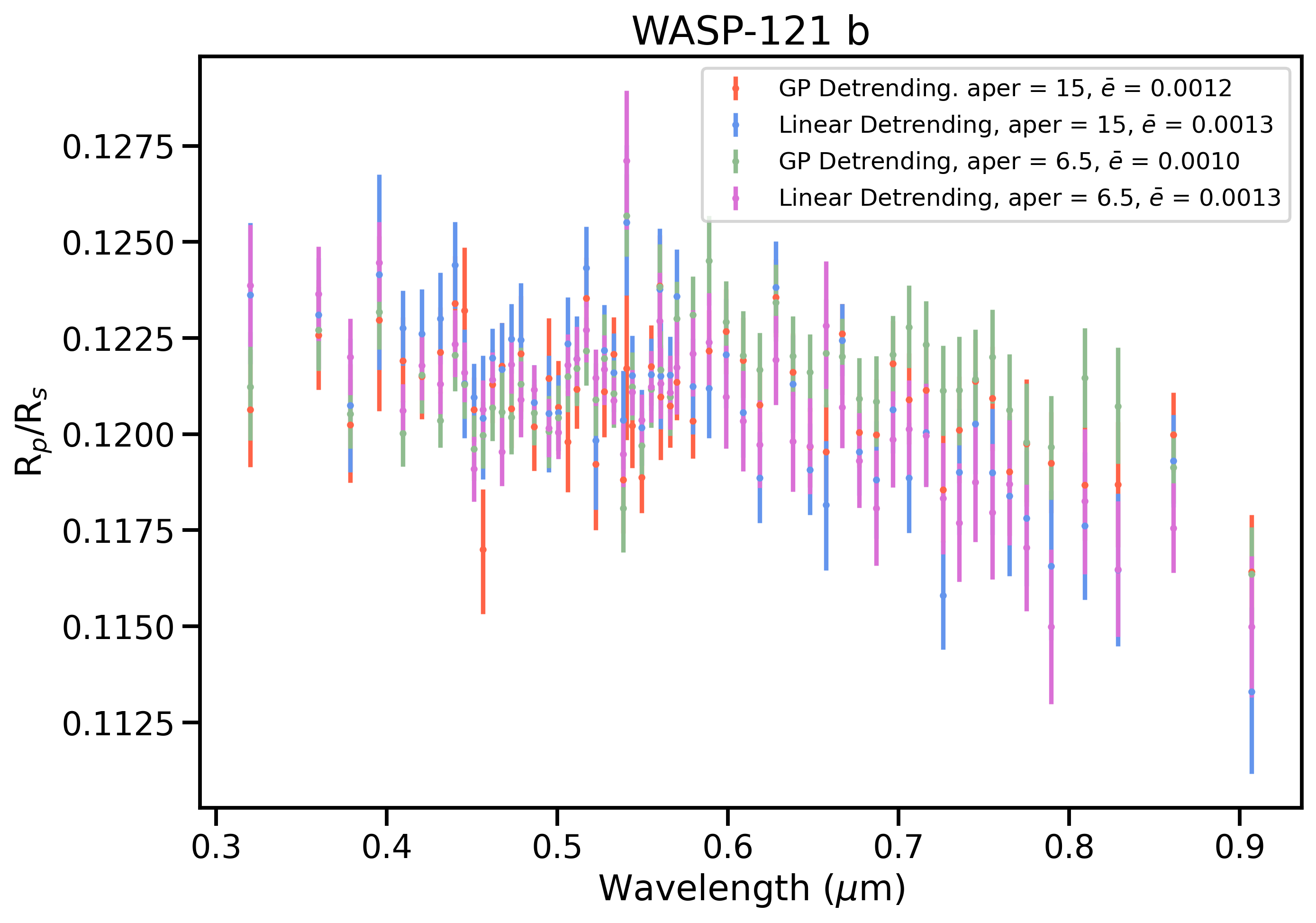}{0.49\textwidth}{}}
\gridline{\fig{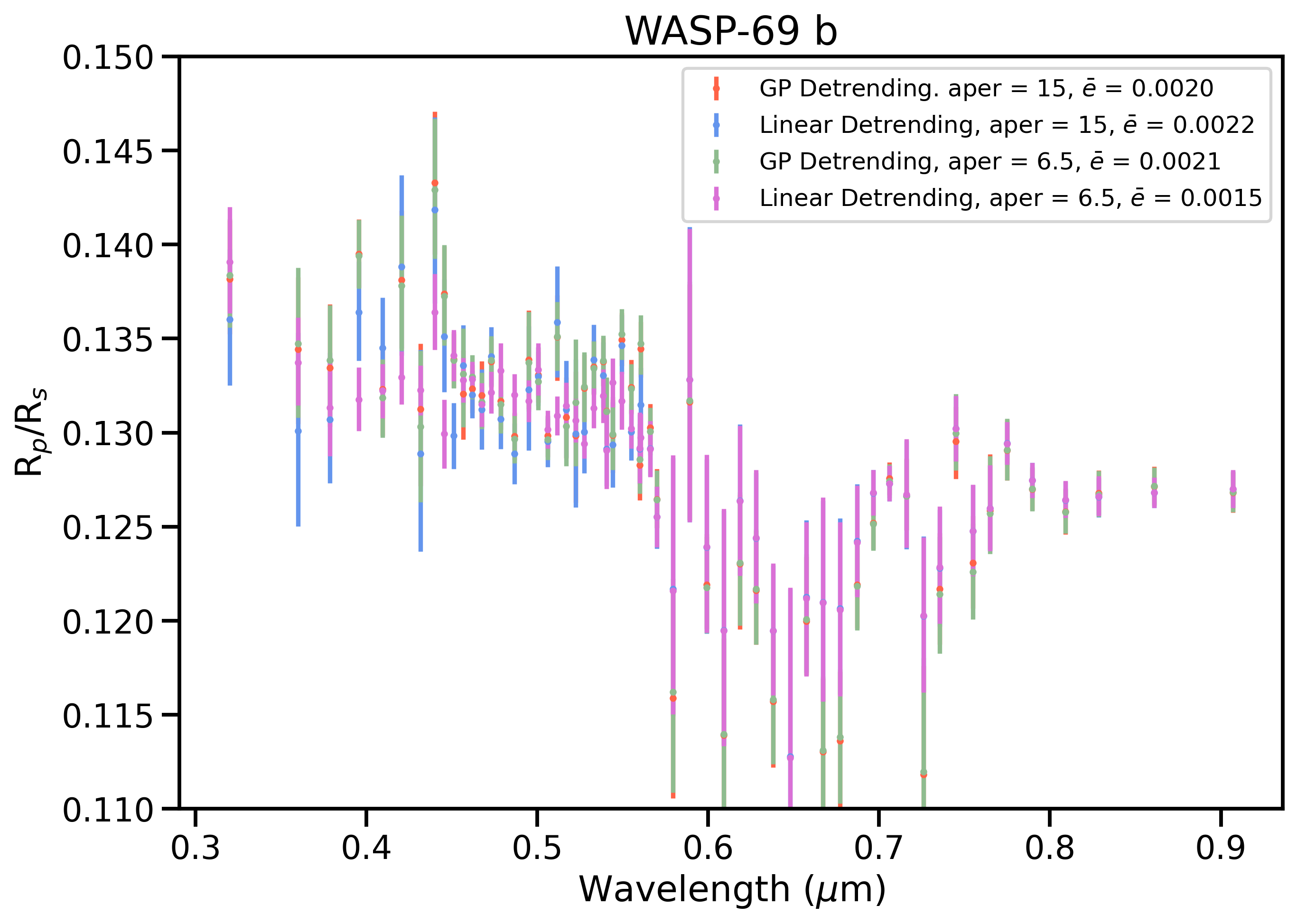}{0.49\textwidth}{}
          \fig{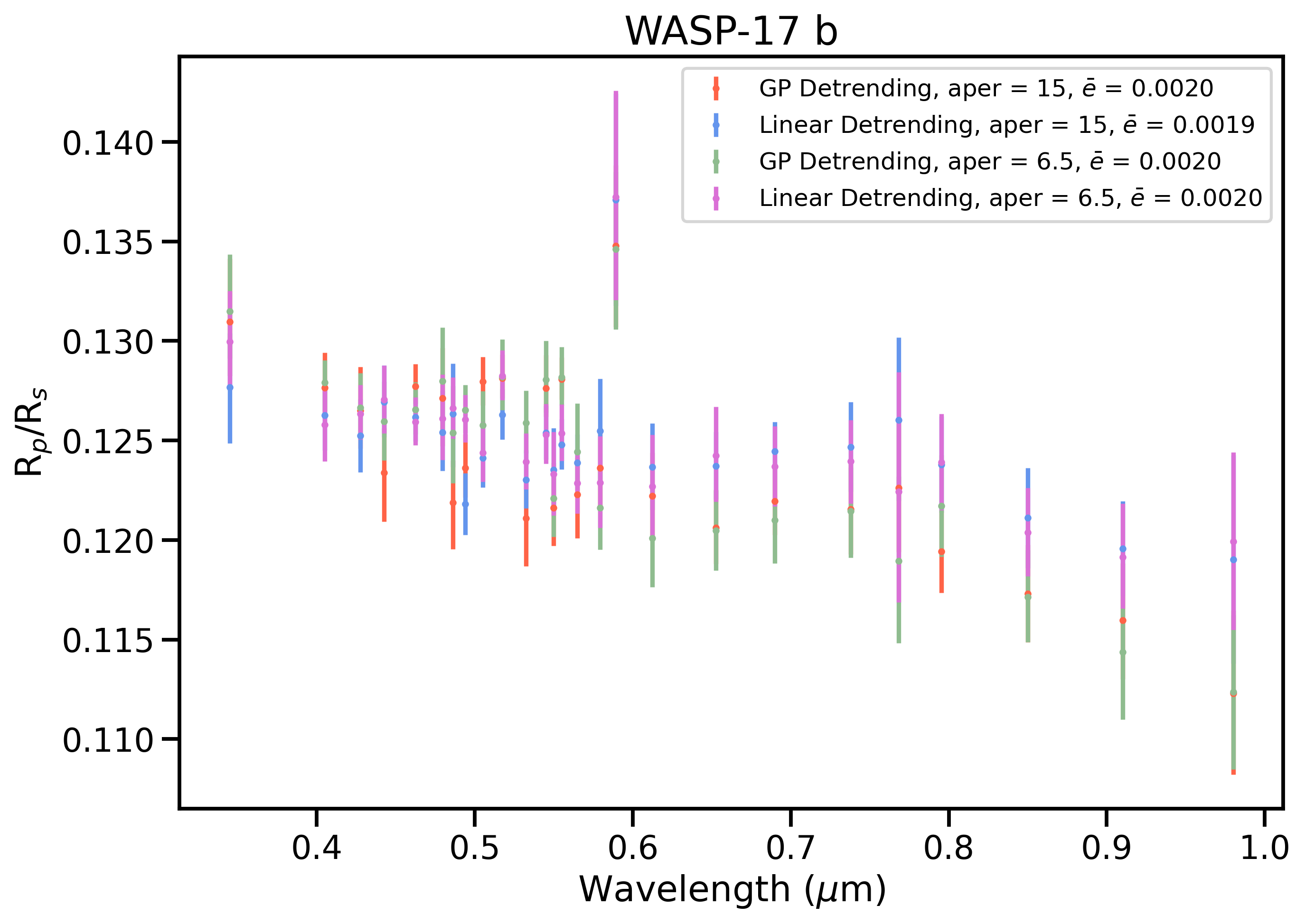}{0.49\textwidth}{}}
    \caption{The result of changing the aperture used for spectral extraction on the final transmission spectrum. It can be seen that the resulting spectra are broadly consistent with each other, but that the median error bar (given in the legend) do change a significant amount. }
    \label{fig:aper}
\end{figure}

To test the effect of the size of the aperture used for spectral extraction, we complete our typical spectroscopic light curve fitting process with a spectra extracted with 6.5 pixels against our nominal 15 pixel aperture. The results are shown in \autoref{fig:aper}, and show that the resulting spectra are typically consistent to within $\sim 1 \sigma$ across the board, although the resulting median error bar size (given in the legend) do vary significantly depending on the aperture used. 

\bibliography{bib}{}
\bibliographystyle{aasjournal}



\end{document}